 \documentclass[useAMS,usenatbib]{mn2e}
\usepackage{psfig}
\usepackage{natbib}
\usepackage{amssymb}
\usepackage{longtable}
\usepackage[utf8]{inputenc}
\usepackage{supertabular}
\usepackage[maxfloats=20]{morefloats}[2012/01/28]

\newcommand{\be}{\begin{equation}}
\newcommand{\en}{\end{equation}}

\newcommand{\q}{$^{\dagger}$}

\def\zabs{$z_{\rm abs}$}
\def\zem{$z_{\rm em}$}
\def\lya{Ly$\alpha$}

\def\mgii{Mg~{\sc ii}}

\def\ni2{Ni~{\sc ii}}
\def\nv{N~{\sc v}}

\def\alii{Al~{\sc ii}}
\def\aliii{Al~{\sc iii}}
\def\civ{C~{\sc iv}}
\def\siiv{Si~{\sc iv}}

\def\nevii{Ne~{\sc vii}}
\def\mgx{Mg~{\sc x}}
\def\nax{Na~{\sc x}}
\def\si2{Si~{\sc ii}}
\def\feii{Fe~{\sc ii}}
\def\h1{H~{\sc i}}
\def\feiii{Fe~{\sc iii}}

\def\kms{km~s$^{-1}$}
\title[Time variability of LoBAL QSOs]{Variability in Low Ionization Broad Absorption Line Outflows}
\author[M. Vivek et al.]{M. Vivek$^{1}$\thanks{E-mail:vivekm@iucaa.ernet.in}, R. Srianand$^{1}$, P. Petitjean$^{2}$, V.Mohan$^{1}$, A. Mahabal$^{3}$ \& S. Samui$^{4}$\\
$^{1}$Inter-University Centre for Astronomy and Astrophysics, Post bag 4, Ganeshkhind, Pune 410007, India\\
$^{2}$UPMC-CNRS, UMR7095,  Institut $d '$Astrophysique de Paris, 98bis Boulevard Arago, 75014 Paris, France\\ 
$^{3}$Caltech, MC 249-17, Pasadena, CA 91125, USA \\
$^{4}$Department of Physics, Presidency University, 86/1 College Street, Kolkata 700073, India}
\begin{document}
\date{Accepted . Received ; in original form }
\pagerange{\pageref{firstpage}--\pageref{lastpage}} \pubyear{2002}
\maketitle

\label{firstpage}
\begin{abstract}
We present results of our time variability studies of \mgii\ and \aliii\ 
absorption lines in a sample of 22 Low Ionization Broad Absorption Line QSOs 
 (LoBAL QSOs) at 0.2 $\leq$ z$_{em}$ $\leq$ 2.1  using the 2m telescope at IUCAA
 Girawali Observatory  over a time-scale of 10 days to 7.69 years in the QSO's 
rest frame.  Spectra are analysed in conjunction with photometric light curves
 from Catalina Real-Time Transient Survey. Long time-scale (i.e $\ge$ 1 year) 
absorption line variability is seen in 8 cases (36\% systems) while only 4 of 
them (i.e 18\% systems) show variability over short time-scales (i.e  $<$ 1 year). 
We notice a tendency of highly variable  LoBAL QSOs to have high ejection velocity, 
low equivalent width and low redshift. The detection rate of variability in  LoBAL 
QSOs  showing Fe fine-structure lines  (FeLoBAL QSOs) is less than that seen in 
non-Fe  LoBAL QSOs. Absorption line variability is more frequently detected in 
QSOs having continuum dominated by Fe emission lines compared to rest of the QSOs. 
Confirming these trends with a bigger sample will give vital clues for understanding 
the physical distinction between different BAL QSO sub-classes. 
We correlate the absorption line variability with various parameters derived from 
continuum light curves and  find no clear correlation between continuum flux and 
absorption line variabilities. However, sources with large absorption line variability
 also show large variability in their light curves. We also see appearance/disappearance
 of absorption components  in 2 cases and clear indications for profile variations  
in 4 cases.  The observed variability can be best explained by a combination of 
process driven by continuum variations and clouds transiting across the line of sight.
\end{abstract}


\section{Introduction}
In recent years, the importance of understanding the nature and origin of QSO outflows  has become widely recognized in the context of  well
 studied  correlations between the Super Massive Black Hole (SMBH) and host galaxy properties \citep{silk98,ferrase00}. The inferred kinetic luminosities of the these outflows suggest that they could in principle be a major contributor to the  Active Galactic Nuclei (AGN) feedback mechanisms \citep{moe09,dunn10,bautista10,aoki11,borguet12,borguet13}. QSO outflows are also an important diagnostic of the physical conditions and chemical enrichment prevailing
 in the inner regions of the AGNs. 

QSO outflows are readily identified as  blue shifted broad absorption lines (BAL) seen in 10-40\%  of
 the total quasars \citep{hewett03,reichard03,trump06,allen11,paris12}. By definition, BAL QSOs are characterized by  absorption troughs which dip by more than 10\% below the continuum for 
a contiguous stretch of at least 2000 \kms and extend up to velocities $\sim$ 0.2 c with respect to the systemic redshift \citep{weymann91}.
BAL QSOs are further divided in three sub-types depending on the type of transition seen in absorption. Most of the known BAL QSOs are high ionization
 BALs (HiBALs) characterized by absorptions from \civ, \siiv, \nv, and \lya. About 15\% are Low ionization BALs  (LoBAL QSOs) which exhibit
  absorption lines of \mgii, \alii, \aliii\ and sometimes \feii\ and \feiii\ in addition to the above listed high ionization absorption lines.  
The rarer ($\sim$ 1\%) iron  LoBAL QSOs (FeLoBAL QSOs), also show absorption from excited fine-structure levels of \feii\ or \feiii. Recently thanks to the cosmic origin spectrograph on-board HST (COS/HST), outflows traced by very highly ionized species like \nevii, \mgx\ and \nax\ have been detected in $\sim$ 40\% of low-z QSOs \citep{muzahid12,muzahid13}.  The observed  BAL incidence is attributed to either an orientational effect where the observer's line of sight passes through the BAL clouds or an evolutionary phase in the QSO lifetime or a combination of the two. 

Even though recent spectroscopic surveys have dramatically increased the number of known BAL QSOs, still there is no commonly accepted  model which explains the geometry and the physical properties of the outflows. Line driven radiative  acceleration is suggested based on spectroscopic signatures \citep{arav94,arav95,anand02}. \citet{dekool95} proposed a model in which gas is clumped in small clouds confined by magnetic pressure and driven by radiation-pressure.  There is another dynamical model which assumes outflows as winds streaming out of the accretion-disk at a distance R $\sim$ 10$^{16}$ cm and driven by the radiative UV line pressure \citep{murray95,proga00,elvis00}. The proposed   "shielding gas"  absorbs the X-rays from the quasar and protects the wind from being completely ionized, also explains the observed X -ray weakness of BAL QSOs \citep[see,][]{green95,gallagher06,fan09,gibson09,stalin11}. 
 Most current models favor the equatorial geometry of the BAL outflows although after the discovery of radio loud BAL QSOs \citep{becker97,becker00}  polar outflows have been suggested for some of them \citep{ghosh07}. A two-component wind model  combining the polar and equatorial components have also been suggested to explain the polar outflows \citep{lamy04,proga04,borguet10}. The exact geometry of the BAL outflows is still under theoretical and observational investigation. 
 \begin{table*}
\caption{Log of observations}
\flushbottom
\begin{tabular}{|c|c|c|c|c|c|c|c|}
\hline
QSO&Instrument& Date&MJD&Exposure Time&$\lambda$ Coverage&Resolution&S/N$^a$\\
&&(DD-MM-YYYY)&&(mins)&($\AA$)&(kms$^{-1}$)&\\
\hline	
                    & SDSS        & 23-09-2000 & 51812 & 15x6 & 3800-9200 & 150 & 34\\
		    & SDSS        & 20-11-2000 & 51868 & 15x5 & 3800-9200 & 150 & 34 \\
		    & SDSS        & 01-12-2000 & 51879 & 15x4 & 3800-9200 & 150 & 34\\	
J004527.68+143816.1 & IGO/IFOSC 7 & 13-12-2007 & 54447 & 40x2 &	3800-6840 & 310 & 11\\
	            & IGO/IFOSC 7 & 18-12-2008 & 54818 & 40x2 &	3800-6840 & 310 & 19\\
                    & IGO/IFORS 1 & 22-12-2009 & 55187 & 35x3 & 3270-6160 & 370 & 12 \\
		    & IGO/IFOSC 8 & 01-01-2009 & 54832 & 45x3 & 5800-8350 & 240 & 17  \\
\hline 
		    & SDSS        & 06-09-2000 & 51793 & 15x3 & 3800-9200 & 150 & 30	\\ 
J014950.96-010314.1 & IGO/IFOSC 7 & 09-01-2008 & 54474 & 45x2 & 3800-6840 & 310 & 11      \\
          	    & IGO/IFOSC 7 & 23-12-2008 & 54823 & 45x5 & 3800-6840 & 310 & 12      \\
		    & SDSS        & 26-08-2010 & 55447 & 15x5 & 3800-9200 & 150 & 23\\
		    & IGO/IFOSC 7 & 13-02-2012 & 55970 & 45x3 & 3800-6840 & 310 & 11      \\
\hline
                    & SDSS        & 29-09-2000 & 51816 & 45x1 & 3800-9200 & 150 & 44    \\
J030000.57+004828.0 & SDSS        & 29-11-2000 & 51877 & 50x1 & 3800-9200 & 150 & 40    \\
 (FeLoBAL QSO)	    & SDSS        & 23-10-2001 & 52205 & 45x1 & 3800-9200 & 150 & 33     \\
                    & IGO/IFOSC 7 & 12-12-2007 & 54446 & 35x2 & 3800-6840 & 310 &  5     \\
                    & IGO/IFOSC 7 & 27-12-2008 & 54817 & 45x8 & 3800-6840 & 310 & 54      \\
\hline
                    & SDSS        & 15-01-2001 & 51924 & 80x1 & 3800-9200 & 150 & 27 \\
                    & IGO/IFOSC 8 & 14-12-2007 & 54448 & 45x3 & 5800-8350 & 240 & 10  \\
                    & IGO/IFOSC 7 & 09-01-2008 & 54474 & 45x3 & 3800-6840 & 310 & 23   \\
J031856.62-060037.7 & IGO/IFOSC 8 & 27-11-2008 & 54797 & 45x3 & 5800-8350 & 240 & 21    \\
    (FeLoBAL QSO)       & IGO/IFOSC 7 & 31-01-2009 & 54862 & 45x4 & 3800-6840 & 310 & 19     \\
                    & IGO/IFOSC 8 & 23-12-2009 & 55188 & 45x4 & 5800-8350 & 240 & 19    \\
                    & IGO/IFOSC 7 & 24-12-2009 & 55189 & 45x4 & 3800-6840 & 310 & 17    \\
\hline
		    & SDSS        & 01-01-2001 & 51910 & 15x3 &	3800-9200 & 150 & 33	\\
J033438.28-071149.0 & IGO/IFORS 1 & 20-01-2010 & 55216 & 45x4 & 3270-6160 & 370 & 28 \\ 
	            & IGO/IFORS 1 & 24-01-2012 & 55950 & 45x2 & 3270-6160 & 370 &  7\\
\hline
                    & SDSS	  & 29-11-2000 & 51877 & 15x3 & 3800-9200 & 150 & 37	\\ 
                    & IGO/IFOSC 7 & 21-12-2006 & 54090 & 30x2 & 3800-6840 & 310 & 6     \\
                    & IGO/IFOSC 7 & 04-12-2008 & 54804 & 45x3 & 3800-6840 & 310 & 29     \\
J073739.96+384413.2 & IGO/IFORS 1 & 24-12-2009 & 55189 & 45x3 & 3270-6160 & 370 & 34     \\
                    & IGO/IFOSC 8 & 05-12-2008 & 54805 & 45x4 & 5800-8350 & 240 & 32    \\
                    & IGO/IFOSC 8 & 31-01-2009 & 54862 & 45x3 & 5800-8350 & 240 & 34    \\
                    & IGO/IFOSC 8 & 25-02-2009 & 54887 & 45x3 & 5800-8350 & 240 & 28    \\
\hline
		    & SDSS        & 24-12-2001 & 52207 & 20x7 & 3800-9200 & 150 & 30      \\
J082319.65+433433.7 & SDSS        & 19-02-2001 & 51959 & 15x3 & 3800-9200 & 150 & 21      \\
                    & IGO/IFORS 1 & 13-12-2007 & 54447 & 40x3 & 3270-6160 & 370 &  7      \\
                    & IGO/IFOSC 7 & 06-12-2008 & 54806 & 45x3 & 3800-6840 & 310 & 20    \\
                    & IGO/IFOSC 7 & 19-04-2009 & 54575 & 45x2 & 3800-6840 & 310 & 2       \\
                    & IGO/IFOSC 7 & 25-01-2010 & 55221 & 45x3 & 3800-6840 & 310 & 8    \\

\hline
                    & SDSS        & 19-11-2001 & 52232 & 48x1 & 3800-9200 & 150 & 26   \\
J083522.77+424258.3 & IGO/IFORS 1 & 14-12-2007 & 54448 & 45x3 & 3270-6160 & 370 & 9     \\
     (FeLoBAL QSO)      & IGO/IFOSC 7 & 05-12-2008 & 54805 & 45x3 & 3800-6840 & 310 & 21   \\
                    & IGO/IFOSC 7 & 21-01-2010 & 55217 & 45x3 & 3800-6840 & 310 & 29   \\	
\hline
                    & SDSS        & 15-02-2002 & 52320 & 50x1 & 3800-9200 & 150 & 40   \\
                    & IGO/IFOSC 7 & 20-12-2006 & 54089 & 45x3 & 3800-6840 & 310 & 8    \\
J084044.41+363327.8 & IGO/IFOSC 7 & 12-12-2007 & 54446 & 45x3 & 3800-6840 & 310 & 23    \\
   (FeLoBAL QSO)        & IGO/IFOSC 7 & 07-12-2008 & 54807 & 45x4 & 3800-6840 & 310 & 26    \\
                    & IGO/IFOSC 7 & 17-12-2009 & 55182 & 45x6 & 3800-6840 & 310 & 50    \\
\hline
                    & SDSS        & 27-11-2002 & 52605 & 15x6 & 3800-9200 & 150 & 34    \\
J085053.12+445122.4 & IGO/IFORS 1 & 19-01-2010 & 55215 & 45x5 & 3270-6160 & 370 & 31     \\
                    & IGO/IFORS 1 & 12-03-2011 & 55632 & 45x5 & 3270-6160 & 370 & 16     \\ 
\hline
%
%
%
                    & SDSS        & 12-03-2003 & 52710 & 15x3 &	3800-9200 & 150	& 45    \\ 
J094443.13+062507.4 & IGO/IFOSC 7 & 03-04-2008 & 54559 & 40x3 & 3800-6840 & 310 & 45    \\
		    & IGO/IFOSC 7 & 30-01-2009 & 54861 & 45x5 & 3800-6840 & 310 & 60    \\
\hline
\end{tabular}
 \begin{flushright}
 \it{ Continued on next page} \\
      \end{flushright}
 \label{log}
\end{table*}
\setcounter{table}{0}
 \begin{table*}
\caption{{ Continued:} Log of observations}
\flushbottom
\begin{tabular}{|c|c|c|c|c|c|c|c|}
\hline
QSO&Instrument& Date&MJD&Exposure Time&$\lambda$ Coverage&Resolution&S/N$^a$\\
&&(DD-MM-YYYY)&&(mins)&($\AA$)&(kms$^{-1}$)&\\
\hline	
\hline
                    & IGO/IFOSC 7 & 24-01-2010 & 55220 & 45x2 & 3800-6840 & 310 & 28    \\ 	 
J094443.13+062507.4 & IGO/IFOSC 7 & 11-04-2010 & 55297 & 45x2 & 3800-6840 & 310 & 20    \\
                    & IGO/IFOSC 7 & 14-03-2011 & 55634 & 45x4 & 3800-6840 & 310 & 50    \\
\hline
                    & SDSS        & 30-12-2000 & 51908 & 15x3 &	3800-9200 & 150	& 27    \\
                    & IGO/IFOSC 7 & 03-04-2008 & 54559 & 40x3 & 3800-6840 & 310 & 22    \\
J095232.21+025728.3 & IGO/IFOSC 7 & 25-02-2009 & 54887 & 45x3 & 3800-6840 & 310 & 25    \\
                    & IGO/IFOSC 7 & 25-01-2010 & 55221 & 45x3 & 3800-6840 & 310 & 10    \\ 	 
                    & IGO/IFOSC 7 & 15-03-2011 & 55635 & 45x5 & 3800-6840 & 310 & 19    \\
\hline 
                    & SDSS        & 12-04-2002 & 52376 & 15x3 &	3800-9200 & 150	& 23    \\	 
J101053.98+451817.0 & IGO/IFORS 1 & 09-01-2008 & 54474 & 45x3 & 3270-6160 & 370 & 14     \\
                    & IGO/IFORS 1 & 26-02-2009 & 54888 & 45x3 & 3270-6160 & 370 & 17     \\
                    & IGO/IFORS 1 & 24-01-2010 & 55220 & 45x4 & 3270-6160 & 370 & 18     \\
\hline
                    & SDSS        & 03-01-2003 & 52642 & 15x3 &	3800-9200 & 150	& 24   \\
J112822.42+482310.0 & IGO/IFOSC 7 & 22-01-2010 & 55218 & 45x3 & 3800-6840 & 310 & 23   \\ 	 
                    & IGO/IFOSC 7 & 14-03-2011 & 55634 & 45x4 & 3800-6840 & 310 & 22   \\
\hline
                    & SDSS        & 04-04-2002 & 52368 & 15x3 &	3800-9200 & 150	& 25    \\ 
J114340.96+520303.3 & IGO/IFORS 1 & 08-01-2008 & 54473 & 40x2 & 3270-6160 & 370 & 11     \\
                    & IGO/IFORS 1 & 28-01-2009 & 54859 & 45x2 & 3270-6160 & 370 & 13     \\
\hline
                    & SDSS        & 25-04-2001 & 52024 & 15x6 &	3800-9200 & 150	& 62    \\
J120813.42+023015.1 & IGO/IFOSC 7 & 23-01-2010 & 55219 & 40x3 & 3800-6840 & 310 & 57  \\
                    & IGO/IFOSC 7 & 13-03-2011 & 55633 & 45x3 & 3800-6840 & 310 & 42   \\ 
\hline
                    & SDSS        & 28-04-2000 & 51662 & 15x3 & 3800-9200 & 150 & 27\\
                    & SDSS        & 15-02-2001 & 51955 & 15x7 &	3800-9200 & 150	& 40     \\
                    & IGO/IFOSC 7 & 03-04-2008 & 54559 & 40x2 & 3800-6840 & 300 & 11\\
J133356.02+001229.1 & IGO/IFORS 1 & 26-02-2009 & 54888 & 45x3 & 3270-6160 & 360 & 21\\
                    & IGO/IFOSC 7 & 26-03-2009 & 54916 & 45x5 & 3800-6840 & 300 & 15\\
                    & IGO/IFORS 1 & 22-01-2010 & 55218 & 45x5 & 3270-6160 & 360 & 26\\
                    & IGO/IFOSC 7 & 06-04-2011 & 55657 & 45x3 & 3800-6840 & 300 & 19\\
\hline
                    & SDSS        & 01-06-2002 & 52426 & 15x3 & 3800-9200 & 150 & 30  \\
J133428.06-012349.0 & IGO/IFOSC 7 & 04-04-2008 & 54560 & 40x3 & 3800-6840 & 310 & 18  \\
                    & IGO/IFOSC 7 & 23-12-2008 & 54823 & 45x5 & 3800-6840 & 310 & 30   \\
\hline
                    & SDSS        & 21-04-2001 & 52026 & 15x8 &	3800-9200 & 150	& 53    \\
J144842.45+042403.1 & IGO/IFORS 1 & 01-04-2010 & 55287 & 45x2 & 3270-6160 & 370 & 9     \\
                    & IGO/IFOSC 8 & 22-03-2010 & 55277 & 45x2 & 5800-8350 & 240 & 12    \\
                    & IGO/IFOSC 7 & 15-03-2011 & 55635 & 45x3 & 3800-6840 & 310 & 22  \\
\hline
                    & SDSS        & 05-05-2003 & 52764 & 15x3 &	3800-9200 & 150	& 33    \\
J161425.17+375210.7 & IGO/IFOSC 7 & 03-04-2008 & 54559 & 40x1 & 3800-6840 & 310 & 19    \\
                    & IGO/IFOSC 7 & 11-04-2010 & 55297 & 45x3 & 3800-6840 & 310 & 25    \\ 	 
                    & IGO/IFOSC 7 & 07-04-2011 & 55658 & 45x1 & 3800-6840 & 310 & 27    \\
\hline

                    & SDSS        & 04-09-2000 & 51804 & 53x1 & 3800-9200 & 150 & 41  \\
                    & VLT/FORS1   & 20-09-2003 & 52902 &  5x1 & 3300-11000& 370 & 62  \\
J221511.93-004550.0 & IGO/IFOSC 7 & 31-10-2008 & 54783 & 45x4 & 3800-6840 & 310 & 38  \\
   (FeLoBAL QSO)        & IGO/IFOSC 8 & 01-11-2008 & 54784 & 45x4 & 5800-8350 & 240 & 39  \\
		    & IGO/IFORS 1 & 04-01-2010 & 55213 & 45x8 & 3270-6160 & 370 & 9   \\
                    & MagE        & 10-08-2010 & 55431 & 15x1 & 3100-10000&  70 & 16  \\
\hline
                    & SDSS        & 12-10-2002 & 52559 & 15x3 &	3800-9200 & 150	& 28    \\
                    & SDSS        & 19-08-2007 & 54331 & 15x5 &	3800-9200 & 150	& 38    \\
J234711.44-103742.4 & IGO/IFOSC 7 & 31-10-2008 & 54770 & 45x3 & 3800-6840 & 310 & 19   \\
                    & IGO/IFOSC 8 & 21-11-2008 & 54791 & 45x4 & 5800-8350 & 240 & 19    \\
                    & IGO/IFOSC 7 & 15-12-2009 & 55180 & 35x2 & 3800-6840 & 310 &  8  \\

\hline
\hline
\end{tabular}
 \begin{flushleft}
 \it{ $^a$ The values  quoted are mean of S/N calculated per pixel over the wavelength range 5800 \AA - 6200 \AA.} 
      \end{flushleft}
 \label{log}
\end{table*}

As the launching radii of the winds in some models are very close to the QSO central engine, instabilities in the accretion disk can result in variations in the outflow properties on time-scales of a few years. Thus, BAL variability studies are important to understand the location and physical conditions in the absorbing gas and the physical mechanisms responsible for these outflows.  BAL troughs are, by and large, not observed to be varying in velocity  at a significant level \citep{wampler95,vilk01,rupke02}. However, there are several reported cases  of BAL troughs varying in strength:  
Q1303+308 \citep{foltz87}, Q1413+113 \citep{turnshek88}, 
Q1246-057 \citep{smith88}, UM 232 \citep{barlow89}, 
QSO CSO 203 \citep{barlow92}, Tol 1037-270 \citep{anand01},
J1054+0348 \citep{hamann08}
 FBQS J1408+3054 \citep{hall11} and SDSS 1333+0012\citep{vivek12}. 

Apart from these individual reports, there are six studies of BAL variability which looked for variations in a BAL QSO sample. \citet{barlow94} studied a sample of 23 BAL QSOs covering time-scales of $\Delta$t$_{rest}$$\leq$ 1 yr, reported optical depth variability in 15 of them and proposed  the variable photo-ionizing continuum as the primary cause for these variations.  \citet{lundgren07} analyzed a sample of 29 BAL QSOs with two-epoch  observations in Sloan Digital Sky Survey (SDSS) with similar time-scales and reported similar variabilities.  \citet{gibson08} studied a sample of 13 BAL QSOs over a longer time-scale ( $\Delta$t$_{rest}$$\sim$ 3-6 yrs). Their study found no evidence  for photoionization driven variability and  indicated variations produced by changes in outflow geometry. \citet{gibson10}, using a sample of  9 BAL QSOs with multi-epoch data, reported that the scatter in the change of absorption equivalent width, $\Delta$W, increases with the elapsed time between two epochs of observations. \citet{capellupo11} studied the variability trends in the short-term ($\le$1~yr in the QSO rest frame) and long-term ($\ge$ 1 yr in the QSO rest frame) data  for a sample of 24 BAL QSOs and reported variations in 39\% and 65\% of the quasars respectively. \citet{capellupo12} compared the variabilities in the \civ\ and \siiv\ lines for the above sample and found that \siiv\ BALs are more likely to vary than \civ\ BALs. \citet{filiz12} studied the multi-epoch observations of 582 quasars in SDSS and reported 21 examples of BAL trough disappearance in 19 quasars. Very recently, \citet{filiz13} studied the variability of 428 \civ\ and 235 \siiv\ BAL troughs identified in multi-epoch SDSS observations and reported a variability fraction of  60\%. 
In all these studies, BALs are found not to vary monotonically over time.  It is  also found that  components at higher outflow velocities are more likely to vary than those at lower velocities, and weaker BALs are more likely to vary than stronger BALs. A co-ordinated variability between absorption lines at different velocities are interpreted as  due to changing ionizing radiations whereas variability in  limited portions of broad troughs are identified as movement of individual clouds across our line-of-sight.  Most of the studies favor the movement of clouds across the continuum source over  the changes in the ionizing continuum. The actual scenario may be a complex combination of both.

Till date, all the BAL variability studies  have focused  mainly on variability in \civ\ and \siiv\ lines. In this paper, we study the nature of time variability in a sample of 22  LoBAL QSOs over the redshift range z$_{em}$$\sim$ 0.3-2.1 covering time-scales from   10 days to 7.69 years in the quasar rest frame. This work is the third paper in a series on  LoBAL QSO variability. The first paper, \citet{vivek12} reported the dynamical evolution of the Mg II BAL in SDSS 1333+0012. The second paper,\citet{vivek12a} studied the time variability of five  FeLoBAL QSOs and reported for the first time, significant variability in the strength of fine structure lines in the source SDSS 2215-0045. Our sample is drawn from the catalog of BAL QSOs from the SDSS 3$^{rd}$  data release \citep{trump06}. Typically each source, has 3-5 epoch spectroscopic data. Unlike all previous studies, we supplement the spectroscopic data with long-term photometric light curves. This enables us to discuss the connection between the absorption line variability with the continuum flux variability of the QSOs.
\newcommand{\ag}{$^{\dagger}$}
\newcommand{\ap}{$^{\ddag}$}

 \begin{table*}
\caption{ Source parameters measured from the data.   }
\flushbottom
\begin{tabular}{|l|c|c|c|c|c|c|c|c|c|c|c|}
\hline
Name		    & z$_{em}$&	z$_{abs}$&   V$_{max}$ &  $<w>$ &   L$_{Bol}$&    M$_{BH}$&   $\Delta$m &  $\sigma$($\Delta$m)&slope($\Delta$m/$\Delta$t)& Abs. line(s)\\
		    &	      &          &    (\kms)   & (\AA)  & (ergs s$^{-1}$)& (M$_{\odot}$)&(mag)&(mag)         &               ( mag/year) & probed                         \\
(1)		    	& (2)	& (3)  & (4)	  &  (5)	 & (6)			  & (7)      & (8)     &  (9)	&(10)	   & (11)  \\
\hline
\hline     
J0045$+$1438    	&1.992 & 1.948 & 5746.5   &   6.93	 & 3.9$\times$ 10$^{47}$   & 3.2E+09&  -0.04 &   0.04 & -0.032 $\pm$  0.007 & \aliii   \\
J0149$-$0103\ap   	&1.074 & 1.030 & 7910.7   &   5.60       & 8.4$\times$ 10$^{46}$   & 6.8E+08&  -0.01 &   0.13 & -0.046 $\pm$  0.007 & \mgii   \\
J0300$+$0048\ag		&0.568 & 0.864 & 6492.8   &     -        & 3.8$\times$ 10$^{45}$   & 3.0E+07&   0.01 &   0.04 & -0.003 $\pm$  0.024 & \mgii    \\
J0318$-$0600\ag		&1.968 & 1.941 & 7612.2   &  14.57       & 7.3$\times$ 10$^{46}$   & 5.9E+08&   0.07 &   0.05 &  0.075 $\pm$  0.021 & \mgii,\aliii    \\
J0334$-$0711 		&0.634 & 0.600 & 7537.6   &   6.50       & 4.3$\times$ 10$^{46}$   & 3.5E+08&     -  &    -   &          -          & \mgii  \\
J0737$+$3844\ap	  	&1.400 & 1.374 & 5000.2   &   7.16       & 1.8$\times$ 10$^{47}$   & 1.5E+09&  -0.07 &   0.06 & -0.068 $\pm$  0.004 & \mgii,\aliii\\
J0823$+$4334 		&1.660 & 1.620 & 11045.0  &   5.44$^a$   & 2.2$\times$ 10$^{47}$   & 1.8E+09&  -0.03 &   0.04 & -0.036 $\pm$  0.009 & \aliii    \\
J0835$+$4242\ag		&0.810 & 0.769 & 1865.9   &   4.96       & 3.9$\times$ 10$^{46}$   & 3.2E+08&  -0.04 &   0.07 & -0.005 $\pm$  0.008 & \mgii    \\
J0840$+$3633\ag		&1.230 & 1.225 & 3880.8   &  32.13       & 8.1$\times$ 10$^{45}$   & 6.6E+07&   0.03 &   0.04 &  0.015 $\pm$  0.008 & \mgii,\aliii   \\
J0850$+$4451 		&0.541 & 0.524 & 5224.1   &  17.22       & 2.1$\times$ 10$^{46}$   & 1.7E+08&   0.01 &   0.05 &  0.014 $\pm$  0.006 & \mgii    \\
J0944$+$0625\ap	   	&0.695 & 0.653 & 8507.7   &  15.55       & 1.3$\times$ 10$^{47}$   & 1.0E+09&   0.04 &   0.04 &  0.010 $\pm$  0.004 & \mgii    \\
J0952$+$0257 		&1.355 & 1.301 & 18209.2  &  16.99       & 1.5$\times$ 10$^{47}$   & 1.2E+09&   0.03 &   0.06 & -0.002 $\pm$  0.007 & \mgii,\aliii   \\
J1010$+$4518\ap   	&1.776 & 1.705 & 12910.7  &   7.21$^a$   & 1.6$\times$ 10$^{47}$   & 1.3E+09&  -0.02 &   0.07 & -0.014 $\pm$  0.012 & \aliii    \\
J1128$+$4823 		&0.543 & 0.536 & 5895.8   &  11.82       & 1.5$\times$ 10$^{46}$   & 1.2E+08&  -0.01 &   0.06 &  0.010 $\pm$  0.009 & \mgii    \\
J1143$+$5203\ap	   	&1.816 & 1.788 & 18209.2  &  11.64$^a$   & 2.9$\times$ 10$^{47}$   & 2.3E+09&   0.06 &   0.11 &  0.041 $\pm$  0.016 & \aliii    \\
J1208$+$0230\ap   	&1.180 & 1.067 & 17836.1  &   4.37       & 3.8$\times$ 10$^{47}$   & 3.1E+09&  -0.04 &   0.03 & -0.032 $\pm$  0.004 & \mgii    \\
J1333$+$0012 		&0.918 & 0.811 & 20746.5  &   7.04       & 9.4$\times$ 10$^{46}$   & 7.6E+08&   0.06 &   0.08 &  0.008 $\pm$  0.005 & \mgii    \\
J1334$-$0123\ap   	&1.876 & 1.833 & 18507.7  &  23.66$^a$   & 2.6$\times$ 10$^{47}$   & 2.1E+09&   0.07 &   0.06 &  0.058 $\pm$  0.012 & \mgii,\aliii    \\
J1448$+$0424 		&1.546 & 1.516 & 5821.1   &   8.02       & 3.8$\times$ 10$^{47}$   & 3.1E+09&  -0.01 &   0.03 & -0.012 $\pm$  0.010 & \mgii,\aliii    \\
J1614$+$3752 		&0.553 & 0.525 & 6492.8   &   3.43       & 3.8$\times$ 10$^{46}$   & 3.1E+08&  -0.04 &   0.08 & -0.026 $\pm$  0.004 & \mgii    \\
J2215$-$0045\ag \ap	&1.478 & 1.360 & 18582.3  &  21.76       & 2.8$\times$ 10$^{47}$   & 2.3E+09&  -0.01 &   0.19 & -0.123 $\pm$  0.009 & \mgii,\aliii    \\
J2347$-$1037    	&1.800 & 1.711 & 15522.6  &  27.46       & 2.6$\times$ 10$^{47}$   & 2.1E+09&  -0.07 &   0.06 & -0.072 $\pm$  0.008 & \aliii    \\
\hline
\hline
\end{tabular}
 \begin{flushleft}
 { Notes: Col. 1: QSO name, $^{\dagger}$ FeLoBAL QSO sources, $^{\ddag}$ Sources with strong Fe emission.; Col. 2: z$_{em}$ is obtained from the fitting of SDSS composite.; Col. 3: z$_{abs}$ corresponds to the maximum optical depth.; Col. 4: V$_{max}$ is calculated for the \mgii\ line from the normalized SDSS spectra. V$_{max}$ is identified as the maximum velocity at which source flux matches with the continuum.; Col. 5: Average \mgii\ equivalent widths, $^a$ corresponds to average \aliii\  equivalent widths.; Col. 6: L$_{Bol}$ is computed using the prescription, L$_{Bol}$ = 7.9 $\times \nu$F$_{\nu B }$,  of \citet{marconi04}.;  Col. 7: M$_{BH}$ is derived from L$_{Bol}$ assuming Eddington accretion.; Col. 8:  median $\Delta$m  values.; Col. 9: standard deviation of $\Delta$m values.; Col. 10: slope of $\Delta$m-$\Delta$t graph.; Col. 11: absorption line(s) probed in this study.}\\
  \end{flushleft}
 \label{source_param}
\end{table*}

The manuscript is arranged as follows. In Section 2, we provide details 
of our sample, observations and data reduction. Section 3 describes the 
BAL variability measurements and Section 4 gives the summary of variability in individual sources.  In Section 5, we explore statistical correlations
between absorption line variability and other measurable parameters 
from spectroscopy and photometric light curves. Discussions and summary
of our results are presented in Section 6. Notes on individual variable BAL sources, Light curves  and spectra of all the objects in our data are presented in the Appendix.   
\begin{figure*}
 \centering
\psfig{figure=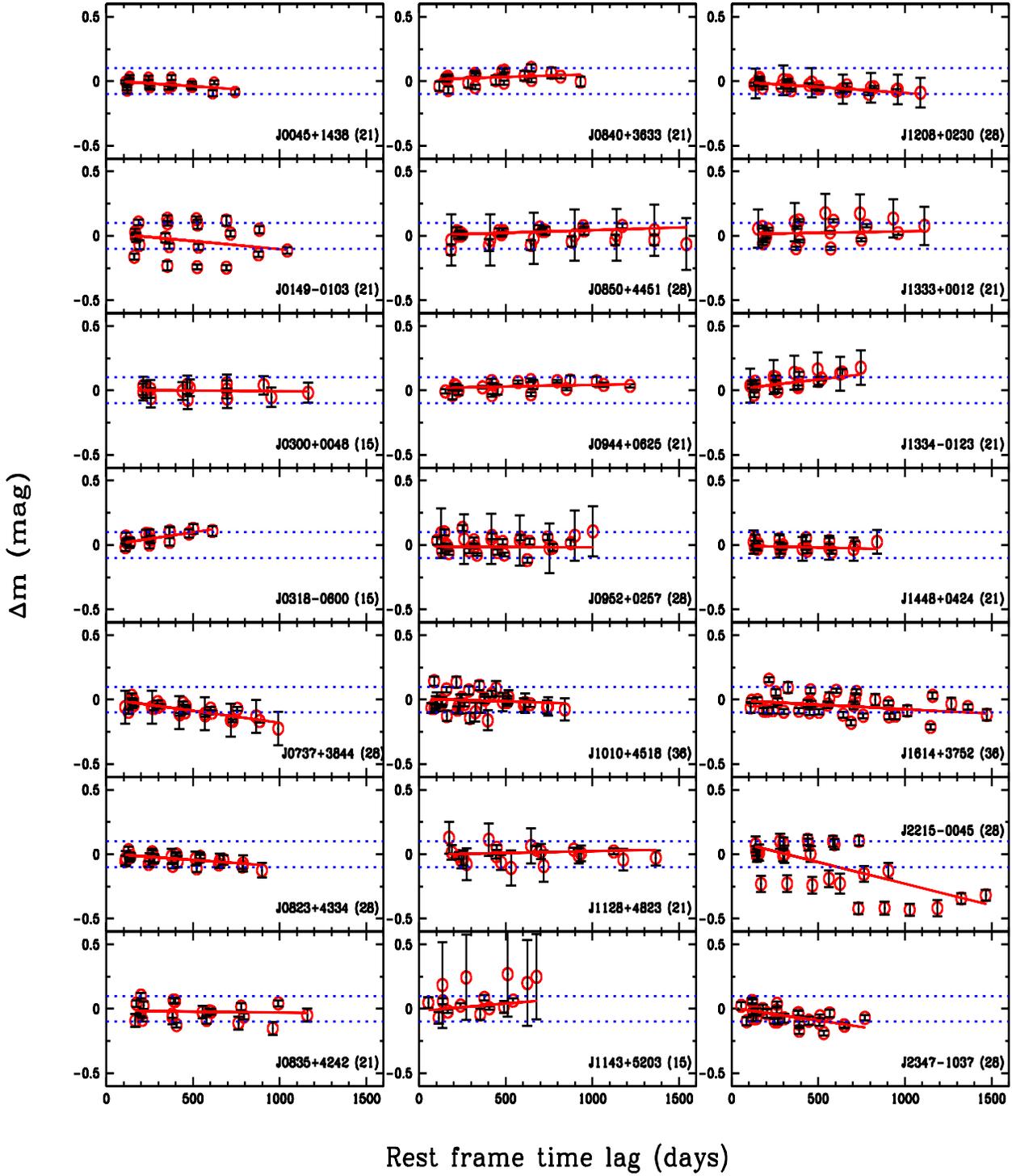,width=1.0\linewidth,height=1.2\linewidth,angle=0}
\caption{ Results of long-term photometric monitoring of sources in our sample from CRTS. The x-axis represents rest-frame pairwise time lags  and the y-axis represents pairwise magnitude differences for each source measured over different time lags.  The  red/grey line marks the least square fit to the data. The number in  parenthesis indicate the number of independent photometric measurements available for the source. { Blue dotted horizontal lines correspond to magnitude variations of 0.1 magnitude. We consider those objects as variable for which $\sigma$($\Delta$m) $>$ 0.1 magnitudes. Variations above this value are seen in the case  of  SDSS J0149-0103 and SDSS J2215-0045. In the case of SDSS J1143+5203, large photometric variations are seen albeit large photometric errors.}}
\label{lc}
\end{figure*}

\section{Observation and Data Reduction}
Our  LoBAL QSO sample  consists of 22 QSOs brighter than  i = 17.5 mag  that are accessible from IUCAA Girawali Observatory (IGO).  Five of these sources show broad Fe II absorption in the resonance lines and in the excited fine-structure lines.

All the new observations presented here were carried out using the 2m telescope at IUCAA Girawali Observatory (IGO).
The spectra were obtained using the IUCAA Faint 
Object Spectrograph (IFOSC). We have been observing the sample  from the year 2006 with the aim of studying the time variability in the BALs over a range of time-scales.   The detailed log of these observations together 
with those of the archival SDSS data and the data from the literature are 
given in Table~\ref{log}. Basic properties of QSOs in our sample are summarised in Table~\ref{source_param}. Spectra were obtained mainly 
using three grisms, Grism 1, Grism 7 and Grism 8 of IFOSC{\footnote { Details of the IGO/IFOSC grisms can be found at http://www.iucaa.ernet.in/$\sim$itp/etc/ETC/help.html\#grism.}} in combination  
with 1.5 arcsec slit. These  combinations cover the wavelength ranges  
3270 - 6160~\AA, 3800 - 6840~\AA\ and 5800 - 8350~\AA\ 
for the above three grisms respectively. Fringing poses a serious issue for Grism 8 observations.  We overcame this by subtracting two 2D science exposures
for which the object was placed at two different locations along the slit \citep[see ][]{vivek09}. Typically  observations 
were split in to exposures of 45 minutes. All the raw frames were 
processed using standard IRAF{\footnote {IRAF is distributed by the 
National Optical Astronomy Observatories, which are operated by the 
Association of Universities for Research in Astronomy, Inc., 
under cooperative agreement with the National Science Foundation.}} 
tasks. One dimensional spectra were extracted from the frames using the  
``doslit'' task in IRAF. We opted for the variance-weighted extraction 
as compared to the  normal one. Wavelength calibrations were performed 
using standard helium neon lamp spectra and flux calibrations were done 
using a standard star spectrum observed on the same night. Air-to-vacuum 
conversion was applied before co-adding the spectra. Individual spectra obtained with in few days, were 
combined using 1/$\sigma^2$ weighting in each pixel after scaling the 
overall individual spectrum to a common flux level within a sliding 
window. The error spectrum was computed taking into account proper 
error propagation during the combining process. Along with IGO data, we also used all the available spectra of the  sample from SDSS database.

   We  also obtained the light curves for all the sources in our sample (except for the source SDSS J0334-0711) in Johnson's V  magnitude from the Catalina Real-Time Transient Survey
\citep[CRTS;][]{Drake09} to probe the variations in the ionizing continuum. The data used in  these light curves are taken between 
April 2005 to July 2010. More discussions on light curve variabilities are given in Section 5.6.  
  The light curve for all the sources  are given in Figs.~\ref{lc_1},\ref{lc_2},\ref{lc_3}\&\ref{lc_4} in the  appendix. CRTS operates with an unfiltered set up and the resulting magnitudes are converted to V magnitudes using the transformation  equation V=V$_{ins}$+a(v)+b(v)*(B-V), where, $V_{ins}$ is the  observed open magnitude, a(v) and  b(v) are the zero point and the slope. The zero point and slope are obtained from three
 or more comparison stars in the same field with the zero point typically being of the order of 0.08. CRTS provides
 four such observations taken 10 minutes apart on a given night. Since we are mainly interested in the long-term variability,  we have averaged these four
 points (or less if one or more of those coincided with bad areas) to get the light curves.
 We also obtained the V magnitudes at the SDSS epochs by convolving the SDSS spectra with Johnson's V band filter function. Comparison of these V magnitudes at SDSS epochs with the CRTS magnitudes will allow us to probe the long-term continuum flux variability. 
%
\begin{figure}
 \centering
\psfig{figure=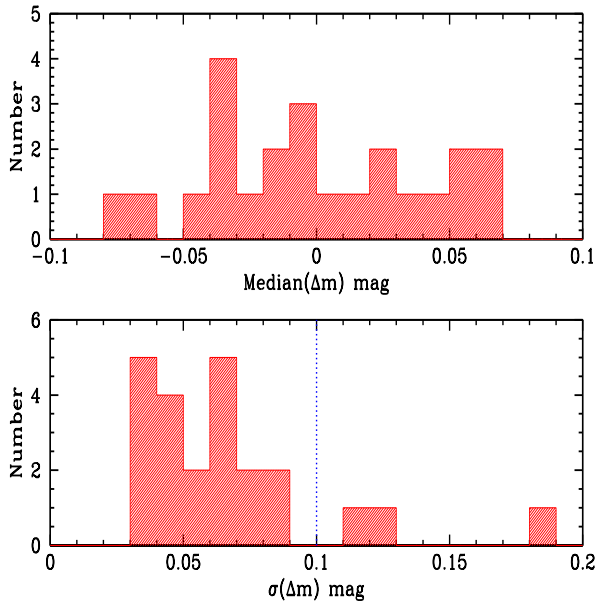,width=1.0\linewidth,height=1.0\linewidth,angle=270}
\caption{  Histogram of the median magnitude differences(top panel) and the standard deviation of the magnitude differences $\Delta$m (bottom panel) for each source. {  Blue/Dotted vertical line corresponds to $\sigma(\Delta m) \sim$ 0.1.  Large photometric variations ($\sim$ 0.1 mag) are only seen in three sources, SDSS J0149-0103, SDSS J1143+5203, SDSS J2215-0045. However, in the case of SDSS J1143+5203, the photometric errors are also large (see Fig.~\ref{lc}). } }
\label{hist_med_std}
\end{figure}

To quantify the variations in the CRTS light curves, we carried out the following procedure. We calculated the differences in magnitudes ($\Delta$m) of all possible pairs of observations and their corresponding time differences ($\Delta$t). The median and standard deviation of the $\Delta$m values characterize the nature of variation in the sources.  A larger standard deviation would imply a source where the magnitudes have changed significantly.   We then fitted a least square minimized straight line to this $\Delta$m vs $\Delta$t data. In Fig.~\ref{lc}, we have plotted the $\Delta$m against the $\Delta$t and the red line marks the least square fit.  The slope of the fitted line together with the standard deviation of the $\Delta$m is used to quantify the continuum variability of the sources.  Columns 8,9 and 10 of Table~\ref{source_param} give the parameters of light curve variations. Fig.~\ref{hist_med_std} gives the histogram distribution of the median and standard deviation of the magnitude changes, $\Delta$m. 
 But large photometric variations i.e., $\sigma$($\Delta$m) $>$ 0.1 magnitudes are seen only in three sources. These sources which exhibit large continuum variations in the CRTS light curves are SDSS J0149-0103, SDSS J1143+5203 and SDSS J2215-0045. From Figs.~\ref{lc_1} \& \ref{lc_4} in appendix, it is clear that in the case of J0149-0103 and J2215-0045, the light curve shows very smooth variation. In the case of J1143+5203, even though large variations are evident in the light curve, the sampling is not that good and the associated photometric errors are large.  As can be seen from Fig.~\ref{lc}, for all  sources $\Delta$m increases or decreases steadily with large time separation, except in the case of SDSS J0149-0103 and SDSS J2215-0045 where there is larger scatter in the $\Delta$m values even with in a short time period. This means that the flux in these two sources is varying appreciably even at smaller time-scales.  Apart from these sources, SDSS J0318-0600, SDSS J0737+3844, SDSS J1334-0123 and SDSS J2347-1037 show larger values for the slope.  As can be seen from the  column 10 of Table~\ref{source_param}, the slopes are significant by more than 3$\sigma$ level.   Except for SDSS J0318-0600, SDSS J1143+5203 and SDSS J1334-0123, all other  sources have negative slopes which means that  with  larger time separation, these objects tend to become fainter.    

 Before we proceed further, we wish to bring up a caveat in using the optical light curves to probe the ionization variability of \feii, \mgii\ and \aliii\ lines. The photons responsible for ionizing Fe$^+$, Mg$^+$ and Al$^{2+}$ are in the far-UV range (i.e. $\lambda_{rest}<$912\AA). Recently, \citet{welsh11} have shown that QSOs in general show higher amplitude variability in FUV compared to NUV and optical wavelengths. Therefore, we need to be cautious while we try to correlate absorption line optical depth variability to the photometric variations seen in CRTS light curves. However, given that we have no detailed FUV light curves available for our objects, we will use the CRTS light curves keeping in mind the caveat described above.
\begin{figure}
 \centering
\psfig{figure=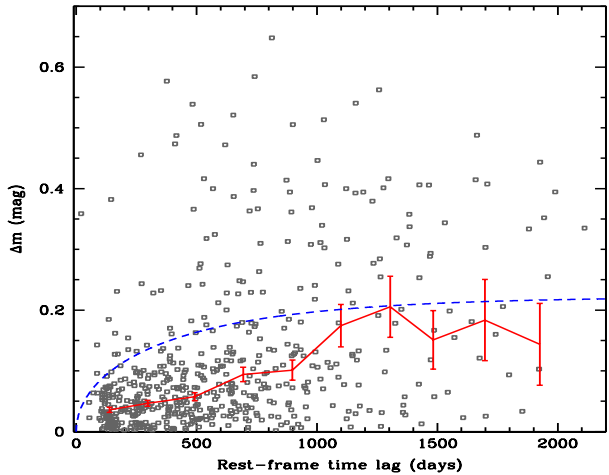,width=1.0\linewidth,height=0.8\linewidth,angle=0}
\caption{ Plot of $\Delta$m versus rest-frame time lags. Grey points represent the pairwise magnitude differences measured over different time lags. The red solid curve represents  the structure function (in mag) of the CRTS magnitude variations of LoBAL QSOs in our sample. The blue dashed line shows the best fit curve to the structure function for SDSS quasars in the wavelength range 4000\AA$< \lambda <$6000\AA\ given by \citet{macleod12}. Clearly, the variability increases over larger time lags. But, sources in our sample are more luminous and have smaller variability at smaller time-scales as compared to SDSS quasars.   }
\label{sf_plot}
\end{figure}

 Studies on QSO continuum light curves using structure function analysis report a monotonically increasing variability amplitude with longer time-scales \citep[for example, see][]{devries05}.  Recently, \citet{macleod12} studied the optical continuum variability of SDSS QSOs with repeated observations and reported that the observed structure function is completely consistent with a damped random walk model. { In their study the structure function is defined as,
\begin{eqnarray}
\centering
SF = 0.74*(IQR);\hspace{0.3cm}SF_{err} = SF*1.15/\sqrt{(N-1)}
\label{eq1} 
\end{eqnarray}
where IQR is the 25\%–75\% interquartile range of the $\Delta$m distribution  and N is the number of $\Delta$m values.} We also looked at the continuum magnitude variations with different time-scales.  Our sample span the rest-frame wavelength range 1540\AA $\leq \lambda_{rest} \leq$ 5430\AA, redshift range 0.2 $\leq$ \zem $\leq$ 2.1 and the luminosity range 4$\times$10$^{45}$ ergs s$^{-1}$ $\leq$ L$_{Bol}$ $\leq$ 4$\times$10$^{47}$ ergs s$^{-1}$ .    Fig.~\ref{sf_plot} shows the plot of absolute V band magnitude variations at different time lags. The red solid curve represents the structure function of the CRTS magnitude variability computed using \ref{eq1}. The blue dashed line shows the best fit curve to the structure function for QSO variability in the wavelength range 4000\AA$< \lambda <$6000\AA\ given by \citet{macleod12}. 
 The plot clearly agrees with the already known result of larger variations at larger time lags. But, the sources in our sample are less variable at shorter time lags  as compared to the non-BAL QSOs. { This could be related to the fact that we picked up bright LoBAL QSOs for our spectroscopic monitoring and bright QSOs tend to vary less and on longer time scales.}

\section{BAL variability measurements}

 We define two metrics to quantify the absorption line variability. First one is the   variation in the equivalent width between different epochs.    Equivalent widths are computed by  normalizing the spectra using  a spectral energy distribution (SED) in the optical-UV region, obtained by fitting the absorption/emission free regions with a template spectrum (see below).  Optical-UV SED is obtained by fitting the observed spectrum $f$($\lambda$) with the template spectrum $f_{\rm t}$($\lambda$) using the following parametrization,
\begin{equation}
f(\lambda) = \left[ a f_t(\lambda) + b \left( \frac{\lambda}{\lambda_0}\right)^{\alpha} \right] e^{-\tau_{\lambda}}
\end{equation}
We use the $\chi^2$ - minimization to get the best values for parameters  
a, b, $\alpha$ and $\tau_{\lambda}$. The second term in the above equation denotes the spectral index differences between the observed and the template QSO spectra. The dust optical depth $\tau_{\lambda}$ is obtained assuming 
a SMC like extinction curve. The fitting method is similar to that described by \citet{anand08}.  In the case of strong Fe emitters, we use the spectrum of a strong Fe emitting source SDSS 0923+5745 as the template \citep[see ][]{vivek12a}. In all other but two cases, we use the SDSS composite spectrum as the template \citep{richards02}.  For SDSS J0045+1438 and SDSS J2347-1037, the fitting did not converge to a good fit. So, we used a simple polynomial connecting line free regions to approximate the continuum. The computed equivalent widths are given in Table~\ref{EW_table}.

 Equivalent width measurements are sensitive to the correct estimation of the continuum. In the case of BAL quasars, continuum measurements are difficult as the spectra are dominated by broad absorption lines. We define the second metric, integrated ratio, to avoid the ambiguities in the continuum.  The ratio spectra for all the epochs with respect to a reference epoch (first epoch) are integrated over the  wavelength range covered by the BAL absorption to get the integrated ratio values. The computed integrated ratio values between different epoch spectra are given in Table~\ref{IR_table}. 

In one source, SDSS J0300+0048, the continuum is completely dominated by absorption lines. Hence, we did not compute the equivalent widths and integrated ratios in this case. But, it can be seen from Fig. 2 of \citet{vivek12a}, this source does not show  any variation in \mgii\ absorption. Hence, we include this source as a non-varying one in our analysis. Equivalent widths and integrated ratios were measured for the rest of the sample.

 We identified  sources showing absorption line variability as those having more than 3$\sigma$ variations in the equivalent widths and 5$\sigma$ in the integrated ratio values between any two epochs. As the equivalent width measurements are prone to uncertainties related to continuum placements, we do not include those sources which  have equivalent width variations, but show no significant changes in  ratios. 
\section{Summary of variability in individual sources}
 \begin{figure*}
 \centering
\begin{tabular}{c c}
\psfig{figure=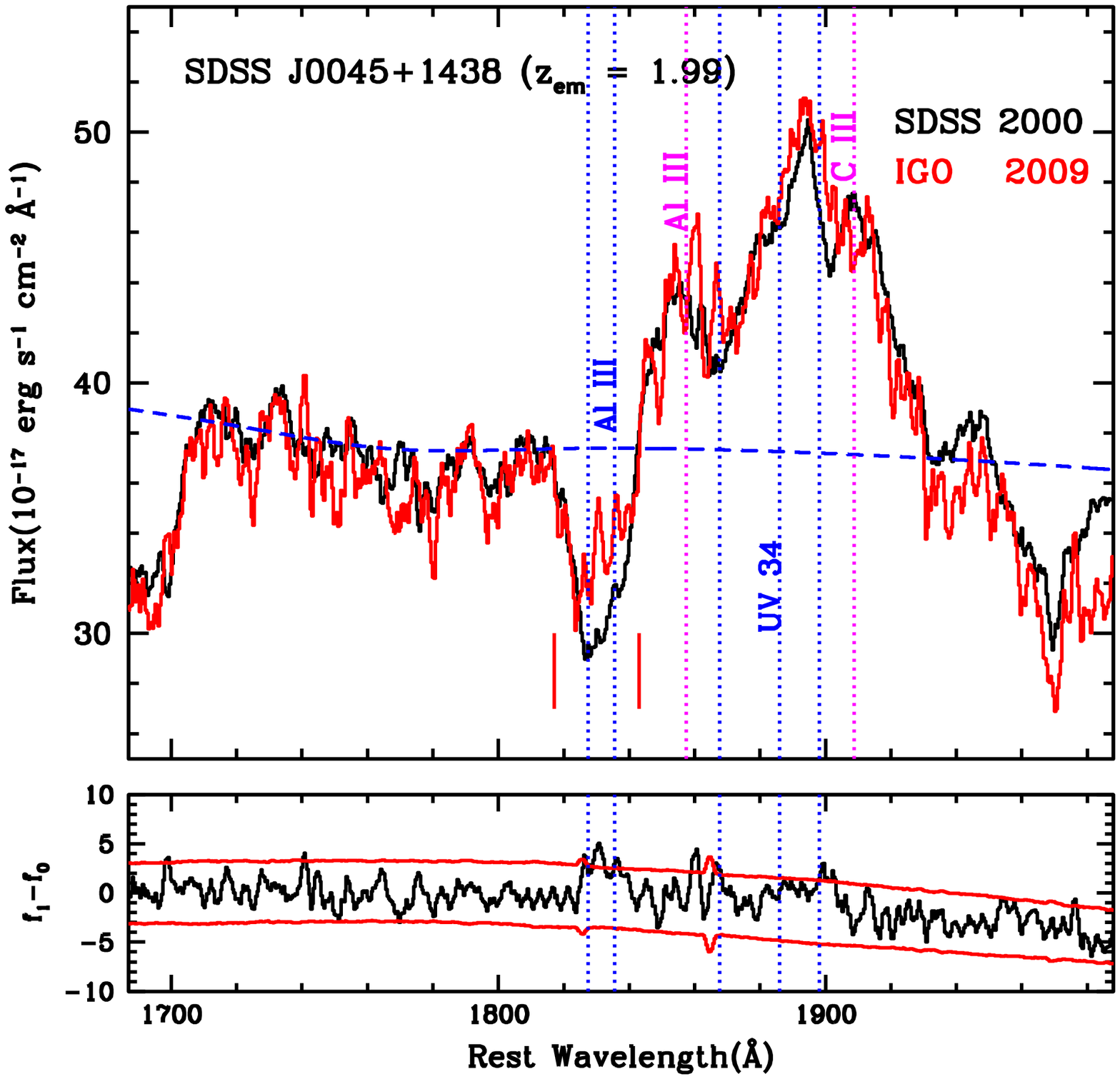,width=0.5\linewidth,height=0.4\linewidth,angle=0}&
\psfig{figure=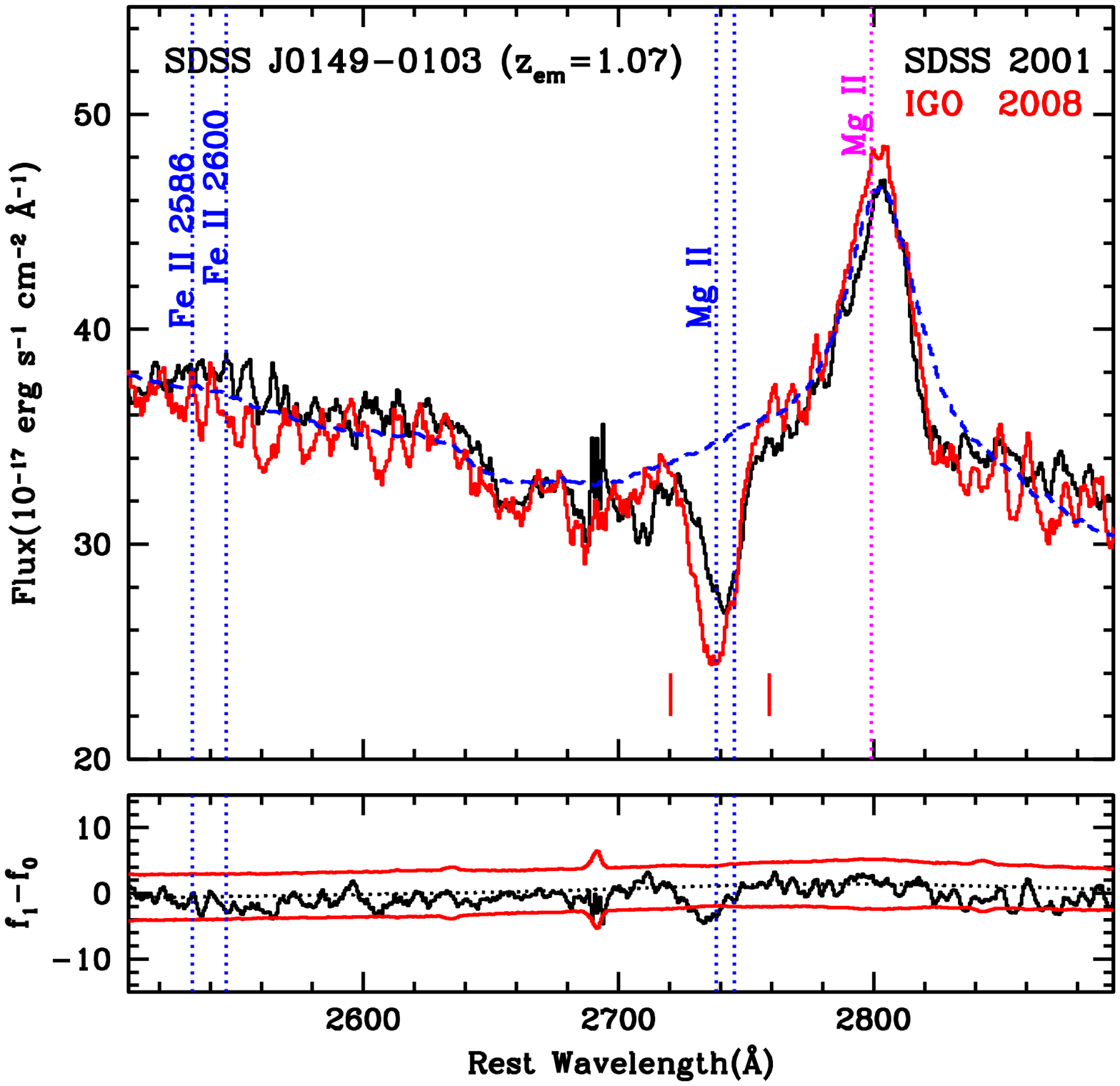,width=0.5\linewidth,height=0.4\linewidth,angle=0}\\
\psfig{figure=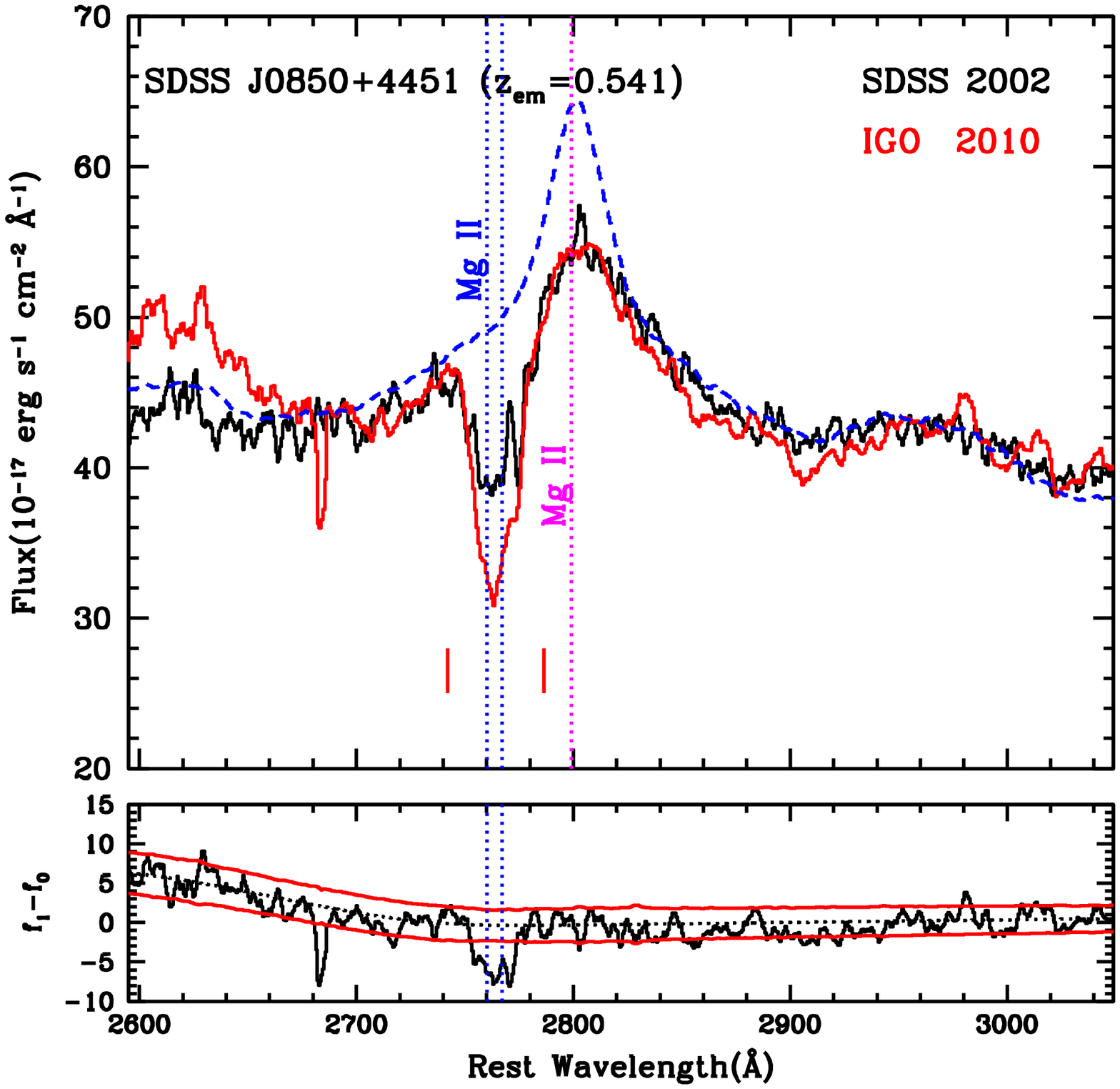,width=0.5\linewidth,height=0.4\linewidth,angle=0}&
\psfig{figure=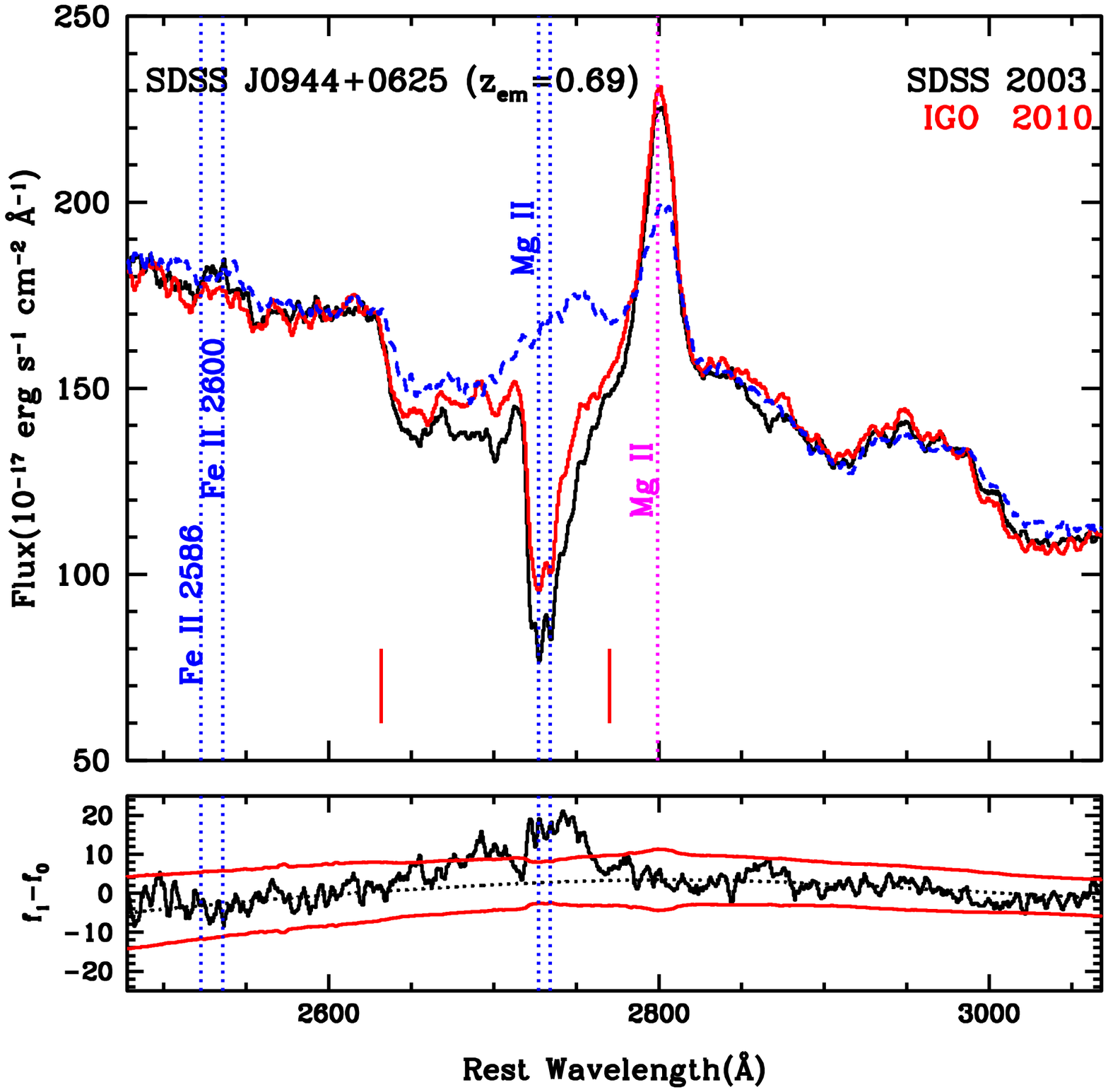,width=0.5\linewidth,height=0.4\linewidth,angle=0}\\
\psfig{figure=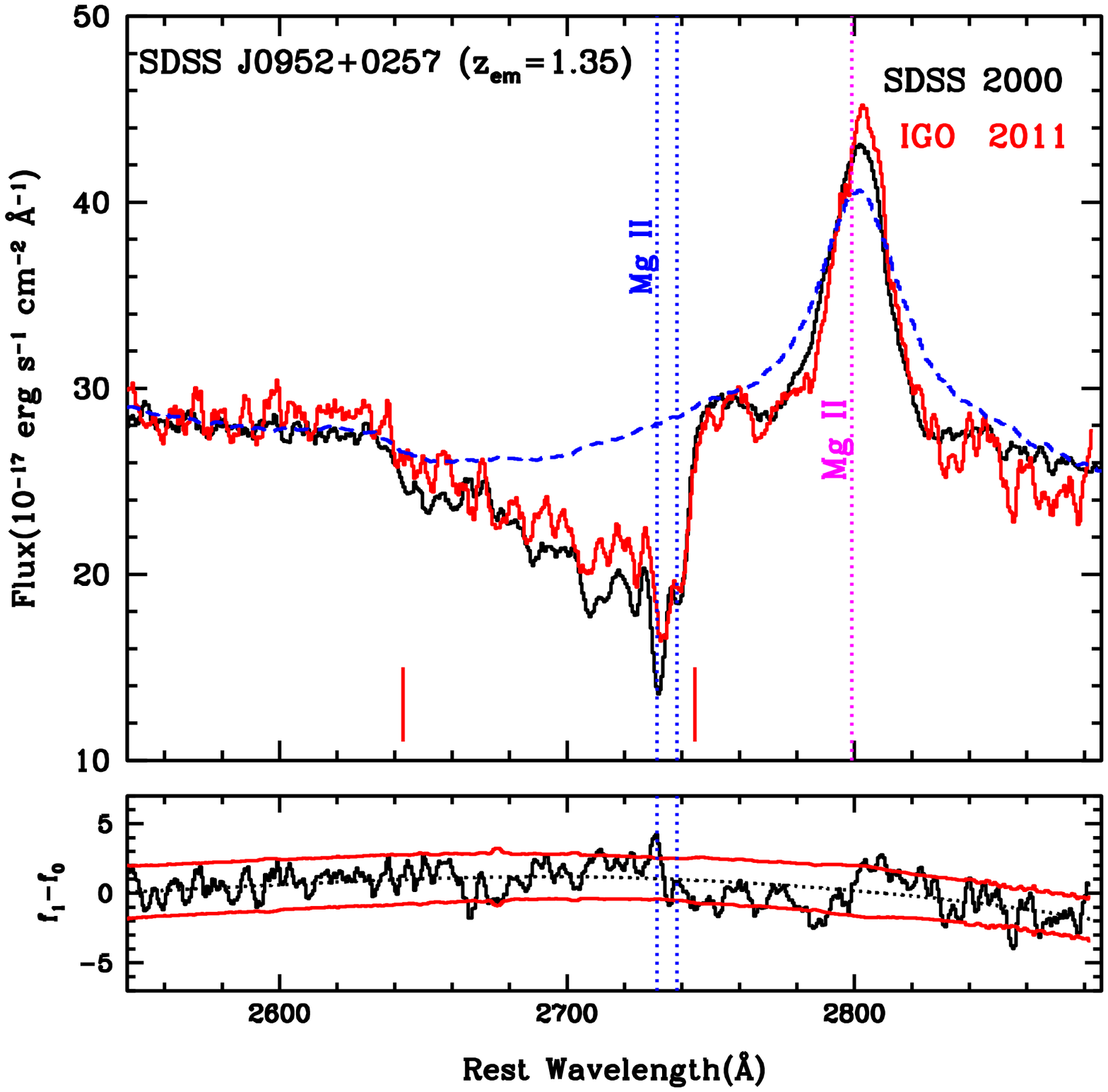,width=0.5\linewidth,height=0.4\linewidth,angle=0}&
\psfig{figure=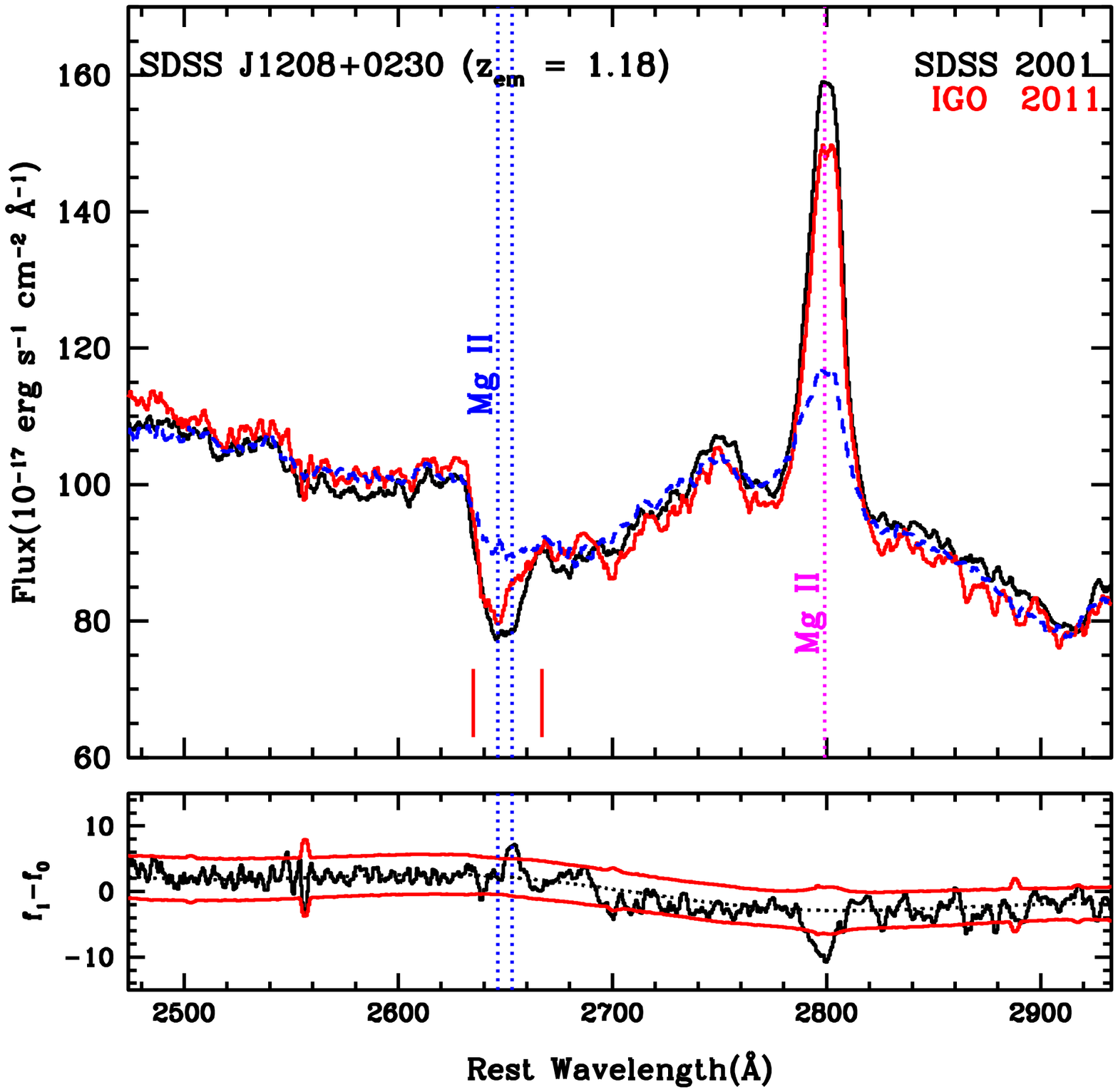,width=0.5\linewidth,height=0.4\linewidth,angle=0}\\

\end{tabular}
\caption{  Comparison of IGO and SDSS spectra for the varying BAL QSOs discussed in Section 4. The flux scale applies to the reference SDSS spectrum and all other spectra are scaled in flux to match the reference spectrum. Blue (long dashed)  spectra represent the continuum used for normalization.  The difference spectrum for the corresponding MJDs is plotted in the lower panel of each plot. 1$\sigma$ error is plotted above and below the mean. Magenta vertical lines (long dashed) mark the locations of emission  lines  and blue vertical lines (dotted) mark the locations of absorption lines (maximum optical depth). Two vertical red lines below the spectra mark the locations of V$_{max}$ and V$_{min}$.}
\label{zoom_pic}
\end{figure*}

\newcommand{\up}{$^{\uparrow}$}
\newcommand{\dn}{$^{\downarrow}$}
\newcommand{\dg}{$^{\it a}$}
\begin{table*}
\caption{Summary of the variation in equivalent width and continuum flux. {  Rest-frame variations occurring with in a year are termed as short time scale. The arrow indicates the direction of variation. \up indicates a positive change (more absorption) and \dn indicates a negative change (less absorption). }}
\begin{tabular}{|c|c|c|c|c|c|}
\hline
\multicolumn{1}{c}{} &Location of& \multicolumn{4}{c}{ Rest-frame variations detected?}\\
\multicolumn{1}{c}{Name}& absorption changes & \multicolumn{2}{c}{equivalent width} &\multicolumn{2}{c}{continuum flux} \\
			&(\kms)&Long time-scale	&Short time-scale&Long time-scale&Short time-scale\\
 \hline   
J0045$+$1438&-3650	& Yes \dn		& 	~~~Yes \dn	& Yes \up	& 	No		\\
J0149$-$0103&-7100	& Yes \up		&       ~~~Yes \dn	& Yes \up	&       ~~~Yes \up		\\
J0850$+$4451&-3490	& Yes \up		&       No		& Yes \dn	&       No		\\
J0944$+$0625&-7300	& Yes \dn		&       ~~~Yes \dn	& Yes \up	&       ~~~Yes \dn		\\
J0952$+$0257&-8500	& Yes \dn		&       No		& Yes \up	&       ~~~~Yes \up\dn		\\
J1208$+$0230&-15100	& Yes \dn		&       No		& Yes \up	&       No		\\
J1333$+$0012&-25500	& ~~~Yes \dn\up\dg	&       ~~~~Yes \up\dn	& Yes \dn	&       ~~~Yes \up		\\
J2215$-$0045&-13500	& Yes \dn		&       No		& Yes \up	&       No		\\
\hline
\end{tabular}
\label{vary_sum}
\begin{flushleft}
  \dg one component increased and another component decreased  \\
      \end{flushleft}
\end{table*}
Detailed distribution of  individual sources showing absorption line variations are given in the Appendix.

We detect BAL variability in 8 sources in our sample of 22 QSOs. All these sources show equivalent width variations at $>$3 $\sigma$ level and the ratio variations at $>$5$\sigma$ level. The summary of the comparison between the equivalent width and continuum flux variations over long and short time-scales is given in Table~\ref{vary_sum}.  Fig.~\ref{zoom_pic} shows the comparison of spectra showing maximum absorption line changes for 6 sources in our sample. Similar plots for the remaining sources, SDSS J1333+0012 and SDSS J2215-0045 are  already published in \citet{vivek12} and \citet{vivek12a} respectively.

Four sources, SDSS J0045+1438, SDSS J0149-0103, SDSS J0944+0625 and SDSS J1333+0012  show absorption line variations at both short ($<$ 1 year) and long ($>$1 year) time-scales.  Except in the case of SDSS J0045+1438, where the variations are seen in \aliii\ BAL, the remaining three sources show variations in the \mgii\ line.  In two cases, SDSS J0149-0103, SDSS J1333+0012, the profile variations are consistent with  dynamical changes  like movement of clouds across the line of sight causing the absorption line variations. In the case of SDSS J0045+1438, \aliii\ variation occurs without appreciable change in the continuum whereas coherent variations are seen in both  continuum and absorption lines for the other three sources.  The role of ionization induced variability cannot be dismissed in these three cases. We note that the shape of the light curves for these continuum varying sources has a curvature (see Appendix B).

The remaining four  sources, SDSS J0850+4451, SDSS J0952+0257, SDSS J1208+0230 and SDSS J2215-0045 show only long-term absorption line variations. The observed variations in SDSS J1208+0230 and SDSS J2215-0045 are again consistent with the dynamical changes in the clouds \citep{vivek12a}. Among these four sources, only SDSS J2215-0045 has shown significant coherent  changes in the continuum light curve.

Moreover, we also identify three sources, SDSS J0318-0600, SDSS J0737+3844 and SDSS J2347-1037, where coherent variations with more than 0.1  magnitudes are seen in the CRTS light curves (with a significant slope in $\Delta$m-$\Delta$t plot; see Fig.~\ref{hist_med_std}) without any absorption line variations (see Fig.~\ref{lc_1}). 
Overall, both the absorption lines and continuum have varied in several sources in the sample. But, there are enough indications to suggest that the two variations need not be related. We check this  in the next section.

\section{Statistical analysis of the variability of  LoBAL QSOs}

 In this section, we study the  correlations between absorption line variability and other properties of  low ionization BAL QSOs. For this purpose, we divide the sample in to several sub-groups.
 
\subsection{Absorption line variability in different groups of  LoBAL QSOs}
 The fraction of varying BALs in our overall sample is small (i.e. 36\%) as compared to \civ\ BALs from previous studies. \citet{capellupo11} used a slightly higher threshold (4$\sigma$) to define the variability. In our sample, only one source show variations less  than 4$\sigma$. A higher threshold  will again reduce the fraction of variable sources in our sample. There are eight BAL QSOs with emission redshift  0.5 $<$\zem$<$ 1.2 where only \mgii\ absorption lines are covered. Based on the absence of \feii\ absorption, we confirm that these  are not  FeLoBAL QSOs. However, the presence of \aliii\ absorption cannot be confirmed as we do not cover the expected wavelength range of \aliii\ absorption lines.  Five out of these eight systems (i.e., 63 \%) show optical depth variation. This fraction becomes 50\% if we include the two low redshift  FeLoBAL QSOs.  Among this 5, three systems show only long term absorption line variability. In the remaining systems, absorption line variability is seen over all time-scales. 

There are 12 sources that show detectable \mgii\ and \aliii\ absorption. They typically have redshift greater than 1.2.
 Out of these, three are  FeLoBAL QSOs. SDSS J2215-0045 is the only  FeLoBAL QSO source which show variations in \mgii\ and \aliii\ lines. 
 Only in four sources, SDSS J0737+3844, SDSS J0952+0257, SDSS J1334-0123 and SDSS J1448+0424,  both \mgii\ and \aliii\ lines are simultaneously covered. In the remaining sources, we use only the \aliii\ to study the absorption line variability.
Out of the nine  non-FeLoBAL QSOs in this sub-sample, two (i.e., 22\%) show variations in the optical depth. The percentage of  sources with variable BALs slightly increase to 25\% (3/12) when  FeLoBAL QSOs are also included. { These results suggest that \mgii\ BAL QSOs at \zem$<$1.2 seem to show larger variability compared to  high redshift \mgii\ BAL QSOs. We notice that the luminosity distribution of the two samples are nearly identical. However, the longest monitoring time-scale (in the QSO's rest frame) is roughly 60\% less for the high-z sub-sample. Moreover, the median long-term variability time-scale ($\sim$ 1200 days) is close to the maximum time-scale probed in some of the high-z QSOs. Therefore, further spectroscopic monitoring of these sources is needed to confirm the apparent excess variability seen in low-z BAL QSOs.}

Next we consider a sub-sample of  FeLoBAL QSOs. Full details of this sub-sample is published in  \citet{vivek12a}. Only one out of 5 sources (i.e., 20\%) considered in this study has shown significant absorption line variability. The fraction of   FeLoBAL QSO systems showing absorption line variations (i.e., 20\%) is less than that seen for the \mgii\ BALs without \feii\ (i.e., 41\%). 
There are indications, in cases where independent constraints on ionization parameter and density can be obtained, that the  FeLoBAL QSOs may originate far away (i.e., $>$ 1 kpc) from the central engine \citep[see ][]{korista08,moe09,dunn10,bautista10,faucher11}. Based on optical-IR observations \citet{farrah07} suggested that  FeLoBAL QSOs depict an evolutionary phase in which a luminous star burst phase in a galaxy is  approaching its end stage and the QSO phase is beginning.  It is also proposed that the observed properties of  FeLoBAL QSOs are readily explained if they are formed in situ in the dense ISM shock heated by QSO blast waves \citep{faucher11}. Our study favors  FeLoBAL QSOs to have a different origin compared to  LoBAL QSOs. If one can associate rapid  absorption line variability to gas close to the QSOs, then our results either favor  { FeLoBAL QSOs originate from a larger distance to the central engine or from a time-steady flow  as compared to  LoBAL QSOs without Fe absorption.}  However, \citet{lazarova12}  do not find differences in the FIR properties of LoBAL and non-LoBAL QSOs. Note that both the results mentioned above are based on small number statistics. It will be good to probe the exact nature of LoBAL QSOs with a larger sample.   

There are 8 sources in the sample which have  optical-UV SEDs dominated by strong Fe emission lines. These sources have a broad Fe emission feature mimicking an absorption slightly blue-ward of the \mgii\ emission line. Some of the variations in  \mgii\ lines can be attributed to the variation in this Fe emission feature. Therefore, we did not include wavelengths covering this feature in our equivalent width and ratio measurements. Four sources, SDSS J0149-0103, SDSS J0944+0625, SDSS J1208+0230 and SDSS J2215-0045 have shown significant variability in \mgii\ BAL equivalent widths. The \aliii\ absorption line is covered only in SDSS J2215-0045, which has varied. We note that iron emitting sources  are among the top ones in the whole sample  showing maximum variations in the absorption lines.  The percentage of  sources with variable BALs in this iron emitter sub-sample is about 50 \% (4/8).   Out of the fourteen remaining sources which do not show
strong iron emission, four(i.e., 29\%) show variations.

\subsection{Correlation with photometric parameters}

\begin{figure}
 \centering
\psfig{figure=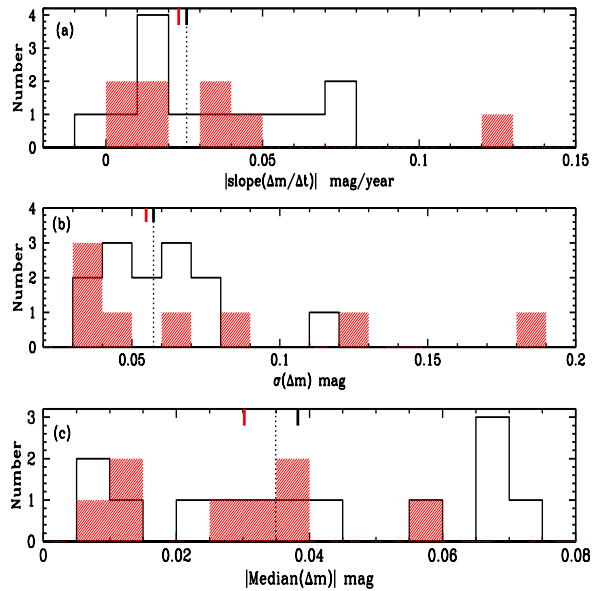,width=1.0\linewidth,height=1.0\linewidth,angle=270}
\caption{The three panels show the histogram distribution of (a) absolute values of slopes ($\Delta$m/$\Delta$t) (b) standard deviation of the magnitude differences and (c) absolute median of the magnitude differences for  sources with variable and non-variable BALs. The line histogram corresponds to  sources with non-variable BALs and the red shaded histogram corresponds to  sources with variable BALs. The median values for  the two distributions are marked at the top in black and red respectively.  It is clear that objects with variable and non-variable BALs appear to be drawn from the same distribution in terms of continuum variability.}
\label{hist_phot}
\end{figure}
   
To understand the correlation of the BAL variations with continuum parameters, we split the whole sample in to two sub-samples of objects with and without BAL variations.   Fig.~\ref{hist_phot} shows the histogram distribution of various photometric parameters for the two groups. The line histogram corresponds to the sample of non-varying BALs and the shaded histogram corresponds to the sample of varying BALs. Panel (a) shows the histogram distribution for the absolute slope values. Panel (b) and (c) show  distributions for the standard deviation and the absolute median values of the magnitude differences.  The median values for the two distributions are marked at the top of each panels and the median value for the whole sample is marked by dotted vertical lines. It is clear from Fig.~\ref{hist_phot} that the distribution of the photometric parameters for the samples with and without absorption line variations of BALs are similar. Two-sided Kolmogorov-Smirnov (KS) test for these distributions results in D and probability values of (0.2, 95\%), (0.3, 69\%) and (0.4, 36\%). This implies both the sub-samples are drawn from the same parent population. We also find the structure function computed for the above two sub-samples to be similar within the measurement uncertainities.  Similarity of photometric parameter distribution for the two samples implies that the   BAL QSOs showing absorption line variations are not associated with a particular range in the values of any of these photometric parameters. 

\subsection{Variation of  equivalent widths with time-scales }
\begin{figure}
\psfig{figure=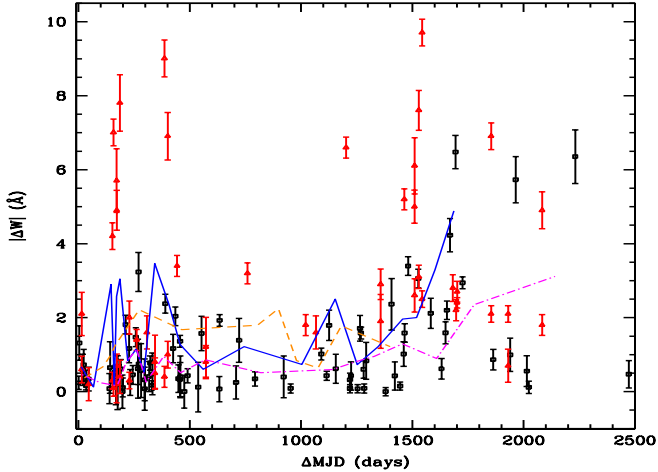,width=1.05\linewidth,height=0.8\linewidth,angle=0}
\caption{ Variation of $\Delta$W with $\Delta$t for \mgii\ absorption. Red triangles mark the two sources with dramatically varying BALs.  The  blue (solid) and magenta (dott dashed) curve represents the variation of rms differences computed in bins containing 6 time-scale values with and without the inclusion of the sources showing dramatic BAL variability respectively.  Orange (dashed) curve represents the variation of rms differences with time-scales for \aliii.  }
\label{dwdt}
\end{figure}

\citet{gibson08},   \citet{capellupo12} and \citet{filiz13}  studied the variation of \civ\ equivalent widths with time and reported that longer the time-scale, larger is the variation in the equivalent widths. The \civ\ variability probability approaches unity when $\Delta$t $\geq$ 10 yrs.   From the discussions presented in section 4, we notice that, even among sources with varying absorption lines, only 50\% of them show variation over shorter time-scales (i.e. $<$ 1 year  in the QSO rest frame). Fig.~\ref{dwdt} shows the plot of the variation of absolute \mgii\ $\Delta$W with $\Delta$t in the QSO rest frame. Though only 8 sources show $>$3$\sigma$ variations, all the variations between epochs are considered here.  As the variation in two sources, namely SDSS J1333+0012 and SDSS J2215-0045 are exceptionally high, they were observed more frequently and hence they contribute more than others in the plot. This will also lead to a clustering of measurements at shorter time-scales. We do not have any preferential sampling of measurements for the rest of the objects.  The red triangles mark the locations of  these two sources in the plot.  The blue (solid) and magenta (dott dashed) curve represents the variation of rms of equivalent width differences computed in 6 time-scale bins with and without the inclusion of two  sources with highly varying BALs respectively. The inclusion of these two sources do not change the overall trend.  It is clear from the figure that \mgii\  LoBAL QSOs also tend to show large amplitude variations over long time-scales. The increased incidence of variability at larger time-scales indicates that the physical process causing the BAL variations is a slow one which dominates over multi-year time-scales.  Orange (dashed) curve in the top panel  shows a similar plot of variation of $\Delta$W for \aliii. There is no clear indication of any trend in the variation over different time-scales.

\subsection{ Variation of equivalent width with mean equivalent width $<$W$>$ }
\begin{figure}
\psfig{figure=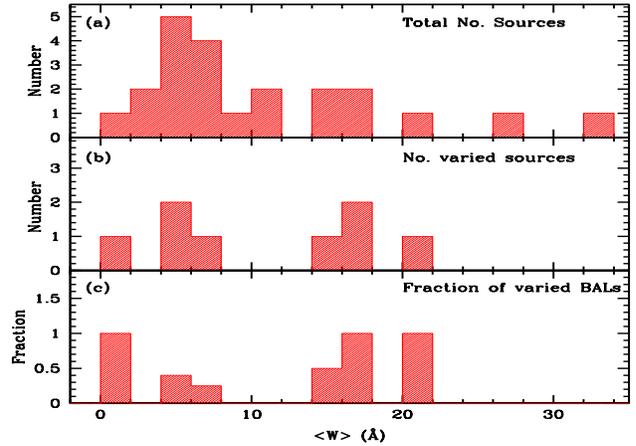,width=1.05\linewidth,height=0.75\linewidth,angle=0}
\caption{  Panels (a) and (b) show the number of QSOs with \mgii\ BAL absorption and the number with \mgii\ BAL variability at each absorption strength. Panel (c) shows the fraction of \mgii\ BALs that have varied at each absorption strength.  Clearly   there is no  evidence for  high fractional variation among BALs with low \mgii\ equivalent widths.}
\label{hist_w}
\end{figure}
\begin{figure*}
 \centering
\begin{tabular}{c c}
\psfig{figure=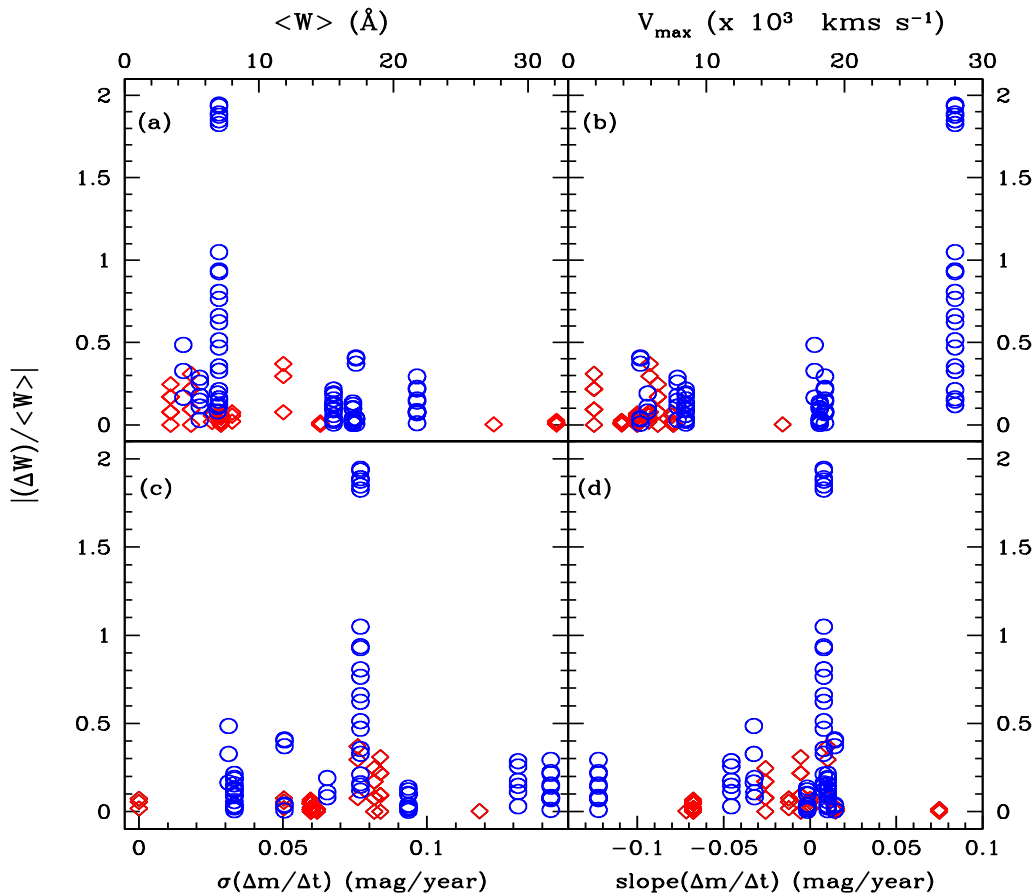,width=0.8\linewidth,height=0.7\linewidth,angle=0}&
\end{tabular}
\caption{Fractional variation of \mgii\ equivalent widths are plotted against $<$W$>$, V$_{max}$, $\sigma(\Delta$m/$\Delta$t) and slope of the $\Delta$m-$\Delta$t curve in panels a,b,c and d respectively. The blue circular  and red diamond points mark the values for the   sources with varying BALs and non-varying BALs.  It is clear that \mgii\ equivalent widths show more scatter at low equivalent widths and high velocities where as no correlation is seen with continuum parameters. }
\label{EW_oth_mg2}
\end{figure*}

 Here we explore the dependence  of \mgii\ BAL variability on the overall strength of the BAL absorption. 
 Previous studies on the correlation between \civ\ BAL equivalent width variation and absorption strength have reported that weaker portions of BAL troughs are more likely to vary than stronger ones \citep[see ][]{barlow94,lundgren07,gibson08,capellupo11}.

In Fig.~\ref{hist_w}, the top  two panels (a) and (b) show the number of QSOs with \mgii\ BAL absorption and the number with \mgii\ BAL variability at each absorption strength.  Panel (c) shows the fraction of \mgii\ BALs that have varied at each \mgii\ equivalent width. Clearly   there is no  evidence for  high fractional variation among BALs with low \mgii\ equivalent widths.  
In Fig.~\ref{EW_oth_mg2}, panel (a) shows the absolute variation of the fractional change in the equivalent widths with the mean equivalent width for \mgii\ BALs. The blue circles   represents the  sources that show BAL variability. Here, we have considered all the possible pairs of equivalent width measurements for each QSOs.   
In Fig.~\ref{EW_oth_al3},  panel (a) shows the absolute fractional variation of the \aliii\ equivalent widths with the average BAL equivalent widths.
 From these plots, it is clear that both \mgii\ and \aliii\ show more scatter at low equivalent widths as has been reported for \civ\ BALs. But, in the case of \mgii\ systems, one source, SDSS J1333+0012 dominates the results. The correlation is more prominent in \aliii\ as compared to \mgii. The wavelength range where \mgii\ absorption is observed is significantly affected by iron emission features where as it is not the case for \aliii.    Strong iron features which contaminate the \mgii\ regions could  be the reason for the weakness of this correlation in \mgii\ BALs.

\begin{figure}
\psfig{figure=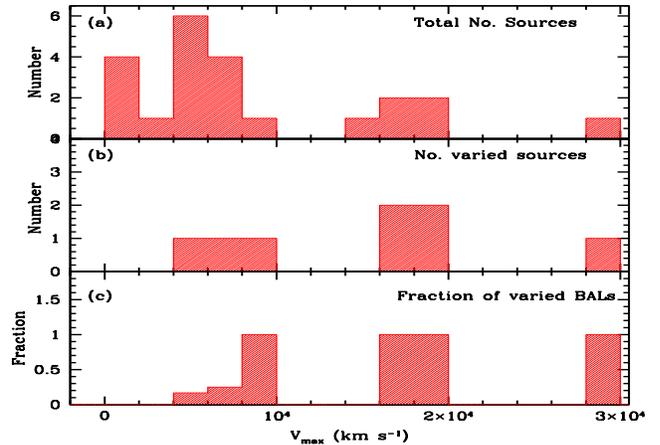,width=1.05\linewidth,height=0.75\linewidth,angle=0}
\caption{The top two panels show the number of QSOs with \mgii\ BAL absorption and the number with \mgii\ BAL variability at each V$_{max}$. The lower panel shows the fraction of \mgii\ BALs that have varied at each V$_{max}$.  The plot clearly points to the correlation between the variations in BAL absorption and the outflow velocities. }
\label{hist_vel}
\end{figure}
\subsection{ Dependence of variation in  equivalent width with  V$_{max}$ }
\civ\ BAL absorption, on an average are found to be shallower at highest velocities. If these systems are less saturated, high velocity BALs can be more responsive to ionization changes. \citet{lundgren07} do not find any linear relationship between the  equivalent width  variability and the BAL velocity. But, they note that the most highly variable BALs in their sample are found at high velocities. \citet{gibson08} also do not find any indications for the variation of  $\Delta$W across the velocity ranges. \citet{capellupo11} find that  components at higher velocity are more likely to vary than those at lower velocities. But, they do not find any statistically significant trend between fractional change in absorption strength and the outflow velocity.

\begin{figure*}
 \centering
\begin{tabular}{c c}
\psfig{figure=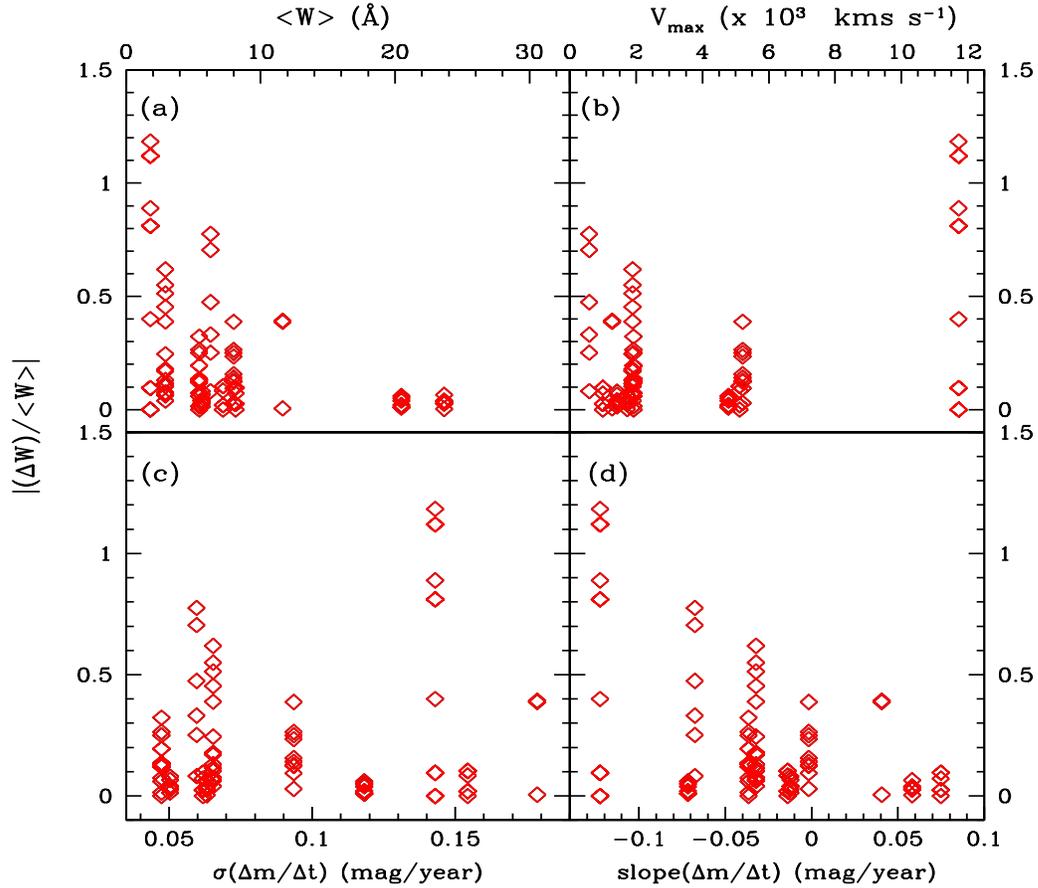,width=0.8\linewidth,height=0.7\linewidth,angle=0}
\end{tabular}
\caption{ Same as in Fig.~\ref{EW_oth_mg2}, but for \aliii.  Here again, \aliii\ equivalent widths show more scatter at low equivalent widths pointing to a correlation between them. But the correlations with V$_{max}$ and continuum parameters are not seen. }
\label{EW_oth_al3}
\end{figure*}

\begin{figure}
\psfig{figure=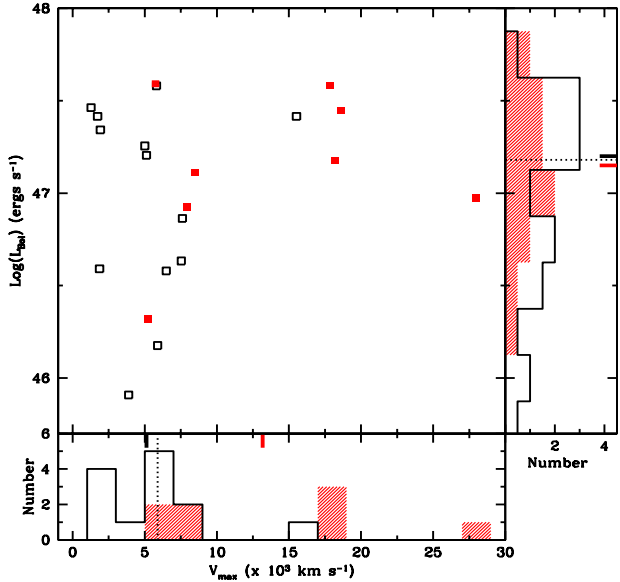,width=1.05\linewidth,height=1.\linewidth,angle=0}
\caption{Bolometric luminosity is plotted against the maximum outflow velocity. Open and filled squares mark the sources with non-varying and varying BALs. Velocity histogram distribution for the samples with varying (shaded histogram) and non-varying (line histogram) BALs is shown in the lower panel and luminosity histogram distribution for the same samples are shown on right panel.  Note that high fraction of varying BAL objects occupy the high velocity high luminosity quadrant.}
\label{hist_vel_lum}
\end{figure}

In Fig.~\ref{hist_vel}, the  two panels (a) \& (b) show the number of quasars with \mgii\ BAL absorption and the number with \mgii\ BAL variability at each velocity.  Panel (c) shows  the distribution of  variable sources  with outflow velocities.  Though the smaller velocity bins have more number of sources, the varying fraction is small. This clearly points to the correlation between the variations in BAL absorption and the outflow velocities. The observed correlation is again reiterated by the panel (b) of Fig.~\ref{EW_oth_mg2} where the  absolute fractional variation of the \mgii\ equivalent widths is plotted against the velocities.
 Panel (b) of  Fig.~\ref{EW_oth_al3},  shows the absolute fractional variation of the \aliii\ equivalent widths with the V$_{max}$.  Our \aliii\ BAL sub-sample do not contain many sources at high velocities. Only one source, SDSS J2215-0045,  has varied significantly at high velocity. Here, there is no apparent correlation evident between the two parameters with in the available \aliii\ velocity ranges.
Overall, this study also points to higher incidence of  sources with variable BALs at high outflow velocities, which has been observed in the  case of HiBALs.
 If the broad absorption troughs originate in a radiatively driven wind, a relation between luminosity and V$_{max}$ is expected. Studies of \citet{laor02}, \citet{ganguly07} and \citet{stalin11} on \civ\ absorption systems confirm a luminosity-dependent  envelope to the maximum velocity of absorption.  In Fig.~\ref{hist_vel_lum}, bolometric luminosity for each source is plotted against its outflow velocity. Open and filled squares mark the sources with non-varying and varying BALs respectively. Similar to what is seen in \civ\ BALs,   LoBAL QSOs in this study also show a lower envelope. The left and bottom side panels show the histogram distributions of velocities and luminosities for the samples with varying (red) and non-varying (black) BALs. The dotted  line in these side panels marks the median of the whole sample.  Black and red/grey ticks at the top of the histogram panels mark the median value of the non-varying and varying samples respectively. From the velocity distributions in the lower panel, it is clear that the varying and non-varying samples are different. The median velocity of the varying sample is at a higher velocity as compared to the non-varying sample.   However, we do not find difference between the luminosity distributions of the varying and non-varying samples (see left panel of Fig.~\ref{hist_vel_lum}).  This could mean that the variations in BALs are not affected by the strength of QSO luminosity.
However, it is interesting  to note that a high fraction of objects in the high velocity high luminosity quadrant of Fig.~\ref{hist_vel_lum} show variability.

\subsection{Dependence of  BAL variation on the continuum }
  One natural possibility that has been invoked to explain absorption line  variability in BALs is the variations in the continuum. Changes in the continuum fluxes can be correlated with  the variations in BALs mainly in three ways. Changes in the far UV continuum flux  can directly change  the ionizing conditions of the outflowing gas and it will reflect as variations in the BAL optical depths. In the simplest form of photoionization models, one expects an overall absorption profile change. Therefore, BAL variations across the complete absorption profiles are usually attributed  to ionization changes. Another possibility is the luminosity induced dynamical changes in the absorbing gas. If radiation pressure is driving the outflows, it is natural to think that the changes in the luminosity can shift the absorptions to higher/lower velocities \citep{anand00}.  \citet{vivek12} discussed about another possibility where the structural changes in the accretion disc result in changes in the continuum and BAL absorptions. This could  mean the ejection being triggered by some events in the accretion disc that caused reduction in the accretion efficiency. Appearance/disappearance of new BAL components can be thought of resulting from this structural changes in the accretion disk.  

The inverse correlation between the BAL variability and the $<$W$>$ suggests a mechanism driven by changes in the continuum ionization. Since larger equivalent width BAL profiles are more likely to be saturated, changes in the continuum may not affect the strong BAL absorption significantly.  Previous studies on continuum variability in BAL QSOs have found no conclusive evidence for the dependence  of the absorption line variations on continuum flux changes. \citet{barlow94} found some marginal evidence of correspondence between these parameters. But, they also reported high frequency of exceptions to this correlation.
 \citet{lundgren07}  found that there is no correlation between BAL variations and the 1549\AA\ flux variability. \citet{gibson10} took the \civ\ emission line flux as a proxy to the continuum and found significant correlations between continuum flux and absorption line equivalent widths. However, none of these studies could use long-term photometric light curves.

In our study, the continuum and absorption line observations almost span the same time period.   From the CRTS light curves, we obtained the V magnitudes of each sources at the same epoch of spectroscopic measurements. These magnitudes are used to compute the differences. 
In Fig.~\ref{EW_oth_mg2} and Fig.~\ref{EW_oth_al3}, the absolute values of \mgii\ and \aliii\ fractional equivalent widths are plotted against the different  parameters of continuum variability in the lower two  panels. In  panel (c) of both the figures, we have plotted fractional variation in equivalent widths against the standard deviation of time averaged  differences in magnitudes, i.e., $\Delta $m/$\Delta$t, between various epochs. A larger value for the standard deviation means  a higher rate for continuum variation. In  panel (d) of the two figures, we have plotted the equivalent width fractional variation against the slope of the $\Delta$m-$\Delta$t curve.   We do not find any trend in the variation of the \mgii\ and \aliii\ BAL equivalent widths with the continuum parameters.

Although we do not detect any strong correlation between the continuum and BAL variations in the sample, some of the  sources are promising candidates for the scenarios discussed in the beginning of this section. As discussed earlier, in SDSS J0149-0103, SDSS J1208+0230  SDSS J1333+0012 and SDSS J2215-0045, the observed variations are consistent with dynamical changes in the absorbing clouds. SDSS J1333+0012 is a good example for the scenario of structural changes in the accretion disk. The observed variations in SDSS J0850+4452 and SDSS J0952+0257 are consistent with the scenario of changing ionization.  In reality, it can be a mixture of all these effects contributing to the variations in the absorption profiles.

\section{Results \& Conclusions}
We have presented a detailed analysis of a sample of 22  LoBAL QSOs that are monitored both spectroscopically and photometrically  over a period of few days to few years in the QSO rest frame.  Most of the results are based on  equivalent widths, but we show that they are consistent with results from a spectral ratio analysis that avoids problems related to the continuum fitting.  The main results from the study are summarized below.
\begin{enumerate}
\item  Out of 22 BALs monitored, 8 are showing absorption line variability. We define a variable BAL as the one with variations in rest equivalent width $>$ 3$\sigma$ level and ratio $>$ 5$\sigma$. \citet{gibson08} and \citet{capellupo11}, in their studies on \civ\ BALs, reported the fraction of varying BALs to be 92\% and 67\% respectively. Very recently, \citep{filiz13} have found 50-60\% of \civ\ BALs showing variability over a time-scale of 1 - 3.7 years using an extended sample from SDSS.  As compared to the \civ\ BALs, the fraction of varying BALs in our overall sample is small (i.e., 36\%).  \citet{capellupo11} used a slightly higher threshold (4$\sigma$) to define the variability and their sample  span the  redshift range 1.2$<$z$<$2.9. In this redshift range, the variability detection is even less (i.e., 25\%) for our  LoBAL QSO sample.    
\item  We  find a greater incidence of variability at long time-scales ($>$ 1 year) than short time-scales ($<$ 1 year). \citet{capellupo11} find that 39 \% of \civ\ BALs varied in the short-term data, whereas 65 \% varied in the long-term data.  In this study of \mgii\ BALs, we find the respective variable fraction to be 18\% and 36\% for short-term and long-term data. This could mean that the physical process causing the BAL variations dominates over multi-year time-scales.
\item  In two sources, SDSS J1333+0012 and SDSS J2215-0045, there is appearance/disappearance of BAL components.  The observed frequency of 9\% for BAL appearance/disappearance is different from the value, 3\% found for \civ\ BALs \citep[][]{filiz12}. Confirming this possible difference is important for understanding the different BAL QSO classes  in the frame work of  "orientation" or "evolutionary" scenario.
\item As has been reported for \civ\ BALs,  we also find a trend for  larger fractional change in absorption equivalent widths to occur  in shallower and higher velocity features. We do not find any correlation between BAL variations and the QSO luminosity. However, it is interesting to note that there is a higher  fraction of objects in the high velocity high luminosity range. 
\item The observed fraction of sources with varying BALs in  FeLoBAL QSOs (i.e, 20\%) is small compared to  LoBAL QSOs (44\%) and HiBALs (67\%).  Our observation suggests that   FeLoBAL QSOs could be a different population compared to  LoBAL QSOs not showing \feii\ absorption. Such a difference is expected based on dust emission and distance estimates of clouds from photoionization models. However, as the predictions for the variability time-scales are not available, we are not in a position to distinguish between the two proposed scenarios of  FeLoBAL QSOs. 
\item We find a larger fraction of sources with varying BALs in a sub-sample of Fe emitting sources. \citet{zhang10} have found that the  LoBAL QSO fraction is higher when the strength of \feii\ emission is high using SDSS-DR5 data. \citet{xiao11} have found a strong correlation between \feii\ emission strength and Eddington ratio. Confirming the higher variability detection in BAL QSOs with \feii\ dominated continuum in a larger sample will establish the connection between Eddington ratio (i.e. accretion efficiency) and line variability.
\item We do not find any trend in the variation of the \mgii\ and \aliii\ BAL equivalent widths with the continuum parameters. This is consistent with what is seen in  previous studies on \civ\ BAL systems.  Wide range in equivalent width differences, $\Delta$W,  seen for a given magnitude difference, $\Delta$m and vice versa imply that absorption line variations are not dominantly driven by photo ionization changes induced by continuum variations.  Emergence and profile variations of BALs in some sources imply that  dynamical changes like movement of clouds across the line of sight, may be important. 
\end{enumerate}

 
\section*{acknowledgements}
{ We thank the anonymous referee for a thorough review and helpful comments which led to an improved manuscript. We wish to acknowledge the IUCAA/IGO staff for their support during our observations. RS and PPJ gratefully acknowledge support from the Indo-French Centre for the Promotion of Advanced Research ((IFCPAR)  under Project N.4304-2. MV gratefully acknowledges IFCPAR for the financial support to visit IAP, Paris. }

{Funding for the SDSS and SDSS-II has been provided by the Alfred P. Sloan Foundation, the Participating Institutions, the National Science Foundation, the US Department of Energy, the National Aeronautics and Space Administration, the Japanese Monbukagakusho, the Max Planck Society and the Higher Education Funding Council for England. The SDSS website is http://www.sdss.org/.} 
\begin{table*}

\caption{Equivalent width measurements}
\begin{tabular}{|c|c|c|c|c|c|c|}
\hline
Name&Ion&Epoch ID&MJD&$\lambda$ range&W$_{rest}$&Comments\\
    & &    & & (\AA)        &   (\AA)  &\\
\hline
\hline
		& \mgii	& sdss1 & 51810      &	       		   &       7.0  $\pm$   0.3& \\
		& \mgii	& sdss2 & 51868      &7911.9 -      8280.8  &       7.6  $\pm$   0.3& \\
		& \mgii	& sdss3 & 51879      &			   &       6.3  $\pm$   0.3& \\
		& \aliii& sdss1 & 51810      &			   &       3.5   $\pm$   0.1& \\
		& \aliii& sdss2 & 51868      &			   &       3.3   $\pm$   0.1& \\
		& \aliii& sdss3 & 51879      & 5439.9  -     5514.1 &       3.1   $\pm$   0.1& \\
SDSS J0045+1438	& \aliii& 2007 & 54447      &			   &       3.0   $\pm$   0.3& Variable \\
		& \aliii& 2008 & 54818      &			   &       2.8   $\pm$   0.2& \\
		& \aliii& 2009 & 55187      &                      &       1.9   $\pm$   0.3& \\  
		& \civ	& sdss1 & 51810      &       		   & 	   53.7  $\pm$   0.2& \\
		& \civ	& sdss2 & 51868      & 			   &	   55.2  $\pm$   0.3& \\
		& \civ	& sdss3 & 51879      & 4378.2       4579.6  &       54.2  $\pm$   0.2& \\
		& \civ	& 2007 & 54447      & 			   &	   47.9  $\pm$   0.8& \\
		& \civ	& 2008 & 54818      &			   &	   50.1  $\pm$   0.4& \\
		& \civ	& 2009 & 55187      &       		   &	   48.3  $\pm$   0.6& \\
\hline		
		& \mgii	& sdss & 51793 & 			   &       5.3 $\pm$      0.2 &		\\
SDSS J0149-0103 & \mgii	& 2008 & 54474 &       5642.1  -     5722.3&       6.1 $\pm$      0.5 &	Variable\\
		& \mgii	& 2009 & 54823 &	  		   &       6.3 $\pm$      0.2 &		\\
		& \mgii & 2010 & 55447 &			   &	   5.8 $\pm$	  0.1 &		\\	
		& \mgii	& 2012 & 55970 & 			   &       4.7 $\pm$      0.4 &		\\
\hline
                & \mgii & sdss & 51924 &			   &       14.8 $\pm$    0.1 & \\	
                & \mgii & 2007 & 54448 &			   &       14.6 $\pm$    0.2 & \\
SDSS J0318-0600 & \mgii & 2008 & 54797 &8087.3   -   8250.7        &       14.5 $\pm$    0.1 & \\
                & \mgii & 2009 & 55188 &                           &       14.6 $\pm$    0.1 & \\
\hline
		& \mgii	& sdss & 51910 &      			   &       6.7  $\pm$   0.1 &		\\
SDSS J0334-0711	& \mgii	& 2010 & 55216 &      4458.5  -     4486.1 &       6.6  $\pm$   0.1 &		\\
		& \mgii	& 2012 & 55950 &      			   &       6.2  $\pm$   0.3 &		\\
\hline
		& \mgii	& sdss & 51877 & 		    	   &       7.3 $\pm$    0.1&\\
		& \mgii	& 2006 & 54090 &     		           &       6.9 $\pm$    0.6&\\
		& \mgii	& 2008 & 54804 &     6609.2   -    6676.4  &       7.0 $\pm$    0.2&\\
		& \mgii	& 2009 & 55189 &			   &       7.3 $\pm$    0.1&\\
SDSS J0737+3844	& \mgii	& 2010 & 54887 &			   &       7.4 $\pm$    0.1&\\
		& \aliii& sdss & 51877 &    			   &	   5.8 $\pm$	0.2& \\
		& \aliii& 2006 & 54090 &			   &	   3.6 $\pm$    1.2& \\
		& \aliii& 2008 & 54804 &     4374.4   -    4442.0  &	   6.3 $\pm$	0.3& \\
		& \aliii& 2009 & 54887 &			   &       5.9 $\pm$    0.1&\\
\hline
		& \aliii& sdss1 & 52207 &        		   &	5.7  $\pm$   0.1  &\\
		& \aliii& sdss2 & 51959 &        		   &	5.0  $\pm$   0.3  &\\
SDSS J0823+4334 & \aliii& 2007 & 54447 &     4779.1 -      4901.4  &    6.5  $\pm$   0.7  &\\
                & \aliii& 2008 & 54806 &        		   & 	5.1  $\pm$   0.3  &\\
                & \aliii& 2009 & 54940 &       			   & 	4.7  $\pm$   2.2  &\\
                & \aliii& 2010 & 55221 &       			   &	5.7  $\pm$   0.7  & \\
\hline
		& \mgii	& sdss & 52232 &                            &	    4.4  $\pm$	0.2   &		\\
SDSS J0835+4242 & \mgii	& 2007 & 54448 &			    &	    4.8  $\pm$	0.3   &		\\
		& \mgii	& 2008 & 54805 &     5027.7 -     5066.5    &       4.8  $\pm$	0.3   &		\\
		& \mgii	& 2010 & 55217 &		      	    &	    5.9  $\pm$	0.2   &		\\
\hline
                & \mgii & sdss & 52320 &     			    &	    32.1 $\pm$   0.1	& \\
                & \mgii & 2006 & 54089 &     			    &       31.8 $\pm$   0.2  & \\
SDSS J0840+3633 & \mgii & 2007 & 54446 &   6158.3   -    6242.5     &       32.0 $\pm$   0.1   & \\
                & \mgii & 2008 & 54807 &     			    &       32.5 $\pm$   0.1   & \\
                & \mgii & 2009 & 55182 &    			    &       32.2 $\pm$   0.1  & \\
\hline
\hline
\end{tabular}
\label{EW_table}
\begin{flushleft}
  $^{\dagger}$ 7 and 1 in the parenthesis refers to IFOSC 7 and IFORS 1 data respectively \\
      \end{flushleft}
\end{table*}
\setcounter{table}{3}
\begin{table*}
\caption{Equivalent width measurements {\it continued}}
\begin{tabular}{|c|c|c|c|c|c|c|}
\hline
Name&Ion&Epoch ID&MJD&$\lambda$ range&W$_{rest}$&Comments\\
    & &    & & (\AA)        &   (\AA)  &\\
\hline
\hline
		& \mgii	& sdss & 52605 &      			   &       12.6 $\pm$     0.4 &		\\
SDSS J0850+4451 & \mgii	& 2010 & 55215 &      4225.3  -     4294.0 &       19.1 $\pm$     0.3 &	Variable\\
		& \mgii	& 2011 & 55632 &      			   &       18.3 $\pm$     0.5 &		\\
		& \mgii	& 2012 & 56044 &      			   &       18.9 $\pm$     0.6 &		\\
\hline
		& \mgii	& sdss & 52710 &    			   &       17.6 $\pm$     0.1 &		\\
		& \mgii	& 2008 & 54559 &    			   &       16.5 $\pm$     0.1 &		\\
SDSS J0944+0625 & \mgii	& 2009 & 54861 &      4603.1  -     4694.7 &       16.0 $\pm$     0.1 &	Variable\\
		& \mgii	& 2010 & 55220 &      			   &	   14.2 $\pm$     0.2 &		\\
		& \mgii	& 2010 & 55297 &      			   &	   14.5 $\pm$     0.2 &		\\
		& \mgii	& 2011 & 55634 &      			   &	   14.6 $\pm$     0.1 &		\\
\hline	

		& \mgii & sdss &  51908 &     			  &       18.6  $\pm$    0.3 & \\
		& \mgii & 2008 &  54559 &     			  &       16.8  $\pm$    0.3 & \\
		& \mgii & 2009 &  54887 &     6279.1 -      6464.1&       16.9  $\pm$    0.3 & \\
		& \mgii & 2010 &  55221 &     			  &       16.2  $\pm$    0.6 & \\
		& \mgii & 2011 &  55635 &     			  &       16.5  $\pm$    0.3 & \\
SDSS J0952+0257	& \aliii& sdss &  51908 & 			  &        9.6  $\pm$    0.2 &\\
 		& \aliii& 2008 &  54559 & 			  &        7.4  $\pm$    0.3 & Variable\\
		& \aliii& 2009 &  54887 &     4211.9  -     4293.1&        8.2  $\pm$    0.2 &\\
 		& \aliii& 2010 &  55221 & 			  &        6.4  $\pm$    0.8 &\\
 		& \aliii& 2011 &  55635 & 			  &        8.4  $\pm$    0.3 &\\
\hline
                & \aliii& sdss &  52376 &			  &        7.1  $\pm$    0.2    &  \\
SDSS J1010+4518 & \aliii& 2008 &  54474 &   4939.1   -    5061.4  &        7.8  $\pm$    0.4    &  \\
                & \aliii& 2009 &  54888 &                         &        7.0  $\pm$    0.3    &  \\
                & \aliii& 2010 &  55220 &                         &        7.1  $\pm$    0.3    &  \\
\hline
		& \mgii	& sdss & 52642 &    			   &       13.5  $\pm$    0.3 &		\\
SDSS J1128+4823 & \mgii	& 2010 & 55218 &      4228.9  -     4330.4 &        9.3  $\pm$    0.4 & \\
		& \mgii	& 2011 & 55634 &     			   &       12.6  $\pm$    0.4 &		\\
\hline
                & \aliii& sdss & 52368 &			   & 	    12.9  $\pm$   0.3 &		\\
SDSS J1143+5203 & \aliii& 2008 & 54473 &     7770.1   -    7848.9  &	    12.1  $\pm$   0.6 & \\
                & \aliii& 2009 & 54859 &                           &	    12.2  $\pm$   0.5 &		\\
\hline
		& \mgii	& sdss & 52024 &			   &       5.6   $\pm$   0.2 &		\\
SDSS J1208+0230	& \mgii	& 2010 & 55219 &       5743.2  -     5813.1&       4.1   $\pm$   0.2 &	Variable\\
		& \mgii	& 2011 & 55633 &			   &       3.4   $\pm$   0.2 &		\\

\hline
		& \mgii	&sdss1 & 51662 &      			   &        2.5   $\pm$   0.2 &		\\
		& \mgii	&sdss2 & 51955 &      			   &        2.8   $\pm$   0.1 &		\\
		& \mgii	& 2008 & 54559 &      			   &        0.1   $\pm$   0.4 &		\\
SDSS J1333+0012	& \mgii&2009(7)\q& 54888 &       5035.7  -    5090.9 &      0.3   $\pm$   0.3 &	Variable\\
R component	& \mgii&2009(1)\q& 54916 &      		   &        0.3   $\pm$   0.2 &		\\
		& \mgii	& 2010 & 55218 &     	 		   &        0.4   $\pm$   0.1 &		\\
		& \mgii	& 2011 & 55657 &     	 		   &        0.7   $\pm$   0.2 &		\\

\hline
		& \mgii	&sdss1 & 51662 &     			    &       0.2  $\pm$    0.3 &		\\
		& \mgii	&sdss2 & 51955 &     			    &       4.4  $\pm$    0.2 &		\\
		& \mgii	&2008  & 54559 &     			    &       6.3  $\pm$    0.7 &		\\
SDSS J1333+0012	& \mgii&2009(7)\q& 54888 &       4790.7 -      5000.9 &      12.0  $\pm$    0.5 &	Variable\\
B component	& \mgii&2009(1)\q& 54916 &			    &      14.1  $\pm$    0.3 &		\\
		& \mgii	&2010  & 55218 &			    &       7.1  $\pm$    0.2 &		\\
		& \mgii	&2011  & 55657 &     			    &       5.1  $\pm$    0.4 &		\\
\hline
                & \aliii& sdss1 & 52426 &			    &  24.5  $\pm$   0.1 & \\
SDSS J1334-0123 & \aliii& 2008 & 54560 &  5142.5  -     5302.9      &  23.6  $\pm$   0.2 & \\
                & \aliii& 2009 & 54823 &                            &  23.6  $\pm$   0.1 & \\
                & \aliii& sdss2 &55272 &                            &  23.0  $\pm$   0.4 & \\
\hline
\hline                                                                           
\end{tabular}
\label{EW_table}
\begin{flushleft}
  $^{\dagger}$ 7 and 1 in the parenthesis refers to IFOSC 7 and IFORS 1 data respectively \\
      \end{flushleft}
\end{table*}
\setcounter{table}{3}

\begin{table*}
\caption{Equivalent width measurements {\it continued}}
\begin{tabular}{|c|c|c|c|c|c|c|}
\hline
Name&Ion&Epoch ID&MJD&$\lambda$ range&W$_{rest}$&Comments\\
    & &    & & (\AA)        &   (\AA)  &\\
\hline
\hline
		& \mgii	& sdss1 & 52026 &			    &      7.8  $\pm$     0.1 &		\\
		& \mgii	& sdss2 & 55706 &			    &      7.9  $\pm$     0.1 &		\\
		& \mgii	& 2010 & 55287 &    6979.6  -     7064.4    &      8.4  $\pm$     0.2  &		\\
SDSS J1448+0424 & \mgii & 2011 & 55635 &                            &      7.9  $\pm$     0.1  &		\\
                & \aliii& sdss1 & 52026 &			    &      5.7  $\pm$     0.1  &		\\
                & \aliii& sdss2 & 55706 &			    &      5.9  $\pm$     0.1  &		\\
		& \aliii& 2010 & 55287 &    4624.1  -     4698.8    &      5.5  $\pm$     0.4  &		\\
                & \aliii& 2011 & 55635 &                            &      5.4  $\pm$     0.2  &		\\
\hline

		& \mgii	& sdss & 52764 &     			   &       4.0  $\pm$    0.1 &		\\
SDSS J1614+3752 & \mgii	& 2008 & 54559 &      4250.4  -     4301.4 &       3.3  $\pm$    0.4 &		\\
		& \mgii	& 2010 & 55297 &      			   &       3.3  $\pm$    0.3 &		\\
		& \mgii	& 2011 & 55658 &      			   &       3.1  $\pm$    0.3 &		\\
\hline

    		& \mgii	& sdss & 51804  &  			  &      25.8  $\pm$    0.2 &  \\
                & \mgii	&  vlt & 52902  &  			  &      22.4  $\pm$    0.2 &  \\
SDSS J2215-0045 & \mgii	& 2008 & 54783  &6512.2  -     6735.2     &      19.2  $\pm$    0.2 & Variable \\
                & \mgii	& mage & 55431  &			  &      20.6  $\pm$    0.2 &  \\
                & \mgii	& 2011 & 55545  &  			  &      20.8  $\pm$    0.4 &  \\
    		& \aliii& sdss & 51804  &     			    &      3.9 $\pm$    0.1 &  \\
                & \aliii&  vlt & 52902  &     			    &      2.6 $\pm$    0.1 & \\
		& \aliii& 2008 & 54783  & 4345.0   -    4419.6      &      1.1 $\pm$    0.1 & \\
	        & \aliii& 2010 & 55213  &     			    &      1.1 $\pm$    0.3 & \\    
                & \aliii& mage & 55431  &     			    &      1.0 $\pm$    0.1 & \\
                & \aliii& 2011 & 55545  &     			    &      1.1 $\pm$    0.1 & \\
\hline
                & \mgii & sdss1 & 52559 &    7448.6   -    7716.4   &       27.4  $\pm$   0.3 &		\\  
                & \mgii & sdss2 & 54331 &                           &       27.5  $\pm$   0.2  &		\\
SDSS J2347-1037 &\aliii & sdss1 & 52559 &                           &       20.2  $\pm$   0.1  &		\\
                &\aliii & sdss2 & 54331 & 	                   &       20.0  $\pm$   0.1  &		\\
 		&\aliii & 2008 & 54770 &   4932.0   -    5110.4    &       20.4  $\pm$   0.3  &		\\
                &\aliii & 2009 & 55180 & 		           &       21.3  $\pm$   0.7  &		\\
\hline                                                                           

\end{tabular}
\label{EW_table}
\begin{flushleft}
  $^{\dagger}$ 7 and 1 in the parenthesis refers to IFOSC 7 and IFORS 1 data respectively \\
      \end{flushleft}
\end{table*}


\begin{table*}
\caption{Variability based on ratio between two epochs}
\begin{tabular}{|c|c|c|c|c|c|c|}
\hline
Name&Ion&Epoch ID 1& Epoch ID 2  & $\lambda$ range& Ratio& Comments\\
    &   &    &          &    (\AA)       &      &            \\
\hline
\hline
		& \mgii & sdss1	& sdss2 &                        &     2.9   $\pm$   1.4&    \\     
		& \mgii & sdss1	& sdss3 &  7911.9   -    8280.8  &    -2.2   $\pm$   1.3&     \\    
		& \aliii & sdss1 & sdss2 &                        &    -1.0  $\pm$   0.5&  \\
		& \aliii & sdss1 & sdss3 &                        &     0.4  $\pm$   0.5&  \\
		& \aliii & sdss1 & 2007  &  5439.9   -    5514.1  &    -5.9  $\pm$   1.1&  \\
SDSS J0045+1438	& \aliii & sdss1 & 2008  &                        &    -4.8  $\pm$   0.7& Variable \\
		& \aliii & sdss1 & 2009  &    			 &    -3.4   $\pm$   0.9&  \\
		& \civ & sdss1 & sdss2 &             		 &     1.2   $\pm$   2.8&     \\
		& \civ & sdss1 & sdss3 &            		 &     5.6   $\pm$   2.7&     \\
		& \civ & sdss1 & 2007  &  4378.2    -   4579.6  &   -26.4    $\pm$   5.8&     \\
		& \civ & sdss1 & 2008  &            		 &   -19.3   $\pm$   3.6&     \\    
		& \civ & sdss1 & 2009  &            		 &   -47.9   $\pm$   4.9&     \\
\hline                                                                                  
		& \mgii & 2008 & 2009 &       			   &  -1.1   $\pm$      1.3   &          \\
		& \mgii & 2008 & 2012 &       			   &  -7.6   $\pm$      1.5   &          \\
		& \mgii & 2009 & 2008 &       			   &   0.7   $\pm$      1.2   &          \\
SDSS J0149-0103 & \mgii & 2009 & 2012 &     5642.1  -     5722.3   &  -8.4   $\pm$      1.2   &   Variable      \\
		& \mgii & sdss & 2008 &       			   &   4.6   $\pm$      1.1   &          \\
		& \mgii & sdss & 2009 &      			   &   4.5   $\pm$      1.0   &          \\
		& \mgii & sdss & 2010 &				   &   3.1   $\pm$      0.5   &           \\ 	
		& \mgii & sdss & 2012 &       			   &  -2.0   $\pm$      1.0   &           \\
\hline
		& \mgii & sdss & 2007 &				   &1.7      $\pm$     1.6  & \\ 
SDSS J0318-0600	& \mgii & sdss & 2008 &	     8087.3  -     8250.7  &1.3      $\pm$     0.6  & \\ 
		& \mgii & sdss & 2009 &  			   &1.9      $\pm$     0.8  & \\
\hline
		& \mgii & sdss & 2010 &     			   & -1.4  $\pm$    0.3     &        \\
SDSS J0334-0711	& \mgii & sdss & 2012 &     4458.5  -    4486.1    & -1.6  $\pm$    0.9     &        \\
		& \mgii & 2010 & 2012 &     			   & -0.3  $\pm$    1.0     &        \\
\hline
 		& \mgii & sdss & 2006 &                           &  2.9  $\pm$   2.0  &	\\
		& \mgii & sdss & 2008 &                           & -0.4  $\pm$   0.5  &       \\
 		& \mgii & sdss & 2009 &   6609.2   -    6676.4    & -0.0  $\pm$   0.4    &     \\
SDSS J0737+3844	& \mgii & sdss & 2010 &                           & -0.1  $\pm$   0.4    &\\
		& \aliii& sdss & 2006 &			   	  & 1.1   $\pm$ 2.5& \\
		& \aliii& sdss & 2008 & 4374.4   -    4442.0	  &-1.8   $\pm$  0.7& \\   
		& \aliii& sdss & 2010 & 			  &-2.8   $\pm$  0.5& \\
\hline
		& \aliii& sdss1 & sdss2 &        		   & -0.9 $\pm$  0.8   &\\
		& \aliii& sdss1 & 2007 &        		   &  1.6 $\pm$   2.0  &\\
SDSS J0823+4334 & \aliii& sdss1 & 2008 &     4779.1 -      4901.4  & -2.2 $\pm$   1.0  &\\
                & \aliii& sdss1 & 2009 &        		   & -1.5 $\pm$   5.5  &\\
                & \aliii& sdss1 & 2010 &       			   & -1.6 $\pm$   2.0  &\\
\hline

		& \mgii & sdss & 2007 &	    			   &  2.6  $\pm$   0.8	   &	\\		
SDSS J0835+4242 & \mgii & sdss & 2008 &     5027.7  -     5066.5   &  2.9  $\pm$   0.6     &	\\
		& \mgii & sdss & 2010 &   			   &  6.9  $\pm$   0.4     &	\\
\hline
                & \mgii & sdss & 2006 &     			    &-5.2  $\pm$     5.2 &\\
                & \mgii & sdss & 2007 &     			    &-5.1  $\pm$     2.7 &\\
SDSS J0840+3633 & \mgii & sdss & 2008 &   6158.3   -    6242.5      &-4.7  $\pm$     1.9 &\\
                & \mgii & sdss & 2009 &     			    &-3.9  $\pm$     1.4 &\\
\hline
		& \mgii & sdss & 2010 &         		   &  4.3   $\pm$   0.4      &       \\
		& \mgii & sdss & 2011 &     			   &  4.2   $\pm$   0.6      &       \\
SDSS J0850+4451	& \mgii & sdss & 2012 &     4225.3  -     4294.0   &  4.5   $\pm$   0.7      &  Variable \\
		& \mgii & 2010 & 2011 &         		   &  0.8   $\pm$   0.6      &       \\
		& \mgii & 2010 & 2012 &         		   & -0.2   $\pm$   0.8      &       \\
		& \mgii & 2011 & 2012 &     			   &  0.5   $\pm$   0.9      &       \\
\hline
		& \mgii & sdss & 2008 &        			   & -2.8  $\pm$     0.4     &        \\
		& \mgii & sdss & 2009 &     			   & -3.1  $\pm$     0.3     &        \\
SDSS J0944+0625 & \mgii & sdss & 2010 &     4603.1  -     4694.7   & -9.8  $\pm$     0.5     &  Variable\\
		& \mgii & sdss & 2010 &        			   & -9.2  $\pm$     0.6     &        \\
		& \mgii & sdss & 2011 &     			   & -9.1  $\pm$     0.4     &        \\
\hline

\end{tabular}
\begin{flushleft}
  $^{\dagger}$ 7 and 1 in the parenthesis refers to IFOSC 7 and IFORS 1 data respectively \\
      \end{flushleft}
\label{IR_table}
\end{table*}
\setcounter{table}{4}


\begin{table*}
\caption{Variability based on ratio between two epochs: Continued }
\begin{tabular}{|c|c|c|c|c|c|c|}
\hline
Name&Ion&Epoch ID 1& Epoch ID 2  & $\lambda$ range& Ratio& Comments\\
    &   &    &          &    (\AA)       &      &            \\
\hline
\hline
	 	& \mgii & sdss & 2008 &        			     &  -12.5 $\pm$  1.1 &  \\
 	   	& \mgii & sdss & 2009 &       			     &  -12.3  $\pm$ 1.0 &  \\
	 	& \mgii & sdss & 2010 &        6279.1   -    6464.1  &  -11.9  $\pm$ 2.0 &  \\
SDSS J0952+0257	& \mgii & sdss & 2011 &        			     &  -11.0  $\pm$ 1.2 &Variable  \\
	 	& \aliii &sdss & 2008 &        			     &   -5.4  $\pm$ 0.9 &  \\
 	  	& \aliii &sdss & 2009 &        4211.9  -    4293.1   &   -4.8  $\pm$ 0.8 &  \\
	 	& \aliii &sdss & 2010 &                              &   -8.5  $\pm$ 2.0 &  \\
	 	& \aliii &sdss & 2011 &                              &   -4.4  $\pm$ 1.1 &  \\
\hline
                & \aliii& sdss &  2008 &			  & 2.2 $\pm$   1.1 & \\
SDSS J1010+4518 & \aliii& sdss &  2009 &   4939.1   -    5061.4   & 0.9 $\pm$   1.1 & \\
                & \aliii& sdss &  2010 &                          & 3.3 $\pm$   1.1 & \\
\hline
		& \mgii & sdss & 2010 &     			   &   -3.7  $\pm$     0.9     &       \\
SDSS J1128+4823 & \mgii & sdss & 2011 &     4228.9  -     4330.4   &   -1.4  $\pm$     0.9     &       \\
		& \mgii & 2010 & 2011 &     			   &    2.7  $\pm$     1.1     &       \\
\hline
                & \aliii& sdss & 2008 &			          &  1.7  $\pm$ 1.7  & \\
SDSS J1143+5203 & \aliii& 2008 & 2009 &     7770.1   -    7848.9  & -0.5  $\pm$   1.4  & \\
\hline 
		& \mgii & sdss & 2010 &     			   &   -2.3  $\pm$      0.3   &         \\
SDSS J1208+0230	& \mgii & sdss & 2011 &     5743.2  -     5813.1   &   -1.9  $\pm$      0.3   &  Variable       \\
		& \mgii & 2010 & 2011 &     			   &    0.2  $\pm$      0.3   &         \\
\hline

		& \mgii & 2000 & 2001 	&        		   &   	-1.1   $\pm$   0.2      &      \\
		& \mgii & 2000 & 2008 	&        		   &   	-3.6   $\pm$   0.7      &      \\
SDSS J1333+0012 & \mgii & 2000 & 2009(7)\q&     5035.7  -   5090.9 & 	-3.8   $\pm$   0.3      &  Variable    \\
R component	& \mgii & 2000 & 2009(1)\q&      		   &   	-5.6   $\pm$   0.5      &      \\
		& \mgii & 2000 & 2010 	&     			   &   	-6.3   $\pm$   0.3      &      \\
		& \mgii & 2000 & 2011 	&     			   &   	-6.9   $\pm$   0.5      &     \\
\hline
		& \mgii & 2000 & 2001 	&     	       		   &     4.3   $\pm$   0.4      &     \\
		& \mgii & 2000 & 2008 	&     			   &    18.9   $\pm$   1.3      &     \\
SDSS J1333+0012 & \mgii & 2000 & 2009(7)\q&     4790.7  -   5000.9 &    31.5   $\pm$   0.5      &  Variable   \\
B component	& \mgii & 2000 & 2009(1)\q&        		   &    28.2   $\pm$   1.0      &    \\
		& \mgii & 2000 & 2010 	&        		   &    19.3   $\pm$   0.4      &    \\
		& \mgii & 2000 & 2011 	&        		   &     8.0   $\pm$   0.9      &    \\
\hline
                & \aliii& sdss1 & 2008  &			    &-3.4 $\pm$ 1.3 &\\ 
SDSS J1334-0123 & \aliii& sdss1 & 2009  & 5142.5  -     5302.9      &-1.2 $\pm$ 0.9 &\\
                & \aliii& sdss1 & sdss2 &                           & 0.3 $\pm$ 1.4 &\\
\hline
                & \mgii	& sdss1 & 2010 &			    &     1.9 $\pm$ 0.9 & \\
		& \mgii	& sdss1 & 2011 & 6979.6  -     7064.4	    &     2.6 $\pm$ 0.6 & \\
		& \mgii	& sdss1 & sdss2 &        		    &     1.4 $\pm$ 0.4 & \\
SDSS J1448+0424 & \aliii& sdss1 & 2010 &			    &      1.5 $\pm$ 1.2 & \\
                & \aliii& sdss1 & 2011 &    4624.1  -     4698.8    &      3.2 $\pm$ 0.6 & \\
		& \aliii& sdss1 & sdss2 &        		    &      2.0 $\pm$ 0.4 & \\
\hline
		& \mgii & sdss & 2008 &     			   &    -1.3  $\pm$    0.5      &       \\
SDSS J1614+3752 & \mgii & sdss & 2010 &     4250.4  -     4301.4   &    -0.7  $\pm$    0.4      &       \\
		& \mgii & sdss & 2011 &     			   &    -0.2  $\pm$    0.4      &        \\
\hline
    		& \mgii	& sdss & vlt   &  			  &   -3.0   $\pm$    0.6 &\\
                & \mgii	& sdss & 2008  &  			  &   -8.3   $\pm$   0.6 & Variable\\
SDSS J2215-0045 & \mgii	& sdss & mage  &6512.2  -     6735.2      &   -7.6   $\pm$   1.8 &\\
                & \mgii	& sdss & 2011  &			  &   -7.5   $\pm$   1.0 &\\
    		& \aliii& sdss & vlt  &     			    &  -21.2  $\pm$    0.6 &\\ 
                & \aliii&  vlt & 2008  &     			    &  -37.0  $\pm$    0.8 &\\
		& \aliii& 2008 & 2010  & 4345.0   -    4419.6      &   -37.5  $\pm$    1.3 &\\
	        & \aliii& 2010 & mage  &     			    &  -36.2  $\pm$    0.9 &\\
                & \aliii& Mage & 2011  &     			    &  -36.6  $\pm$    1.2 &\\

\hline
                & \mgii & sdss1& sdss2 &    7448.6   -    7716.4  & 0.8  $\pm$   1.2&\\
SDSS J2347-1037 &\aliii & sdss1& sdss2&                           &-4.0  $\pm$ 0.9 & \\ 
                &\aliii & 2008 & 2008 &  4932.0   -    5110.4	  &-1.0  $\pm$   1.3 & \\
 		&\aliii & 2009 & 2009 &      			  & 2.8  $\pm$   2.7 & \\

\hline
\hline

\end{tabular}
\begin{flushleft}
  $^{\dagger}$ 7 and 1 in the parenthesis refers to IFOSC 7 and IFORS 1 data respectively \\
      \end{flushleft}
\end{table*}

\def\aj{AJ}%
\def\actaa{Acta Astron.}%
\def\araa{ARA\&A}%
\def\apj{ApJ}%
\def\apjl{ApJ}%
\def\apjs{ApJS}%
\def\ao{Appl.~Opt.}%
\def\apss{Ap\&SS}%
\def\aap{A\&A}%
\def\aapr{A\&A~Rev.}%
\def\aaps{A\&AS}%
\def\azh{AZh}%
\def\baas{BAAS}%
\def\bac{Bull. astr. Inst. Czechosl.}%
\def\caa{Chinese Astron. Astrophys.}%
\def\cjaa{Chinese J. Astron. Astrophys.}%
\def\icarus{Icarus}%
\def\jcap{J. Cosmology Astropart. Phys.}%
\def\jrasc{JRASC}%
\def\mnras{MNRAS}%
\def\memras{MmRAS}%
\def\na{New A}%
\def\nar{New A Rev.}%
\def\pasa{PASA}%
\def\pra{Phys.~Rev.~A}%
\def\prb{Phys.~Rev.~B}%
\def\prc{Phys.~Rev.~C}%
\def\prd{Phys.~Rev.~D}%
\def\pre{Phys.~Rev.~E}%
\def\prl{Phys.~Rev.~Lett.}%
\def\pasp{PASP}%
\def\pasj{PASJ}%
\def\qjras{QJRAS}
\def\rmxaa{Rev. Mexicana Astron. Astrofis.}%
\def\skytel{S\&T}%
\def\solphys{Sol.~Phys.}%
\def\sovast{Soviet~Ast.}%
\def\ssr{Space~Sci.~Rev.}%
\def\zap{ZAp}%
\def\nat{Nature}%
\def\iaucirc{IAU~Circ.}%
\def\aplett{Astrophys.~Lett.}%
\def\apspr{Astrophys.~Space~Phys.~Res.}%
\def\bain{Bull.~Astron.~Inst.~Netherlands}%
\def\fcp{Fund.~Cosmic~Phys.}%
\def\gca{Geochim.~Cosmochim.~Acta}%
\def\grl{Geophys.~Res.~Lett.}%
\def\jcp{J.~Chem.~Phys.}%
\def\jgr{J.~Geophys.~Res.}%
\def\jqsrt{J.~Quant.~Spec.~Radiat.~Transf.}%
\def\memsai{Mem.~Soc.~Astron.~Italiana}%
\def\nphysa{Nucl.~Phys.~A}%
\def\physrep{Phys.~Rep.}%
\def\physscr{Phys.~Scr}%
\def\planss{Planet.~Space~Sci.}%
\def\procspie{Proc.~SPIE}%
\let\astap=\aap
\let\apjlett=\apjl
\let\apjsupp=\apjs
\let\applopt=\ao
\bibliographystyle{mn}
\bibliography{sample}
\appendix

\section{Notes on Individual Variable BAL Sources}
In this section, we discuss  individual QSOs that have shown significant variability of the broad absorption lines.   The plots depicting variabilities in these sources are given in Fig.~\ref{zoom_pic}.  The absorption line variabilities can be compared to the variations in the continuum. As the CRTS survey doesn't cover  SDSS epochs for  sources in our sample,  comparison between the absorption line and continuum variations are better done when absorption line variations are seen between IGO epochs.  
\subsection{SDSS J0045+1438}
 As the \mgii\ absorption in this case falls in the wavelength range just outside the 
IFOSC/GR8 coverage, we only study the variability of \aliii\ and \civ\ 
absorption lines using our IGO data. The  optical-UV SED fitting using both the 
SDSS composite and the  Fe template did not give good fits to the continuum. 
Hence, a simple polynomial fit is used to approximate the continuum. 
SDSS has three epochs of data. The \mgii\ line is only covered in the 
SDSS spectrum and has a maximum ejection velocity of $\sim$ 5800 \kms. 
This source doesn't show any variability in any of the absorption lines 
between different SDSS epochs.   But, there are variations seen in \aliii\  
line between  SDSS and IGO epochs (see Fig.~\ref{zoom_pic} and \ref{0045_diff}).   The absorption line has 
decreased in optical depth in the IGO spectrum as compared to those observed
in SDSS. Even within the IGO epochs, the \aliii\ absorption  continued 
to weaken between the years 2007 and 2009. Both the equivalent width 
measurements and the integrated ratio values point to similar trends in 
variations. We detect 5$\sigma$ variation in \aliii\ equivalent widths 
between SDSS and IGO spectrum observed in 2009 (see Table~\ref{EW_table}) 
epoch.  A similar trend in variation is also seen in the \civ\ lines.  
The ratio values point to changes at about 4$\sigma$ level 
(see Table~\ref{IR_table}) between the SDSS and IGO epochs, but not 
between different IGO epochs. The light curve of this source shows a 
small decrease in the magnitudes over the CRTS monitoring period. But, 
no significant variability over the period of IGO spectroscopic 
monitoring (see Fig.~\ref{lc_1}). There is also no difference seen in the 
fiber magnitudes 
for the SDSS epochs. The scatter in the $\Delta$m values is   0.04 around  
a median value of -0.04. The $\Delta$m-$\Delta$t graph has no significant 
slope. All the continuum parameters point to no appreciable variation 
in the continuum flux for this source.   Therefore, it is most unlikely that the noted variation 
in \aliii\ and \civ\ absorption lines is  caused by a  
change in the ionization state of the gas.
%

\subsection{SDSS J0149-0103}
The emission redshift of the source  is 1.0740. The \mgii\ BAL without having associated \feii\ absorption is  at a 
redshift of 1.0320 and a maximum ejection velocity of 7900 \kms.   There is a narrow intervening \mgii\ absorption at a 
redshift of 0.428.   We fitted the continuum using the SDSS quasar composite as the template. The \mgii\ absorption has deepened in 2008 and 2009 IGO 
spectra  as compared to that seen in the SDSS spectrum 
(see Fig.~\ref{zoom_pic} and \ref{0149_diff}).   Between the IGO epochs 2008 
and 2009, the absorption has not changed. SDSS BOSS survey has 
re-observed this source in 2010. The absorption decreased in strength  
in 2010 and continued to weaken in the IGO  spectrum observed in 2012. 
This trend is clearly seen  both in the equivalent width measurements 
(see Table~\ref{EW_table}) and in the integrated ratio measurements 
(see Table~\ref{IR_table}). The maximum variation is seen between the 
epochs  2009 and 2012. We detect more than 3$\sigma$ variation in 
equivalent widths and 7$\sigma$ variation in the integrated ratio 
values. This is a clear case of confirmed variations between the epochs, 
even within short time-scales probed by the IGO observations. 
As can be seen from Fig.~\ref{zoom_pic}, it appears that, 
for the strongest absorption component, the peak optical depth is blue-shifted
by about $\sim$440 \kms\ in IGO observations  as compared to that
seen in the original SDSS data. 
%
However, no such shifts is noted between spectra obtained in different IGO epochs.

\begin{figure*}
 \centering
\psfig{figure=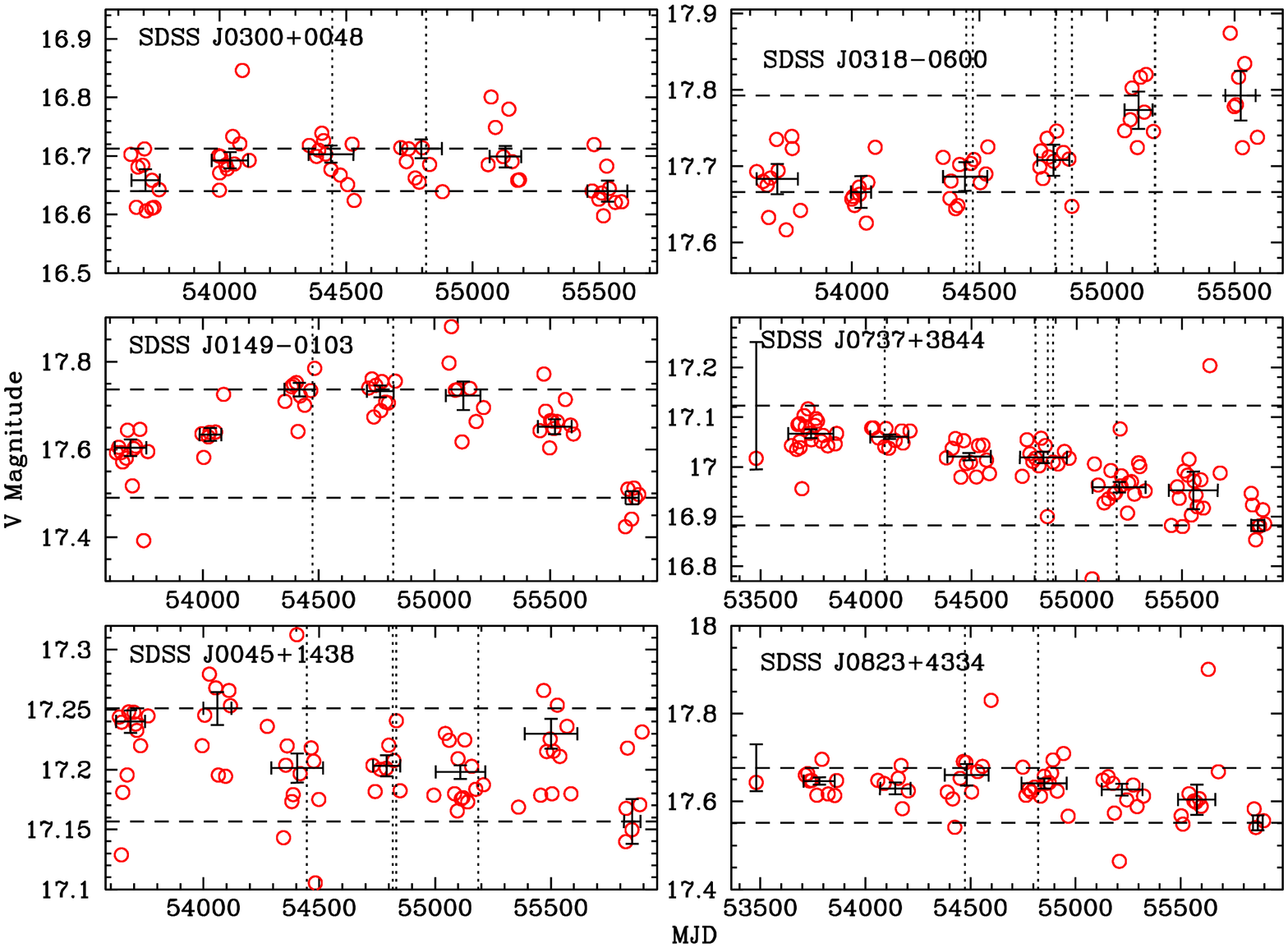,width=1.0\linewidth,height=1.0\linewidth,angle=270}
\caption{  CRTS light curves of  LoBAL QSOs in the sample. The dotted vertical lines mark the epochs (MJDs) of spectroscopic observations. The average magnitudes obtained for closely spaced observations are overplotted as points with error bars for all sources. The x-axis error bars correspond to the time range over which the magnitudes are averaged. }
\label{lc_1}

\end{figure*}

The CRTS light curve clearly shows significant variations in the continuum 
flux. There is a coherent trend of initial decrease in the continuum flux followed
by the brightening of the source (see Fig.~\ref{lc_1}) in the CRTS light
curve. We notice that between SDSS and the first IGO observation, the QSO 
has become faint, in line with the trend seen in  CRTS. 
Our IGO spectroscopic
observations are mostly during the phase when the QSO has brightened.
The standard deviation and the median of the magnitude differences are  
0.13  and -0.01. Large scatter in the magnitude differences implies large variations in the magnitudes at all time-scales. 
The $\Delta$m-$\Delta$t slope is -0.046$\pm$0.007 mag/year.  
This source clearly suggests a possible correlation between the absorption 
line and continuum variations. The \mgii\ absorption line optical depth 
gained in strength  when the source weakened in continuum and vice versa. 
 We have reported a similar variation for the source SDSS J1333+0012 \citep{vivek12} where the source displayed  opposite trends in variations of the QSO flux and the \mgii\ BAL column densities. Like in that case, correlated variability
between optical depth and continuum light suggests a possible connection
between them. However, wavelength shift seen for the peak optical depth noted between SDSS 
and IGO data suggests that  the changes may not purely be triggered by  
ionization changes.



\subsection{SDSS J0850+4451}

The emission redshift of this source is 0.541. The \mgii\ BAL is at a 
redshift of 0.524. The absorption line has an ejection velocity of 
1800 \kms and is spread over 3500 \kms. We fitted the continuum of this source using the SDSS quasar composite spectrum as template.
The equivalent width of the \mgii\ BAL  increased  between the SDSS and IGO epochs with 
more than 10$\sigma$ significance (see Fig.~\ref{zoom_pic} and \ref{0850_diff}). The variation is seen in the strongest part of the absorption. 
There are no variations in the BAL features observed between the IGO epochs 
(see Table~\ref{EW_table} and Table~\ref{IR_table}). The CRTS continuum 
light curve for this source shows a  trend for increase in magnitudes. But, the continuum has not  varied significantly by more than 0.1 mag (see Fig.~\ref{lc_2}). $\Delta$m values for this source has a 
median at  0.01 with a scatter of 0.05 magnitude. But, in comparison with 
the SDSS transformed V magnitudes, the continuum flux  decreased 
significantly ( by 0.5 mag). 
Therefore, it seems even in this source, absorption line variability follows the  variations in the continuum. As the optical depth has varied over the full profile,
one can not rule out the possibility of the optical depth variation being
triggered by the continuum variation of the QSO in this case.     

\subsection{SDSS J0944+0625}
\begin{figure*}
 \centering
\psfig{figure=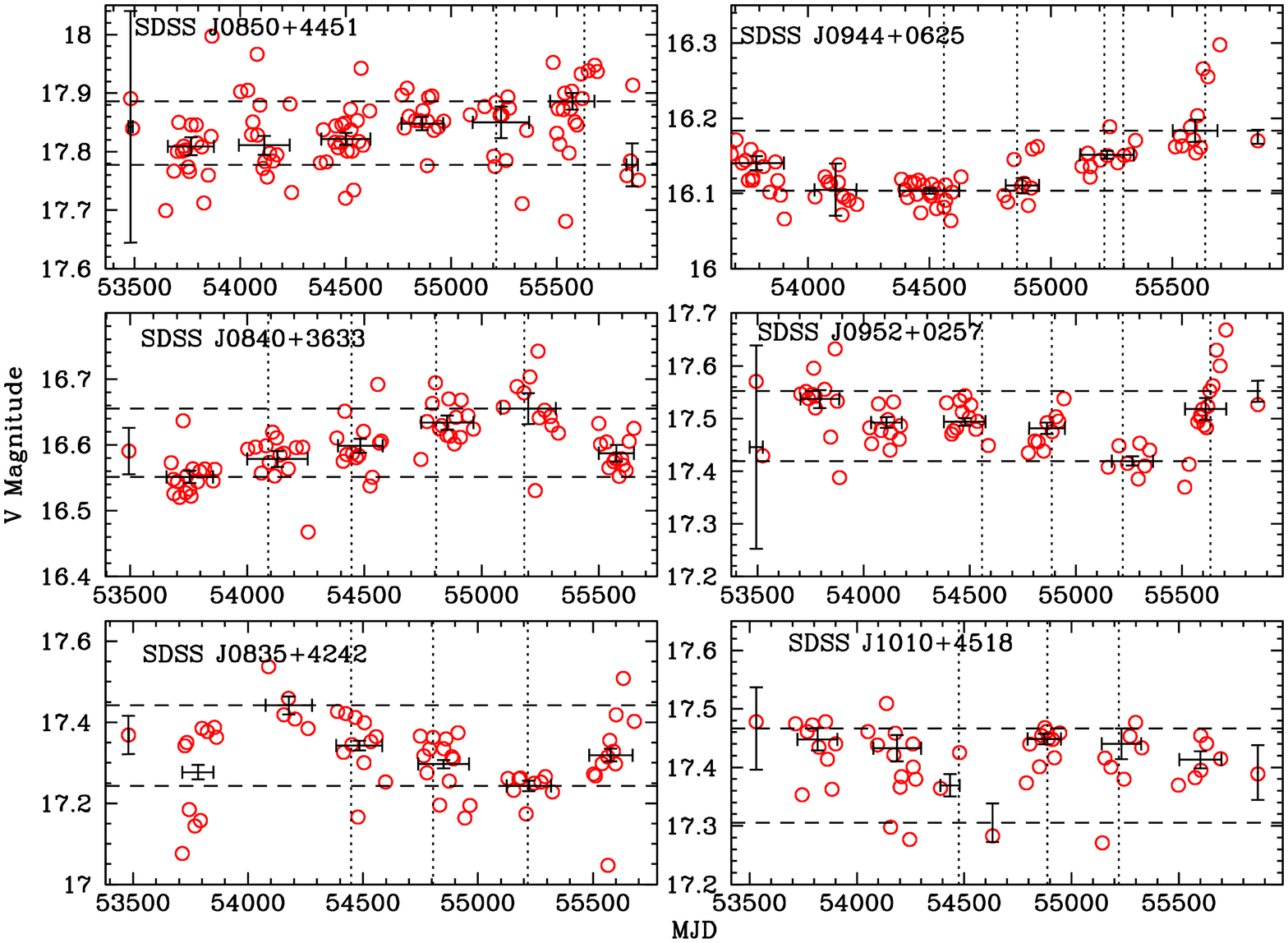,width=1.0\linewidth,height=1.0\linewidth,angle=270}
\caption{CRTS light curves of  LoBAL QSOs in the sample. The dotted vertical lines mark the epochs (MJDs) of spectroscopic observations. The average magnitudes obtained for closely spaced observations are overplotted as points with error bars for all sources. The x-axis error bars correspond to the time range over which the magnitudes are averaged. }
\label{lc_2}
\end{figure*}

 This source at \zem $\sim$ 0.6949 has  emission lines from \mgii, 
H$\beta$ and H$\gamma$. There is a  broad \mgii\ absorption line. 
The strongest BAL absorption line is at a redshift of 0.6529 and has a maximum velocity of 8500 \kms . 
Continuum is dominated by Fe emission.  We fitted the continuum with a 
Fe emission template.  The broad dip (see Fig.~\ref{zoom_pic} and 
Fig.~\ref{0944_diff}) in the higher velocity edge of the \mgii\ absorption 
could well be a part  of the Fe emission  optical-UV SED. So, we do not include this 
region while measuring the equivalent width or integrated ratio to check the variability. 
Here, again there are variations in the \mgii\ absorption line depth between different epochs 
(see Table~\ref{EW_table} and Table~\ref{IR_table}). As compared to the SDSS, 
the \mgii\ BAL weakened in 2008. Between, 2008 and 2009 there is not much 
variation. But, in 2010 and 2011, the BAL weakened further. The maximum 
absorption line variation is seen between the SDSS and the 2010 epochs with 15$\sigma$ 
significance for the equivalent width variation.  The CRTS light curve 
(in Fig.~\ref{lc_2}) shows an initial brightening  followed by a steady 
decrease in the continuum flux by 0.1 magnitude in the IGO epochs. The 
magnitudes did not change between 2008 and 2009, but steadily decreased 
after 2009. As compared to the SDSS  epoch , the QSO has brightened by 0.1 
magnitude in the IGO 2008 epoch.  Magnitude differences have a median of  
0.04  and a standard deviation of   0.04. The  $\Delta$m-$\Delta$t slope 
is negligible.  Between the SDSS and IGO 
epoch, the source brightened whereas the BAL weakened. Between the IGO epochs, 
the source weakened in the continuum flux while the BAL continued to weaken. As the 
equivalent width variation is monotonically decreasing whereas the continuum
light curve is curved, we can probably rule out the  BAL variability being driven by ionization changes.  

\subsection{SDSS J0952+0257}

\begin{figure*}
 \centering
\psfig{figure=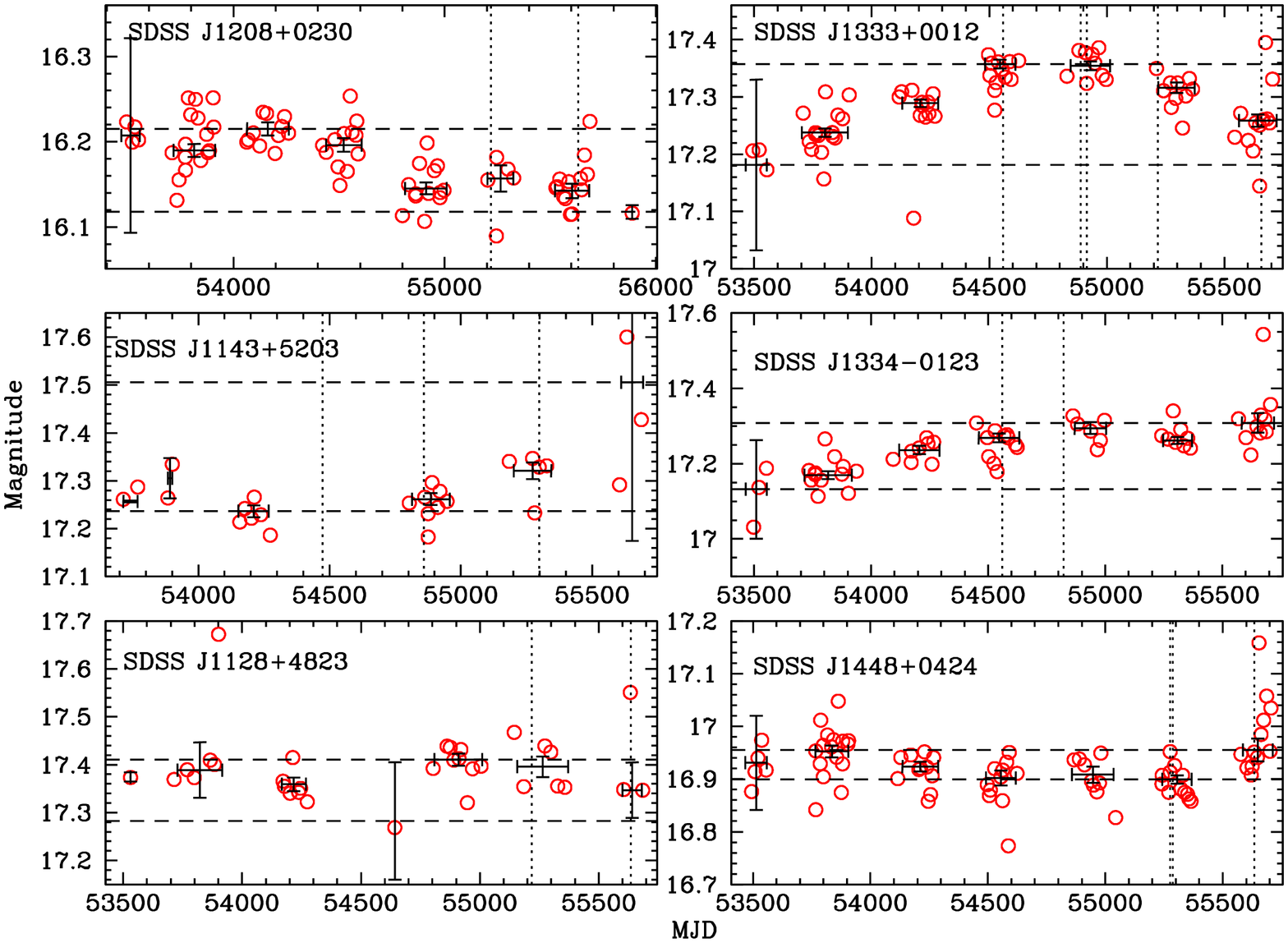,width=1.0\linewidth,height=1.0\linewidth,angle=270}
\caption{CRTS light curves of  LoBAL QSOs in the sample. The dotted vertical lines mark the epochs (MJDs) of spectroscopic observations. The average magnitudes obtained for closely spaced observations are overplotted as points with error bars for all sources. The x-axis error bars correspond to the time range over which the magnitudes are averaged. }
\label{lc_3}
\end{figure*}

This source (\zem $\sim$ 1.355) has broad absorption lines (\zabs$\sim$ 1.300) 
from \aliii\ and \mgii. Continuum has significant contribution from iron 
emission lines. Hence, we fitted the  optical-UV SED with a Fe emission template. 
\mgii\ and \aliii\ lines span a velocity widths of $\sim$ 8500 \kms and 
5700 \kms respectively. Some part of the highest velocity \mgii\ BAL is 
contributed by features in the Fe emission lines. Small variations 
are seen in the \mgii\ and \aliii\ between the SDSS and IGO epochs. 
In all the IGO epochs, both \mgii\ and \aliii\ weakened slightly  
as compared to the SDSS  epoch with more than 3$\sigma$ significance 
(see Table~\ref{EW_table} and Fig.~\ref{zoom_pic}). There are no significant variations within 
the IGO epochs.  Both \mgii\ and \aliii\ varied in the same direction. 
For this QSO, $\Delta$m   has a median of  -0.03 and standard deviation of  0.06. The $\Delta$m-$\Delta$t slope is -0.002$\pm$0.007 mag/year. 
From the light curve in Fig.~\ref{lc_3}, it is clear that the continuum flux has not 
changed much between different IGO observations.
As compared to the SDSS  epoch, the QSO brightened by 0.1 magnitude 
in the IGO epoch 2008. 
In this source, between the SDSS and IGO epochs, the \mgii\ BAL weakened when the QSO flux  increased and between the IGO epochs, the \mgii\ BAL remained constant while the continuum light curve shows initial brightening followed by a mild fading trend. As the optical depth variations are seen
over the full absorption profile, one will not be able to rule out the
ionization changes causing the line variability in this source.


\subsection{SDSS J1208+0230}
 The source contains emission lines from \mgii\ and \aliii\ at an emission redshift of 1.179. In addition to a broad \mgii\ absorption, the spectra contains a narrow \mgii\ absorption  at z$_{abs}$$\sim$0.6679. The associated \feii\ and \aliii\ lines are redshifted below the spectral range covered. Strongest BAL component is at a redshift 1.0671. The BAL is stronger  at the high velocity end and has a maximum ejection velocity and velocity spread of 18000 \kms and 3600 \kms respectively.  As the continuum of this source is dominated by strong \feii\ emission, we fitted the continuum with the template spectrum of a strong Fe emitting quasar. The unusual feature in this source is that the  \mgii\ emission line is comparatively narrower. The \mgii\ absorption strength has reduced in the IGO data as compared to the SDSS (see Fig.~\ref{zoom_pic}). Maximum variation is seen between the SDSS  epoch and the IGO 2011 epoch with 5$\sigma$ significance in the equivalent widths (see Tables~\ref{EW_table} \& \ref{IR_table}). It is also evident from Fig.~\ref{zoom_pic}
that the variation in Mg~{\sc ii} profile is not uniform. The centroid of
the \mgii\ absorption is blue shifted by about 320 \kms between SDSS and IGO 2011 epoch.
The CRTS light curve for this QSO, while showing a gradual decrease in 
magnitudes,  shows that the continuum flux of this source has not varied significantly (i.e. $>$ 0.1 mag)  between our observation period. But, it has brightened by 0.2 magnitude between SDSS and IGO epochs. Magnitude differences has a median of  0.04,  standard deviation of  0.03 and a slope of -0.032 $\pm$0.004 mag/yr.  Although the light curve shows a trend of decrease in the continuum flux within IGO epochs, no continuum variations $\geq$ 0.1 magnitudes are observed. The \feiii\ and \aliii\ emission lines has weakened significantly between the SDSS and IGO epochs.  Again in this source,  the absorption lines weakened when the QSO continuum brightened in flux. However, profile
shape variation noted above suggests that the ionization change may not be
the sole reason for the changes in \mgii\ optical depth in this source.

\subsection{SDSS J1333+0012}
Variation in this source has already been reported in our previous paper \citep{vivek12}. There is dynamical evolution of  two components of \mgii\ absorption in this source. The red component (at a velocity of 1.7$\times 10^4$ \kms) completely disappeared and a blue component (at a velocity of
2.8$\times 10^4$ \kms) emerged and disappeared during our observation campaign. Significant changes are also seen in the continuum light curve. The QSO first 
brightened significantly between SDSS and IGO 2008 epochs, reached a maximum flux in 2009 and dimmed thereafter. A similar but opposite trend is seen in the 
column densities of the BALs \citep[see Fig.~2 of][]{vivek12}.  The observed 
variability is best explained by a motion of absorbing clouds across the line of sight.

\subsection{SDSS J2215-0045}
This source is a  FeLoBAL QSO where dramatic variation in the absorption lines has been reported in our recent paper \citep{vivek12a}. The spectrum has absorption lines from \mgii, \aliii, \civ\ and  \feiii\ fine structure lines. No line from \feii\ is detected. Observed optical depths suggested inverted population ratios for the two \feiii\ fine structure lines, \feiii\ UV 34 and \feiii\ UV 48.  Optical-UV SED fitting using a  quasar template having significant iron emission revealed that the observed  inverted optical depths  result from the over/under estimation of optical depths due to features in the iron emission at these wavelengths. All the  absorption lines, except \civ\ showed significant variations in the optical depth between the SDSS and IGO epochs. No variations are seen within IGO epochs.  The continuum light curve for this source also shows significant changes in the flux. The quasar  brightened significantly by more than 0.4 magnitudes after the SDSS  epoch. The $\Delta$m values are distributed around a median of -0.01 with a scatter of  0.2 magnitude and the slope of magnitude variations with elapsed time is -0.123$\pm$0.009 mag/year. This source shows maximum variations in the continuum flux among our sample. The observed variations in the BAL can be both explained by changes driven by continuum variations or by clouds moving across the line of sight \citep[see][for details]{vivek12a}.

\section{CRTS light curves}
\begin{figure*}
 \centering
\psfig{figure=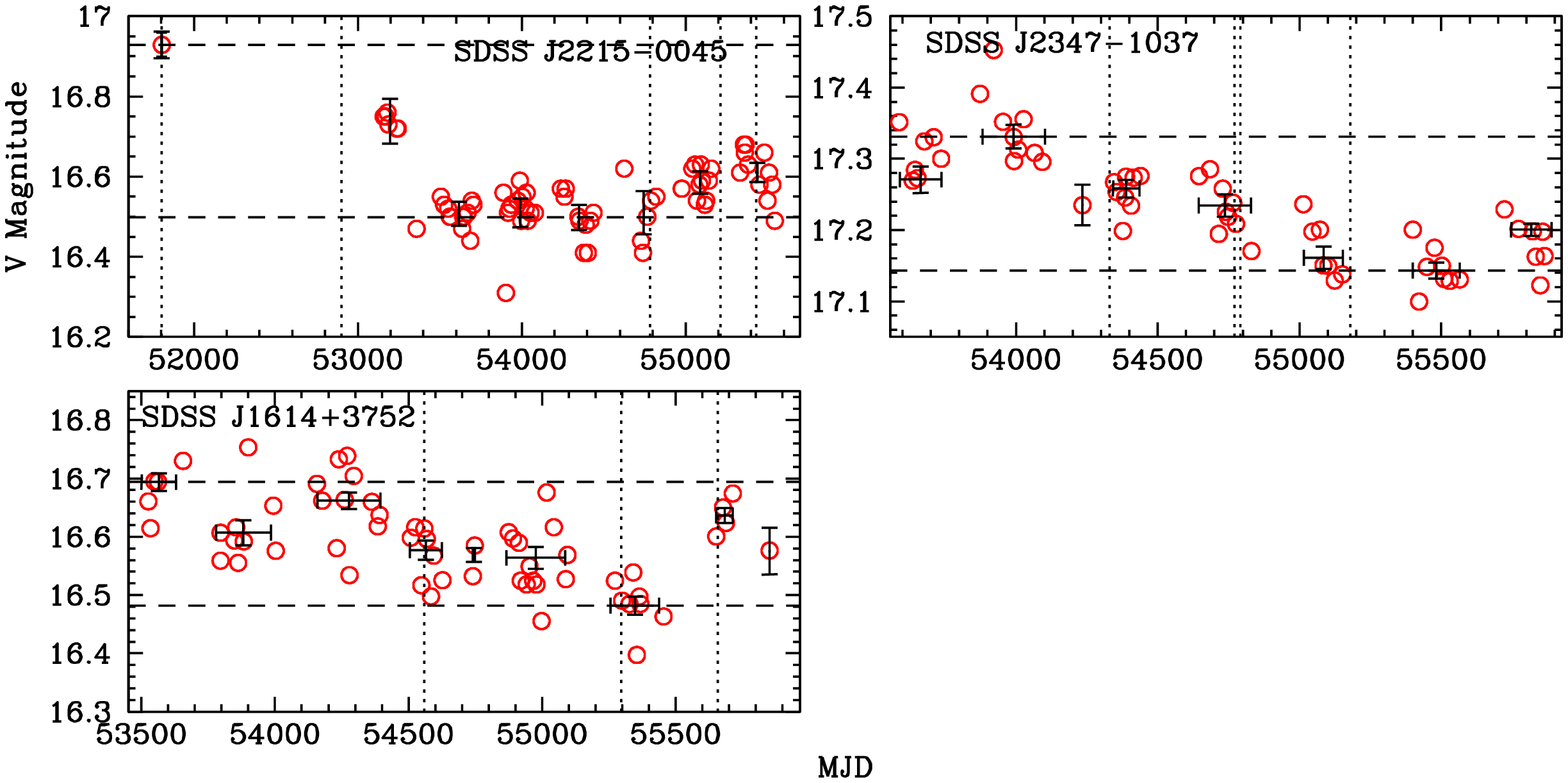,width=1.0\linewidth,height=1.0\linewidth,angle=270}
\caption{CRTS light curves of  LoBAL QSOs in the sample. The dotted vertical lines mark the epochs (MJDs) of spectroscopic observations. The average magnitudes obtained for closely spaced observations are overplotted as points with error bars for all sources. The x-axis error bars correspond to the time range over which the magnitudes are averaged. }
\label{lc_4}
\end{figure*}
\section{Spectral comparison}
\begin{figure*}
 \centering
\begin{tabular}{c c}
\psfig{figure=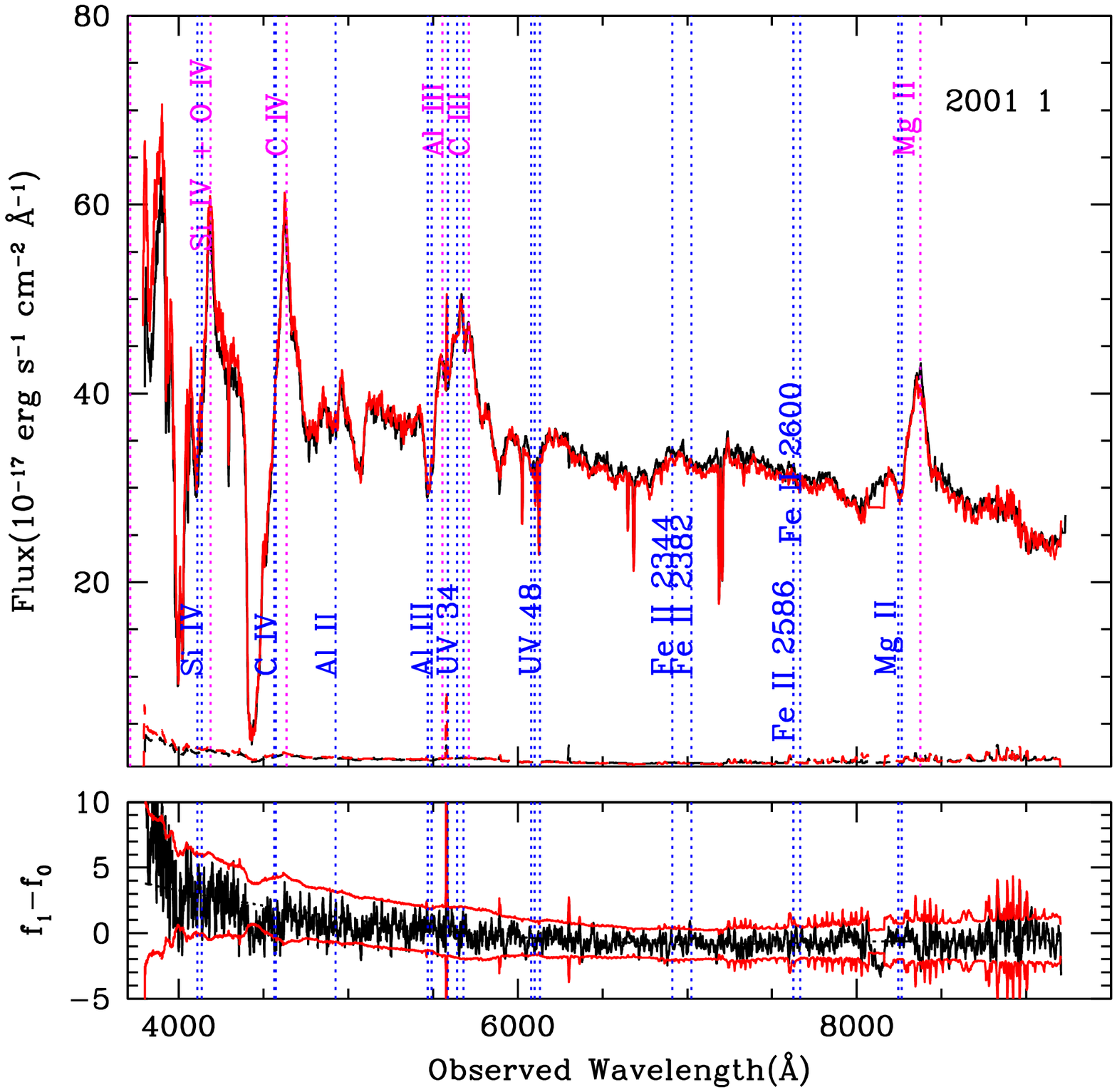,width=0.5\linewidth,height=0.4\linewidth}&
\psfig{figure=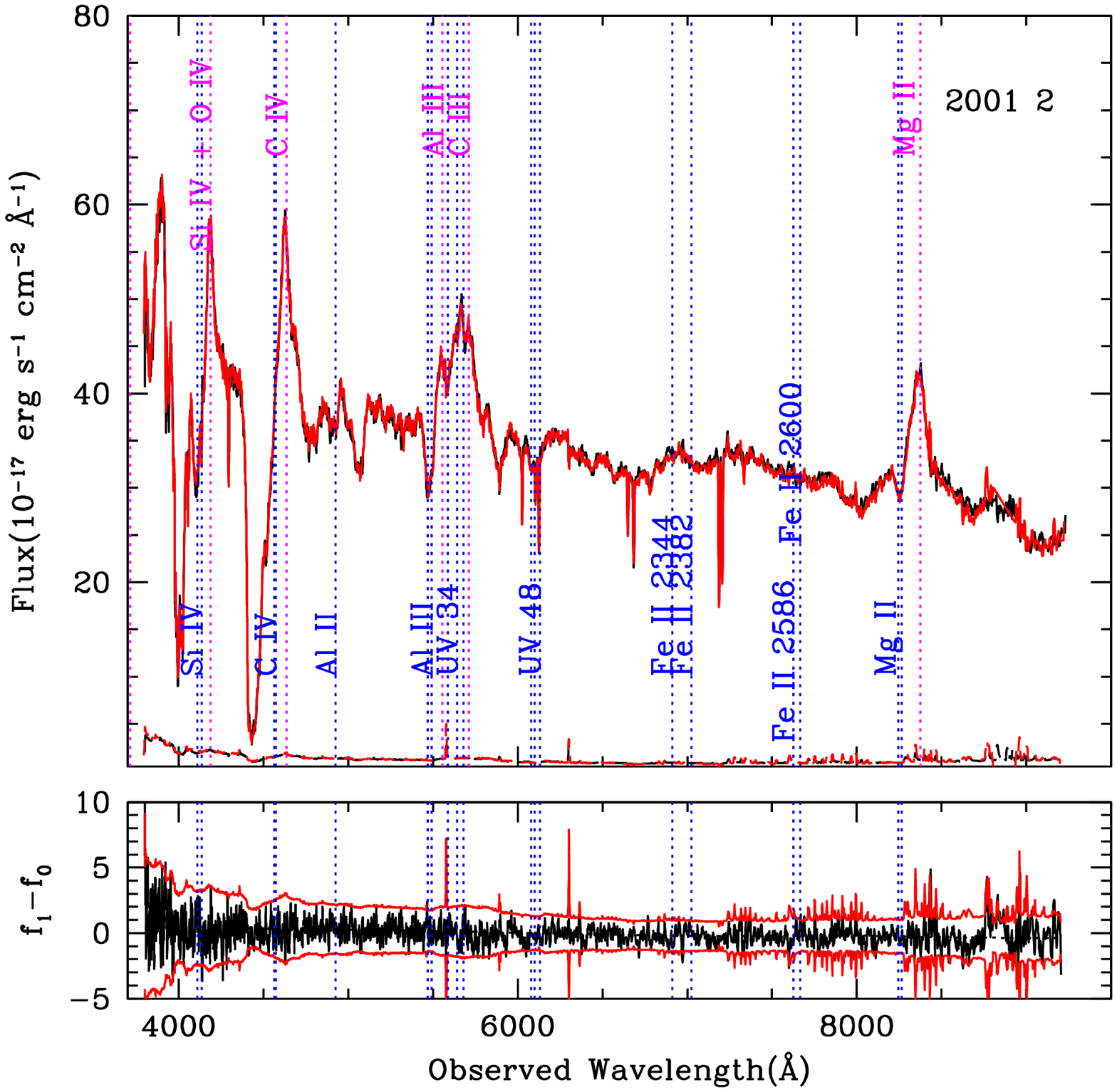,width=0.5\linewidth,height=0.4\linewidth}\\
\psfig{figure=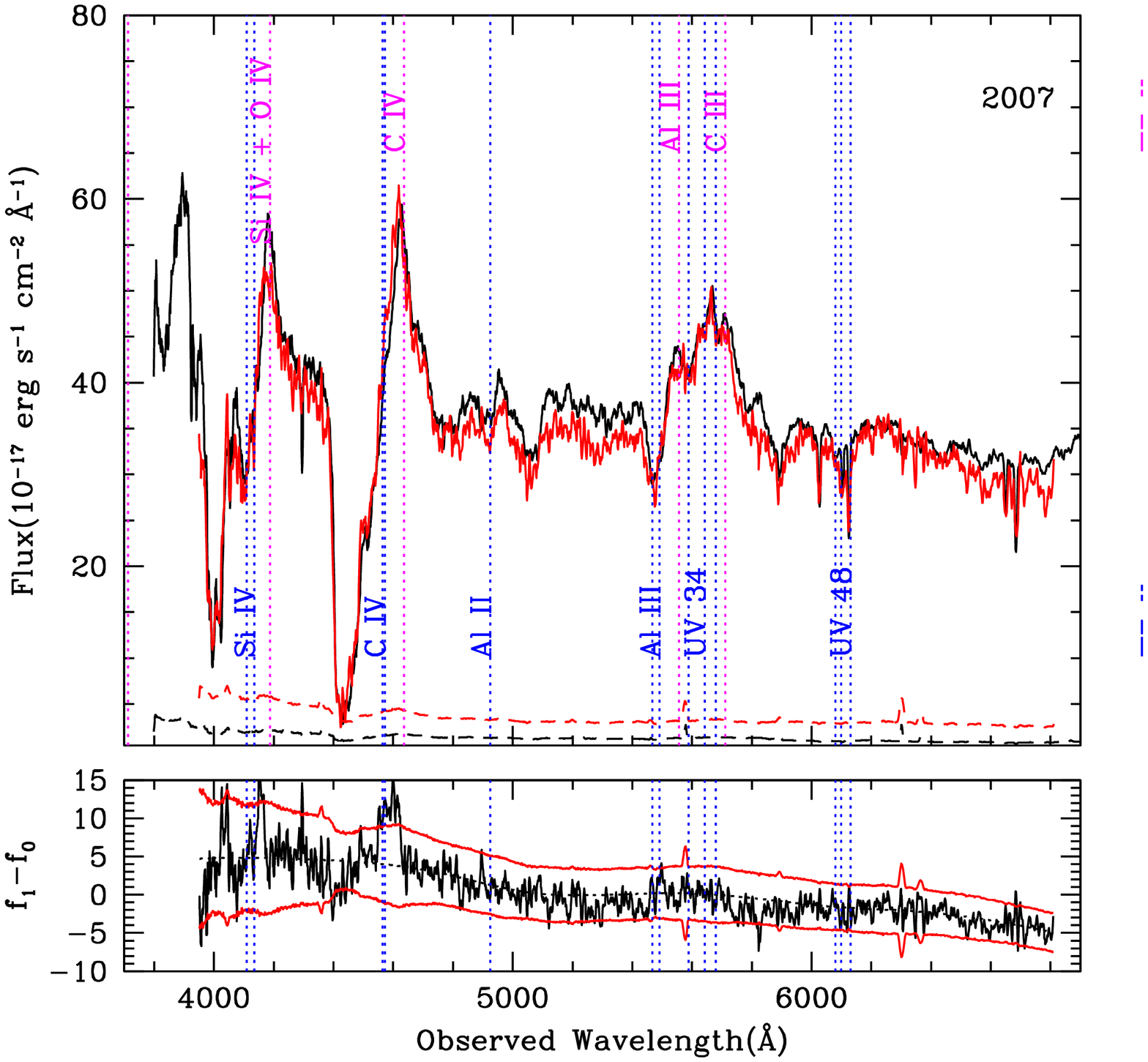,width=0.5\linewidth,height=0.4\linewidth,bbllx=18bp,bblly=144bp,bburx=592bp,bbury=718bp,clip=yes}&
\psfig{figure=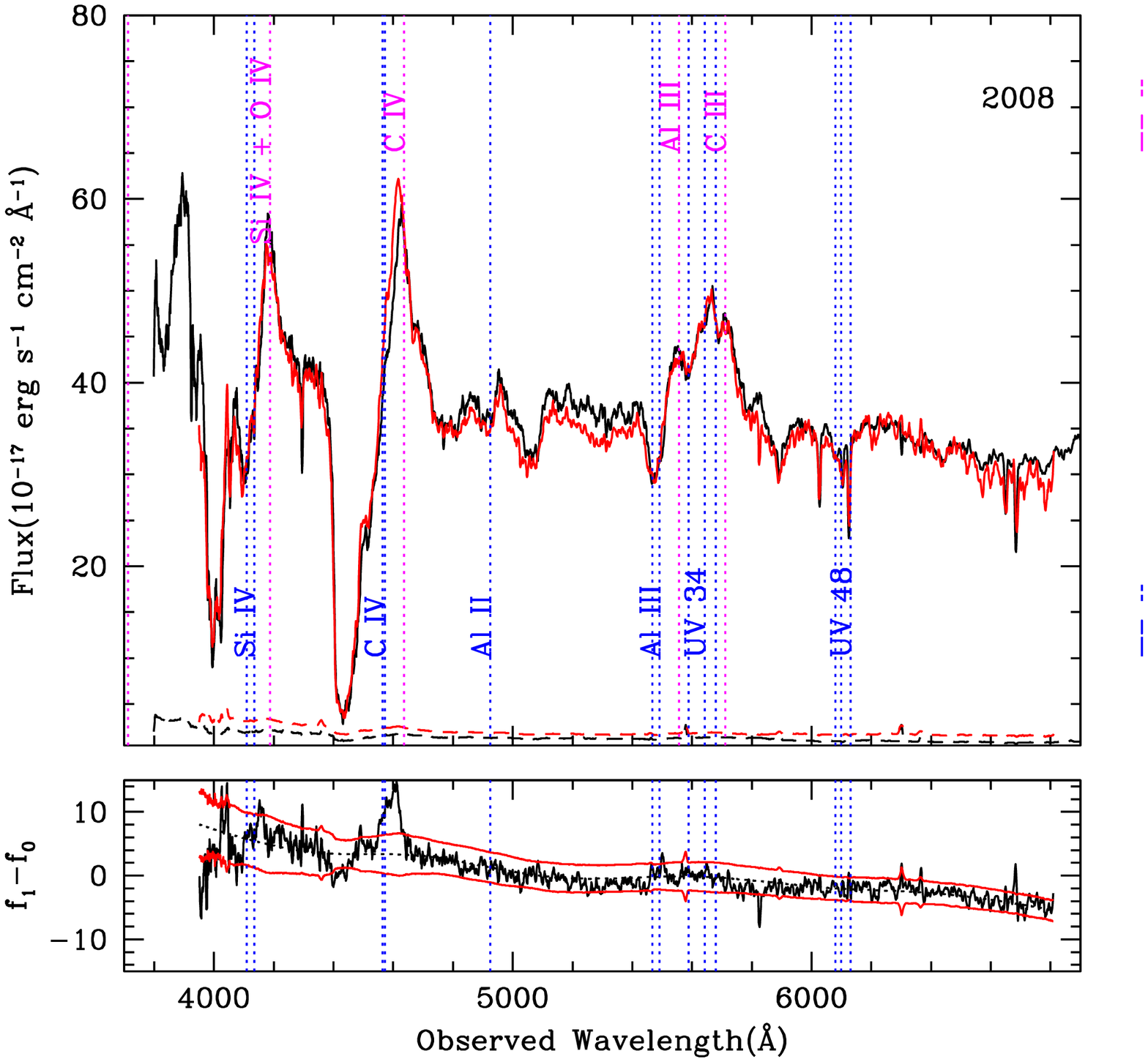,width=0.5\linewidth,height=0.4\linewidth,bbllx=18bp,bblly=144bp,bburx=592bp,bbury=718bp,clip=yes}\\
\psfig{figure=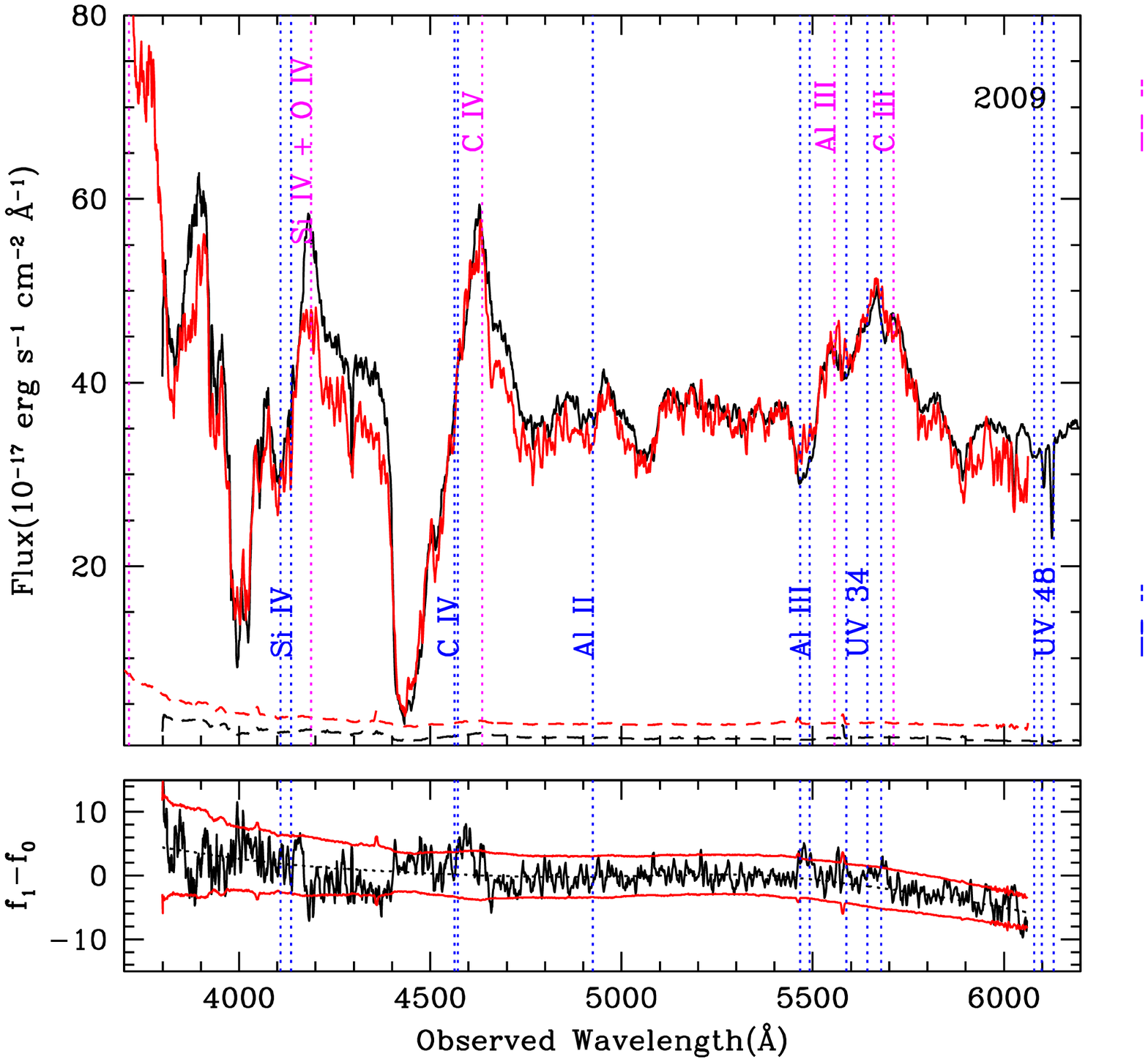,width=0.5\linewidth,height=0.4\linewidth,bbllx=18bp,bblly=144bp,bburx=592bp,bbury=718bp,clip=yes}&\\
\end{tabular}
\caption{IGO Spectra of SDSS J0045$+$1438  observed on MJD  and 54447,54818 and 55187 and SDSS spectra obtained on 51868 and  51879 (in red/grey) are overplotted with the reference SDSS spectrum (black) observed on MJD 51812 . The comparison of two IGO spectra are shown in the lower panel. The flux scale applies to the reference SDSS spectrum and all other spectra are scaled in flux to match the reference spectrum. In each plot, the error spectra are also shown. The difference spectrum for the corresponding MJDs is plotted in the lower panel of each plot. 1$\sigma$ error is plotted above and below the mean. }
\label{0045_diff}
\end{figure*}                                                                                                     
                                                                
\begin{figure*}
 \centering
\begin{tabular}{c c}
\psfig{figure=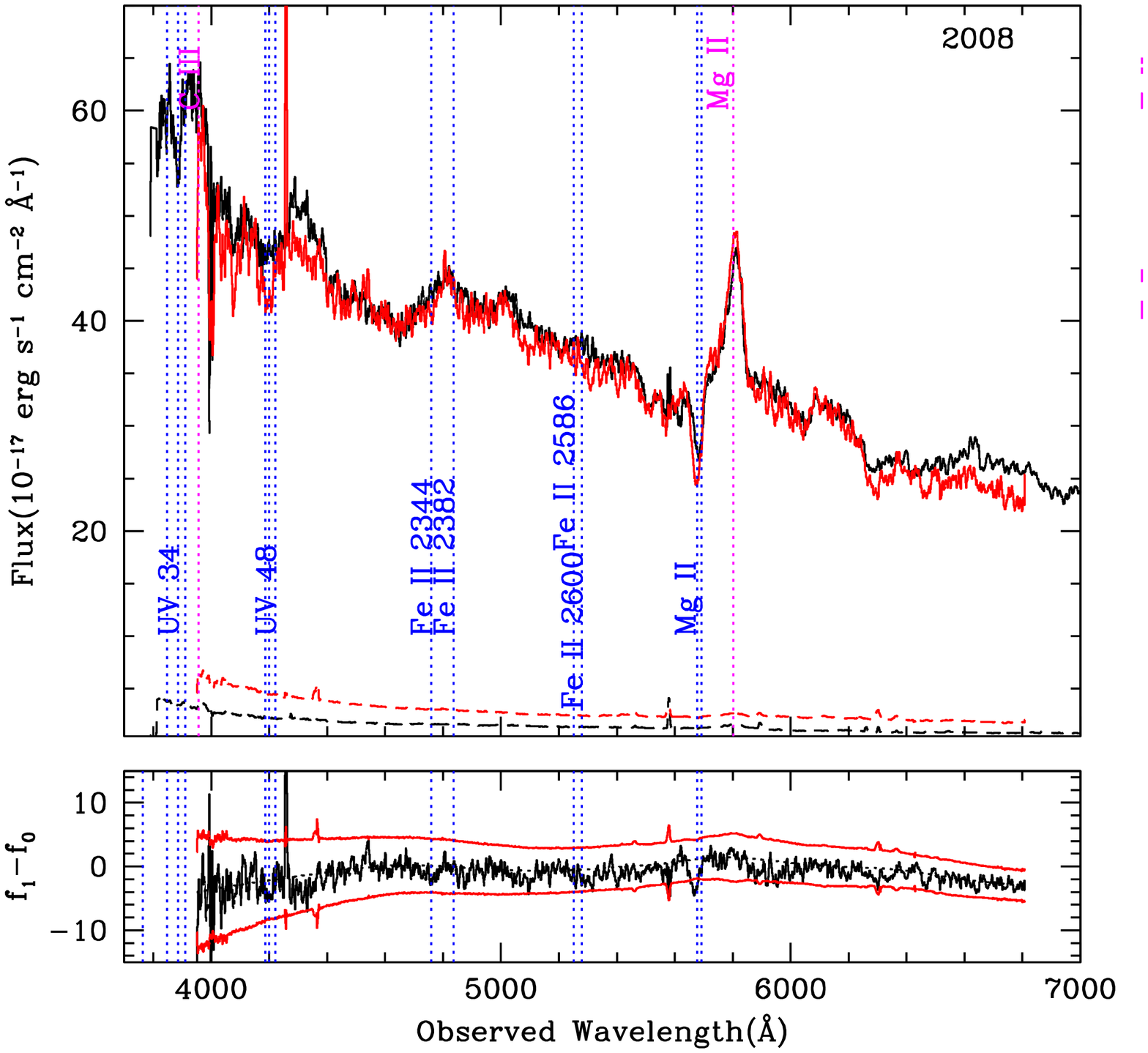,width=0.5\linewidth,height=0.4\linewidth,bbllx=18bp,bblly=144bp,bburx=592bp,bbury=718bp,clip=yes}&
\psfig{figure=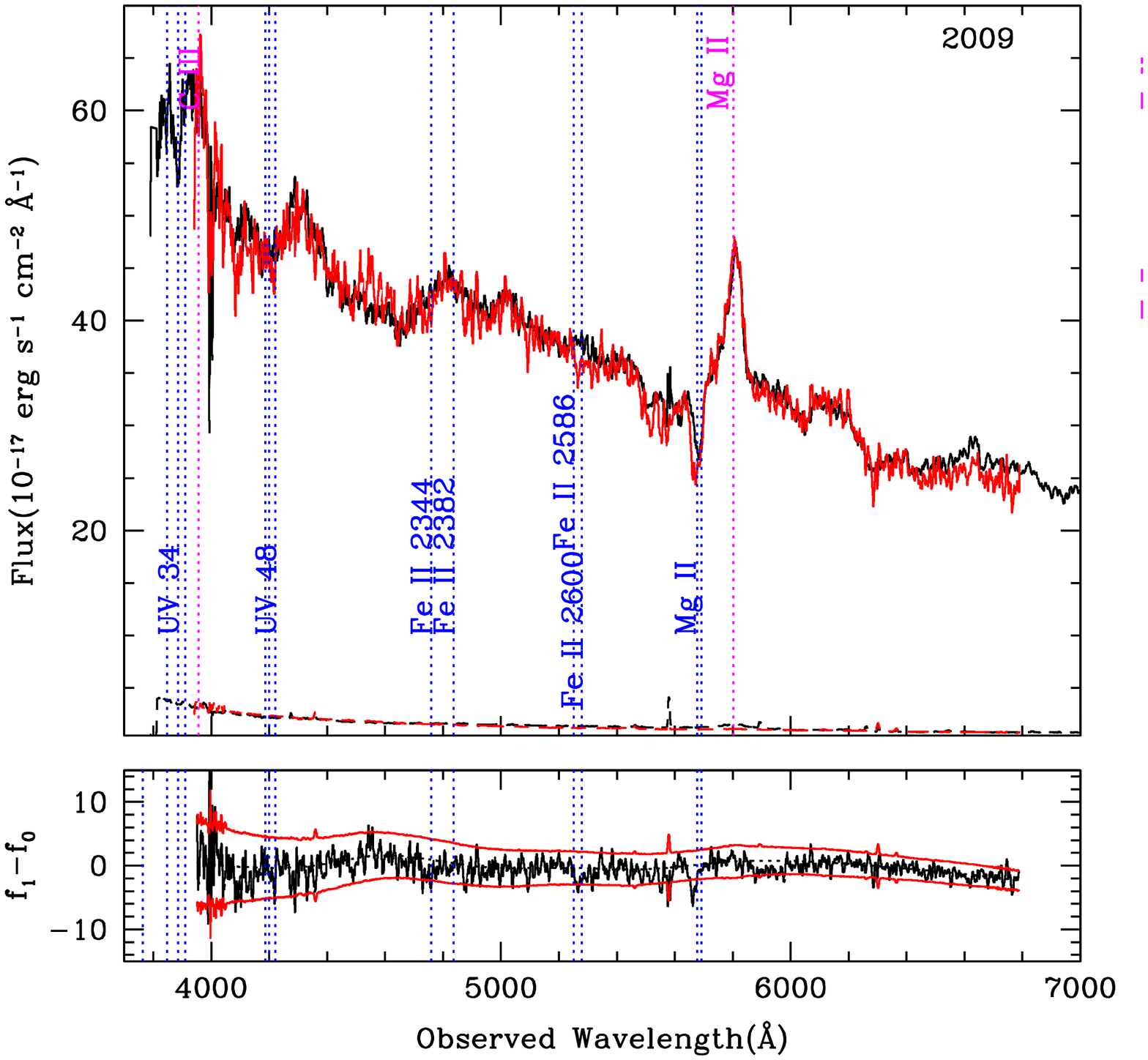,width=0.5\linewidth,height=0.4\linewidth,bbllx=18bp,bblly=144bp,bburx=592bp,bbury=718bp,clip=yes}\\
\psfig{figure=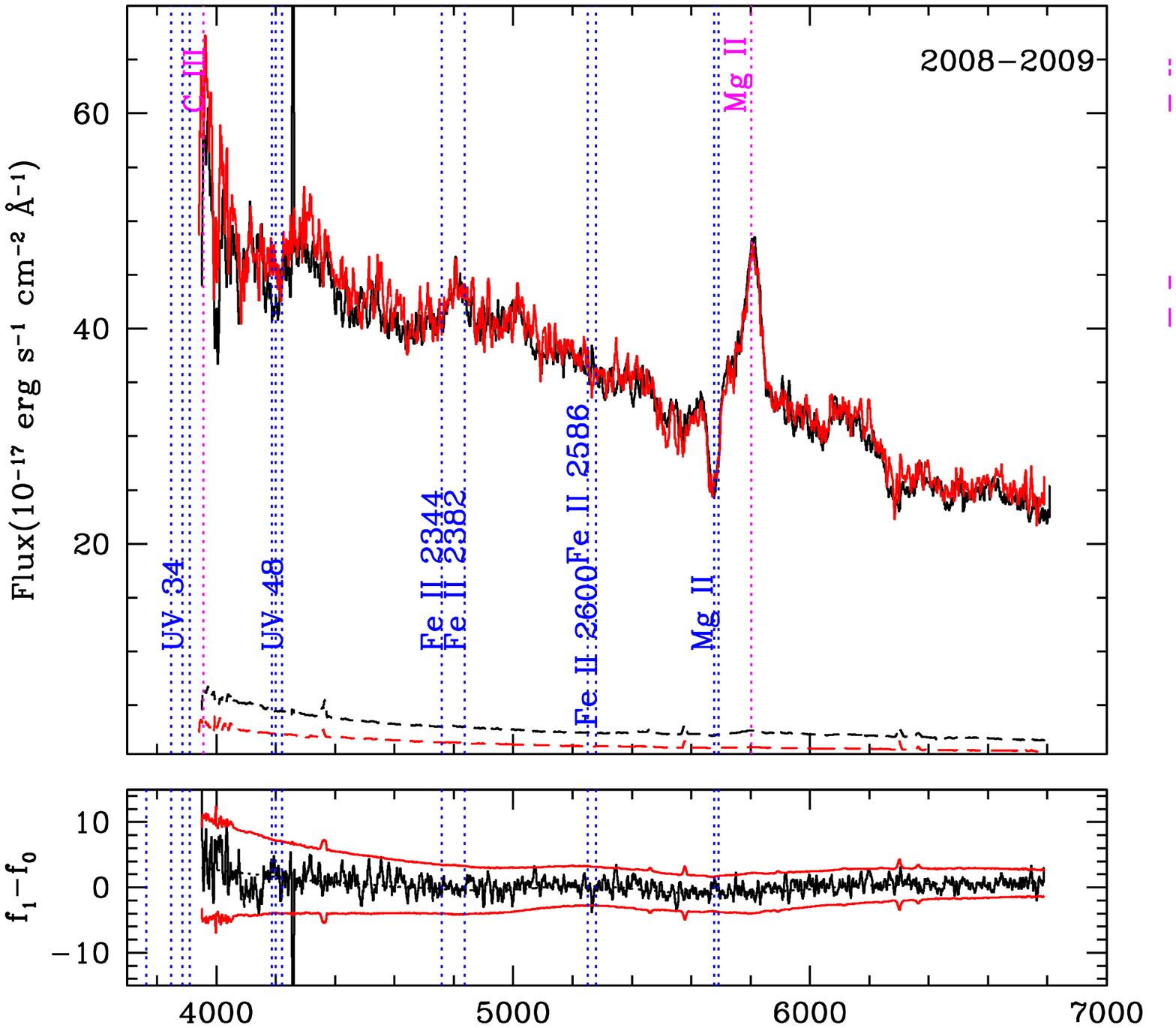,width=0.5\linewidth,height=0.4\linewidth,bbllx=18bp,bblly=144bp,bburx=592bp,bbury=718bp,clip=yes}&
\psfig{figure=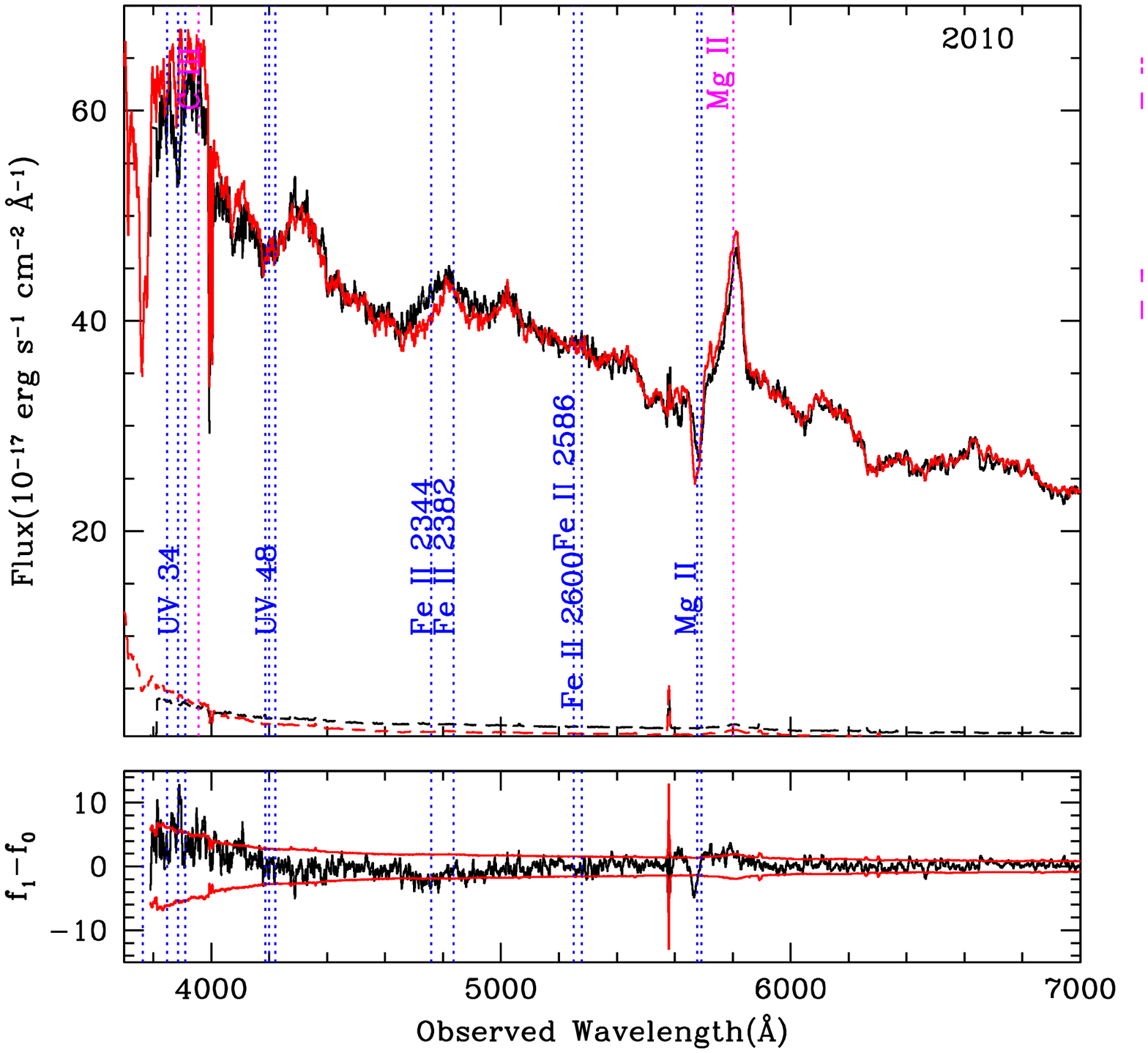,width=0.5\linewidth,height=0.4\linewidth,bbllx=18bp,bblly=144bp,bburx=592bp,bbury=718bp,clip=yes}\\
\end{tabular}
\caption{IGO Spectra of SDSS J0149$-$0103  observed on MJD 54474 and 54823  (in red/grey) are overplotted with the reference SDSS spectrum (black) observed on MJD 51793 . The comparison of two IGO spectra are shown in the lower panel. The flux scale applies to the reference SDSS spectrum and all other spectra are scaled in flux to match the reference spectrum. In each plot, the error spectra are also shown. The difference spectrum for the corresponding MJDs is plotted in the lower panel of each plot. 1$\sigma$ error is plotted above and below the mean. }
\label{0149_diff}
\end{figure*}
\begin{figure*}
 \centering
\begin{tabular}{c c}
\psfig{figure=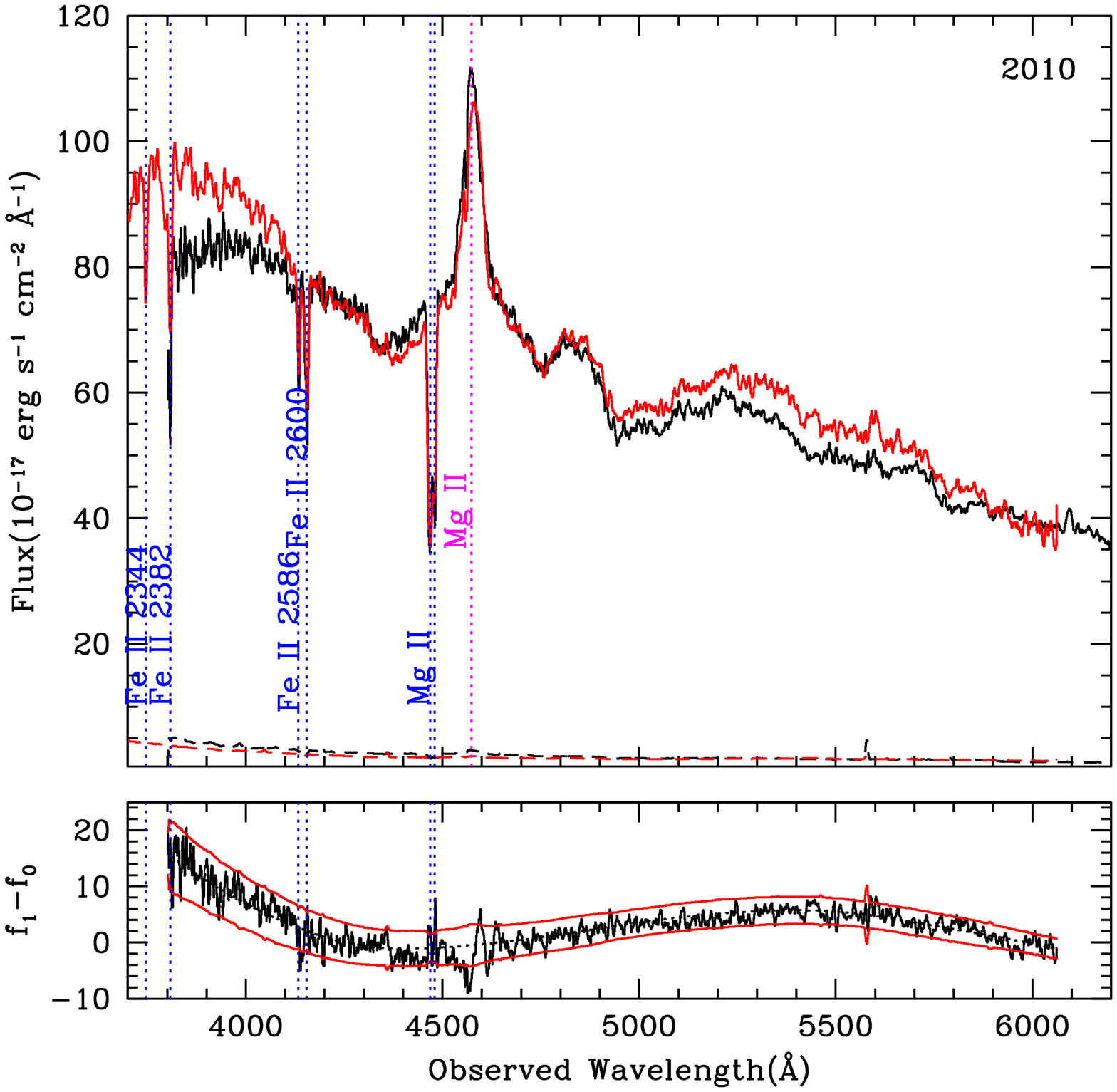,width=0.5\linewidth,height=0.4\linewidth,bbllx=18bp,bblly=144bp,bburx=592bp,bbury=718bp,clip=yes}&
\psfig{figure=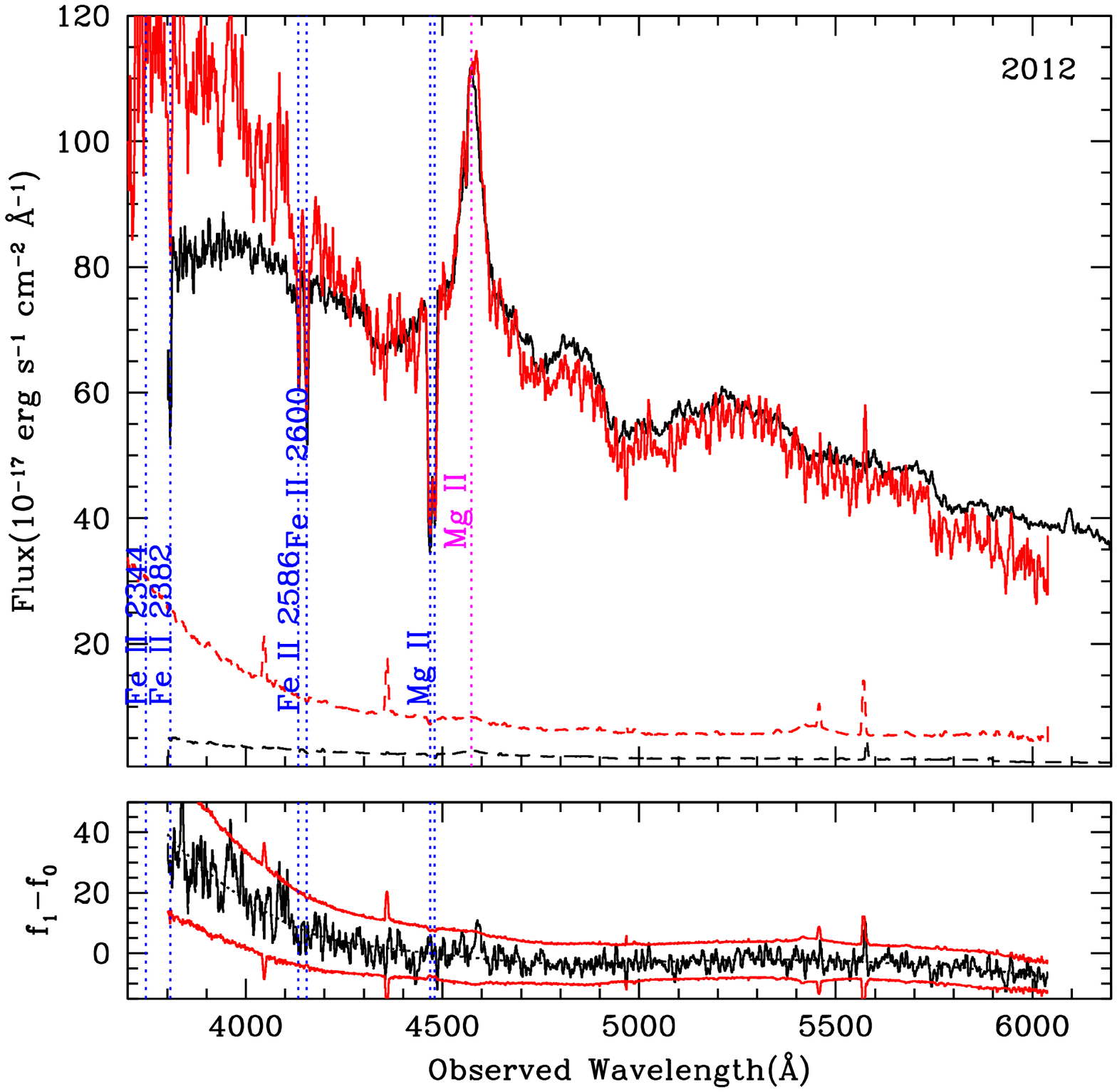,width=0.5\linewidth,height=0.4\linewidth,bbllx=18bp,bblly=144bp,bburx=592bp,bbury=718bp,clip=yes}\\
\end{tabular}
\caption{IGO Spectra of SDSS J0334$-$0711  observed on MJD 55216 and 55950  (in red/grey) are overplotted with the reference SDSS spectrum (black) observed on MJD 51910. The flux scale applies to the reference SDSS spectrum and all other spectra are scaled in flux to match the reference spectrum. In each plot, the error spectra are also shown. The difference spectrum for the corresponding MJDs is plotted in the lower panel of each plot. 1$\sigma$ error is plotted above and below the mean. }
\label{0334_diff}
\end{figure*}
\begin{figure*}
 \centering
\begin{tabular}{c c}
\psfig{figure=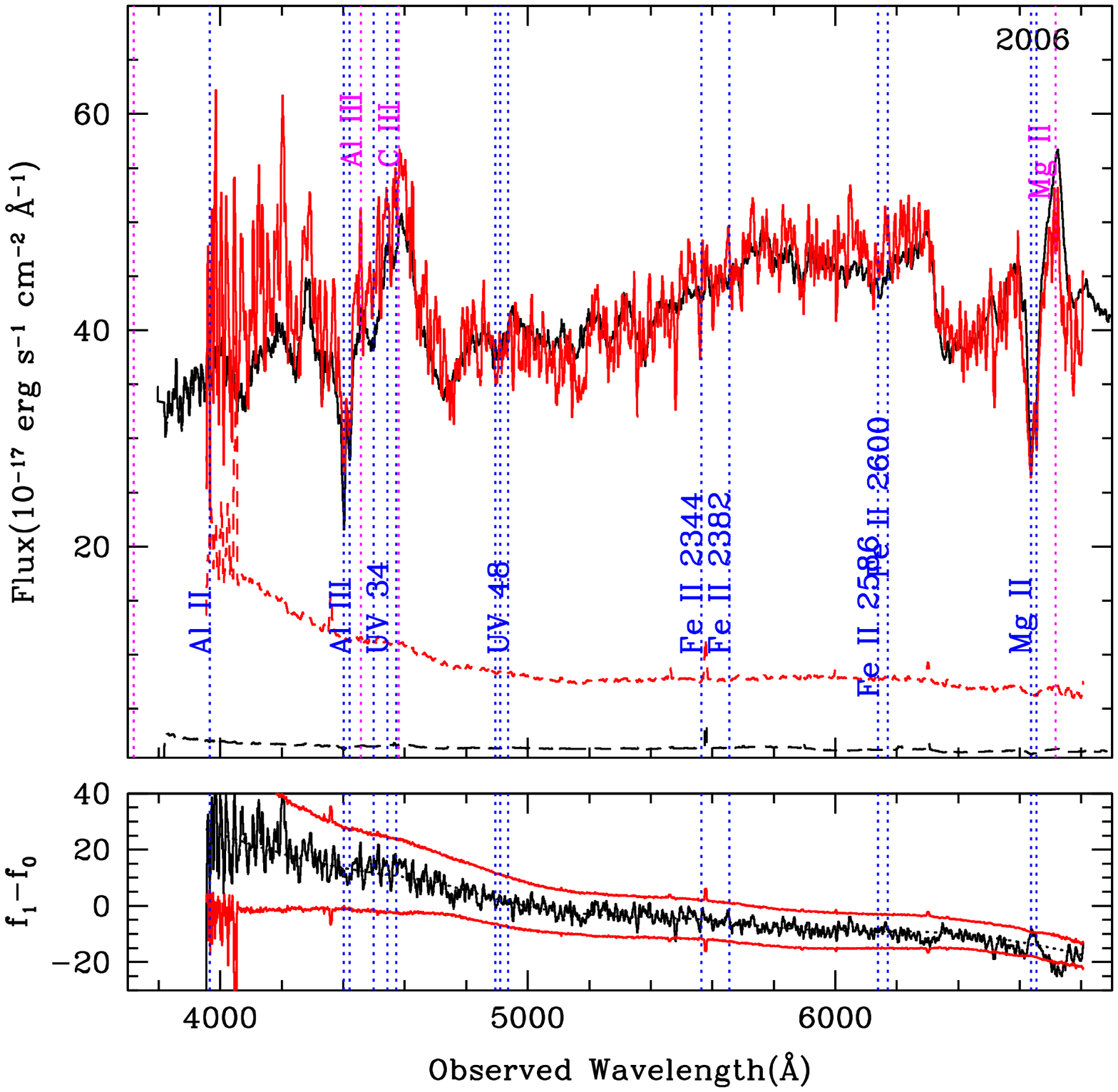,width=0.5\linewidth,height=0.4\linewidth,bbllx=18bp,bblly=144bp,bburx=592bp,bbury=718bp,clip=yes}&%
\psfig{figure=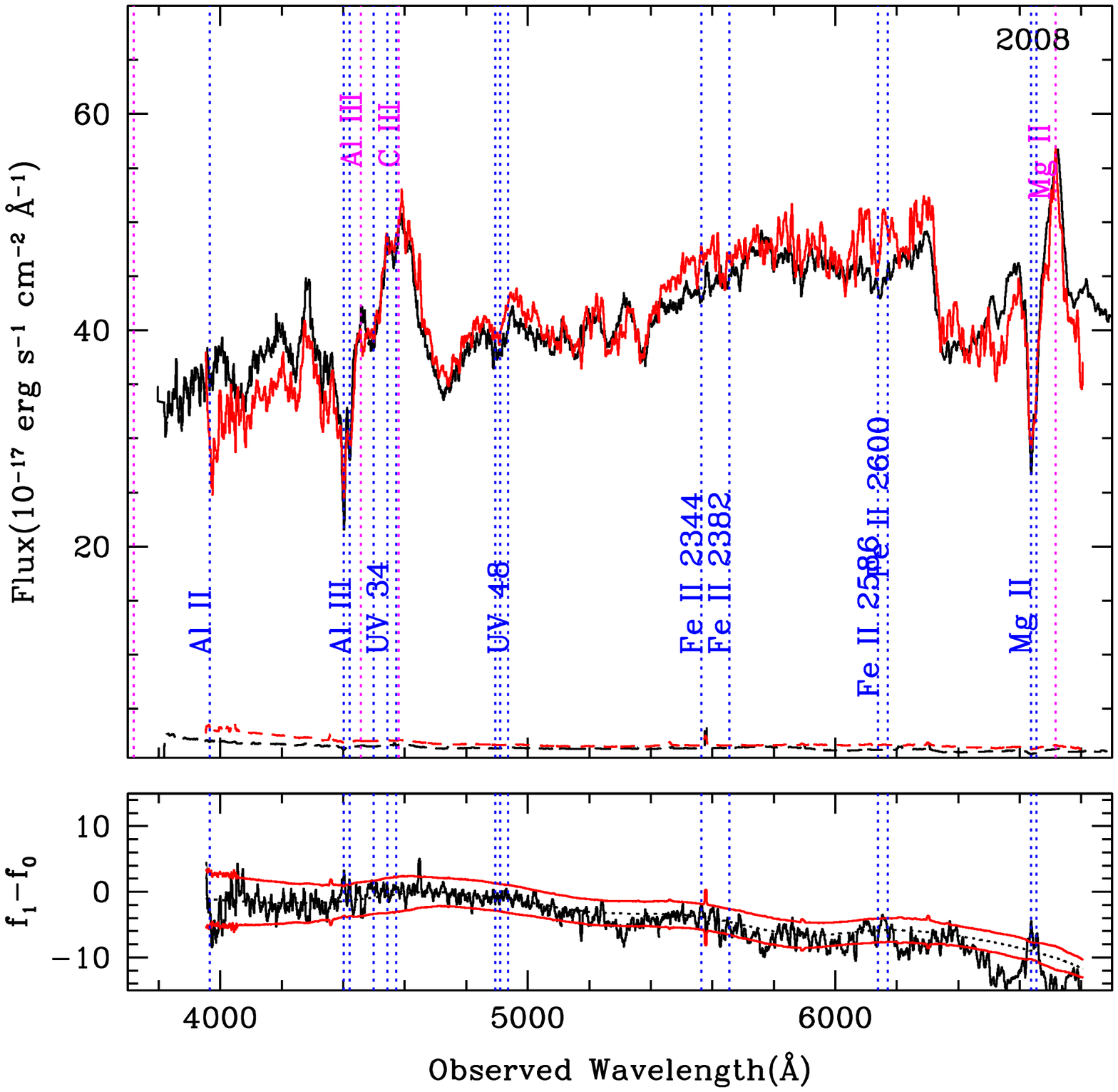,width=0.5\linewidth,height=0.4\linewidth,bbllx=18bp,bblly=144bp,bburx=592bp,bbury=718bp,clip=yes}\\
\psfig{figure=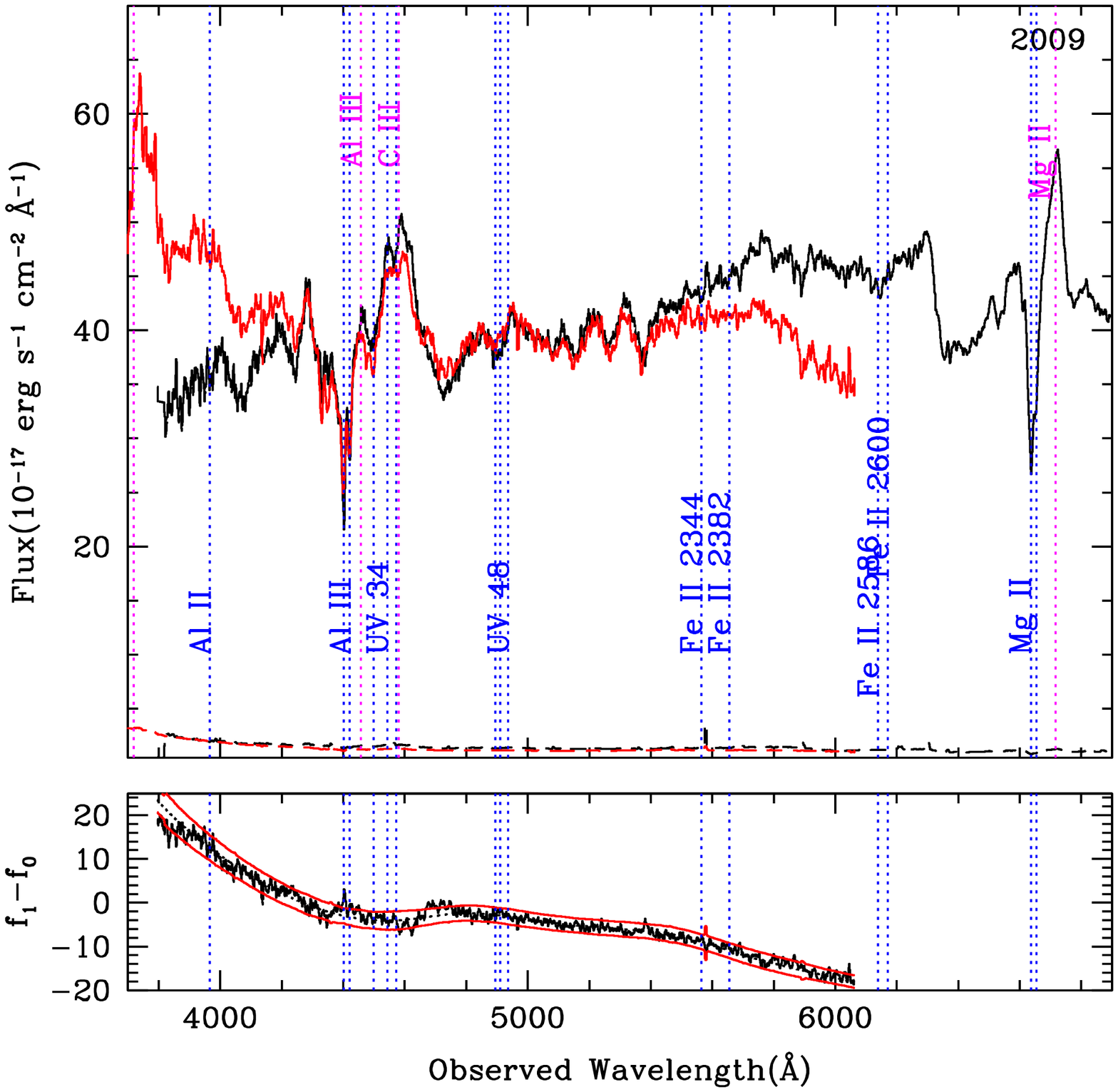,width=0.5\linewidth,height=0.4\linewidth,bbllx=18bp,bblly=144bp,bburx=592bp,bbury=718bp,clip=yes}&
\psfig{figure=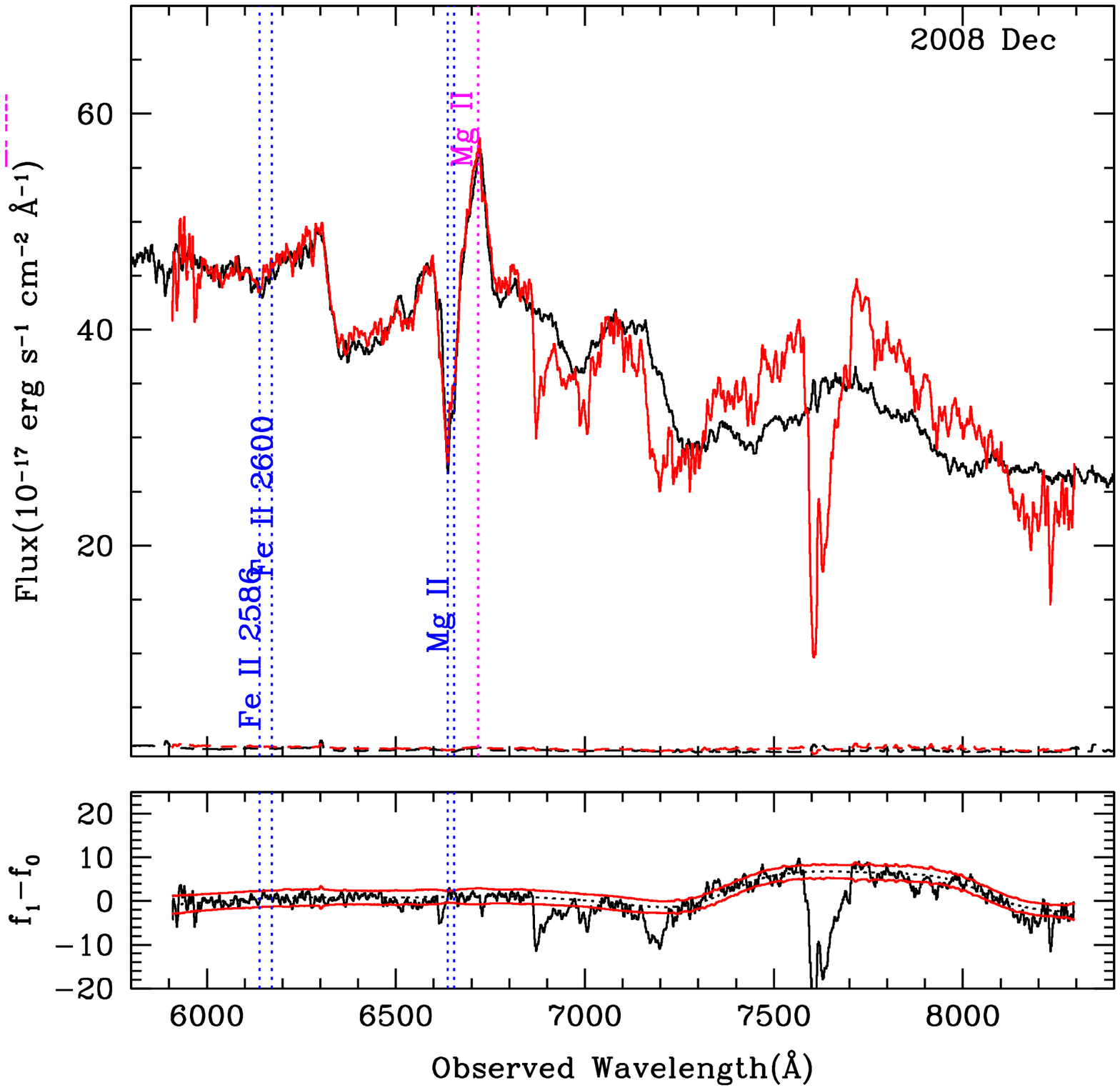,width=0.5\linewidth,height=0.4\linewidth,bbllx=18bp,bblly=144bp,bburx=592bp,bbury=718bp,clip=yes}\\%
\psfig{figure=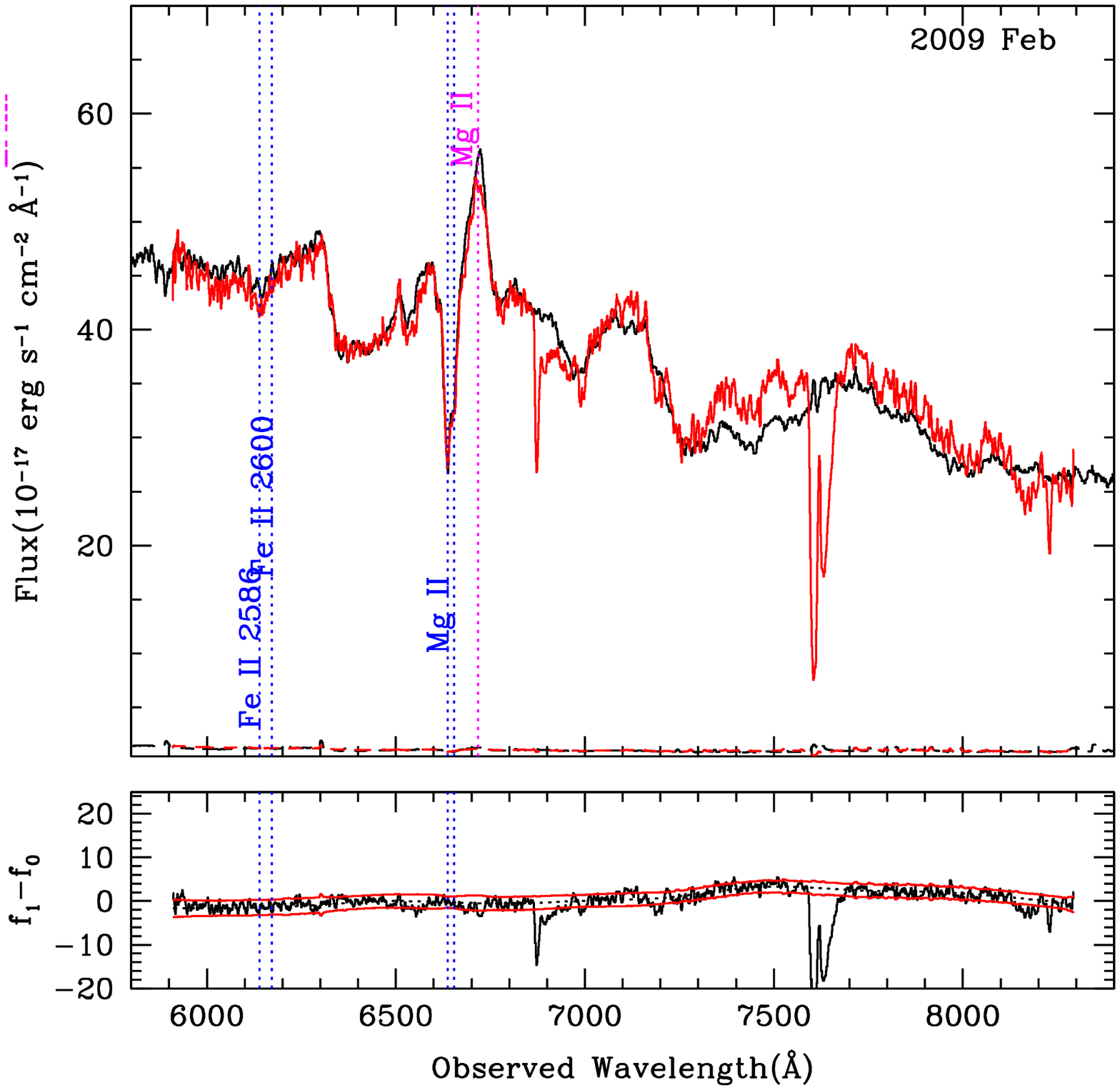,width=0.5\linewidth,height=0.4\linewidth,bbllx=18bp,bblly=144bp,bburx=592bp,bbury=718bp,clip=yes}&
\psfig{figure=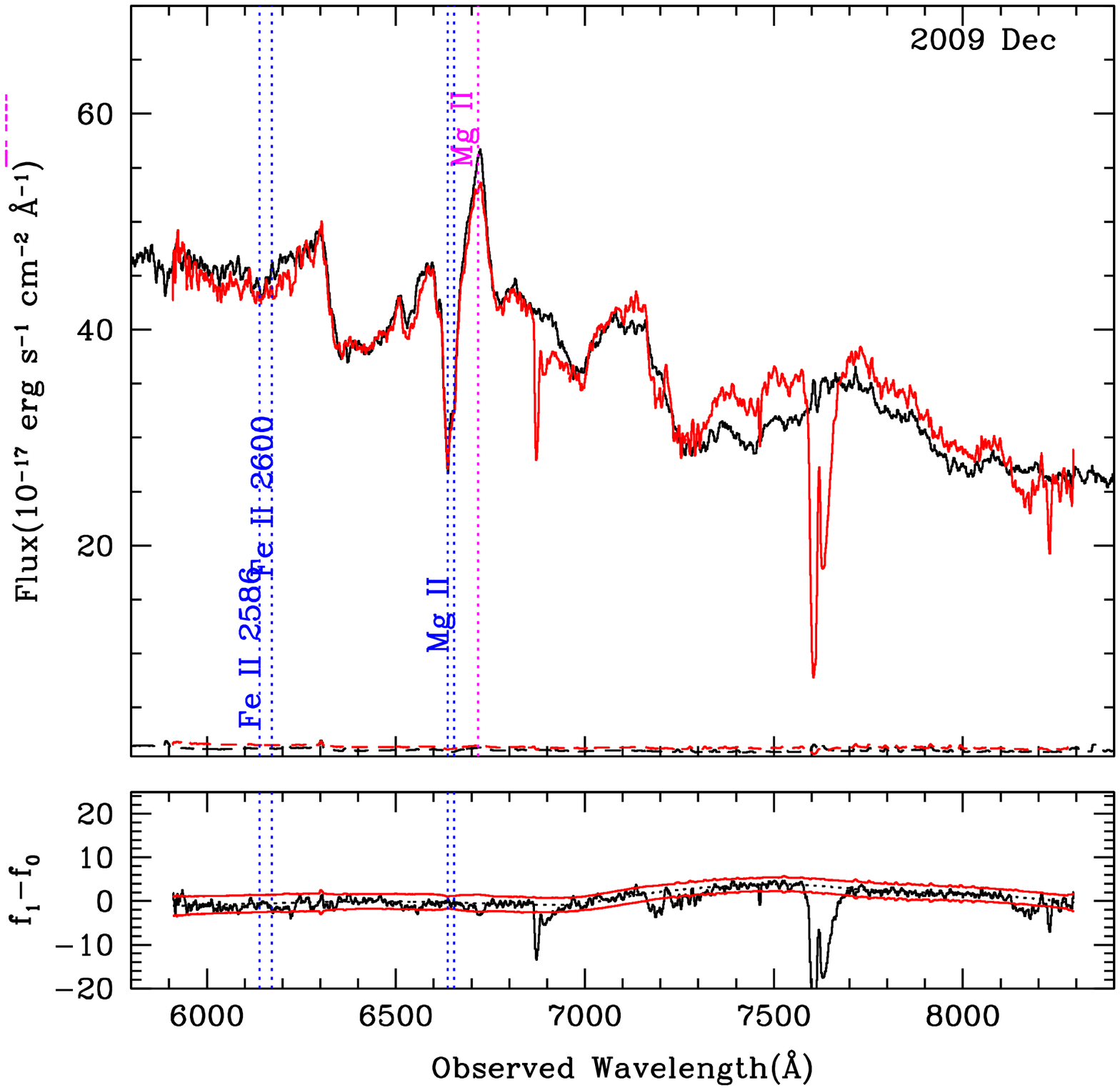,width=0.5\linewidth,height=0.4\linewidth,bbllx=18bp,bblly=144bp,bburx=592bp,bbury=718bp,clip=yes}\\
\end{tabular}
\caption{IGO Spectra of SDSS J0737$+$3844  observed on MJD  54090, 54804, 55189, 54805, 54862 and 54887  (in red/grey) are overplotted with the reference SDSS spectrum (black) observed on MJD 51877. The comparison of two IGO spectra are shown in the lower panel. The flux scale applies to the reference SDSS spectrum and all other spectra are scaled in flux to match the reference spectrum. In each plot, the error spectra are also shown. The difference spectrum for the corresponding MJDs is plotted in the lower panel of each plot. 1$\sigma$ error is plotted above and below the mean. }
\label{0737_diff}
\end{figure*}
\begin{figure*}
 \centering
\begin{tabular}{c c}
\psfig{figure=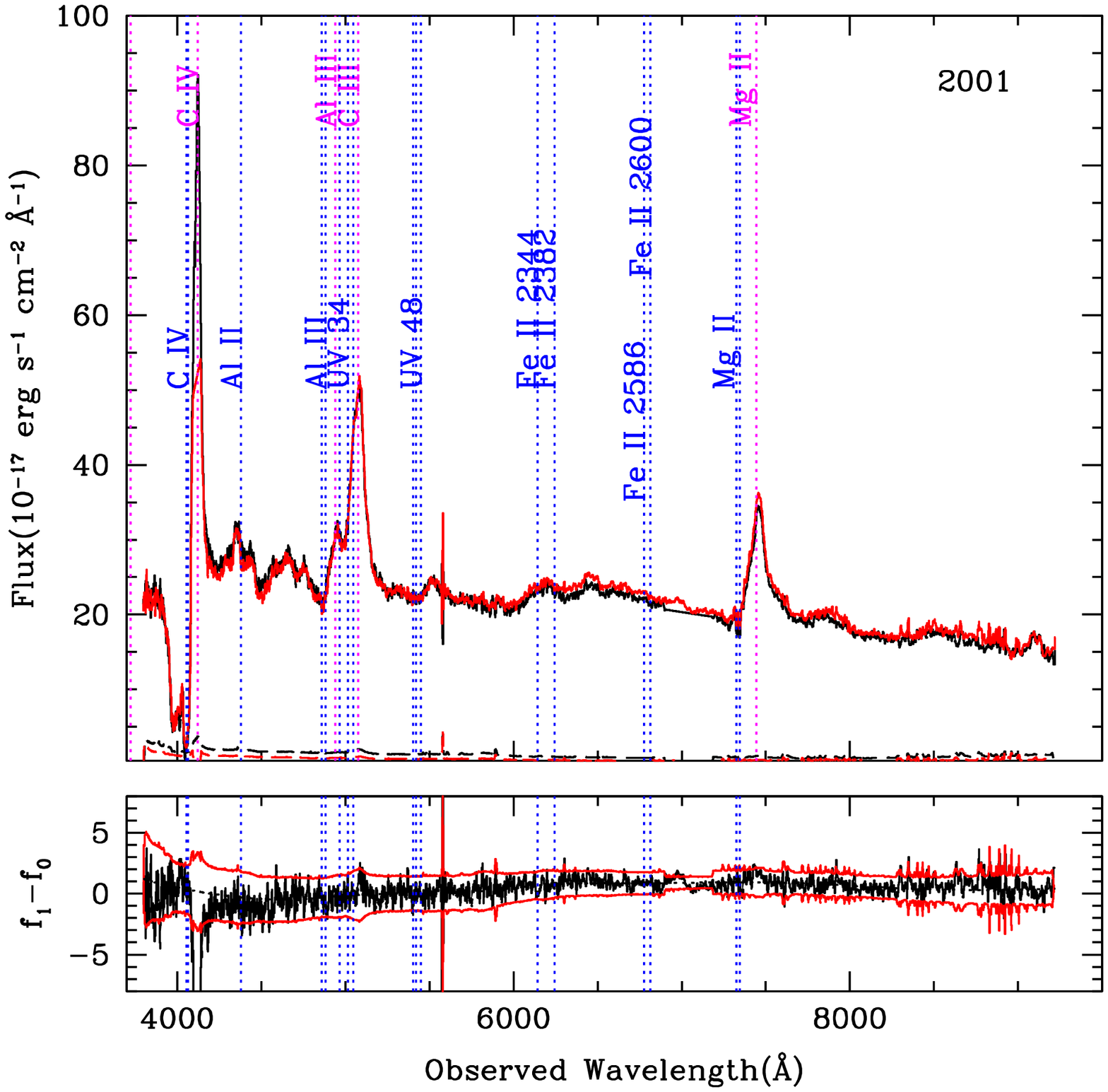,width=0.5\linewidth,height=0.4\linewidth,bbllx=18bp,bblly=144bp,bburx=592bp,bbury=718bp,clip=yes}&
\psfig{figure=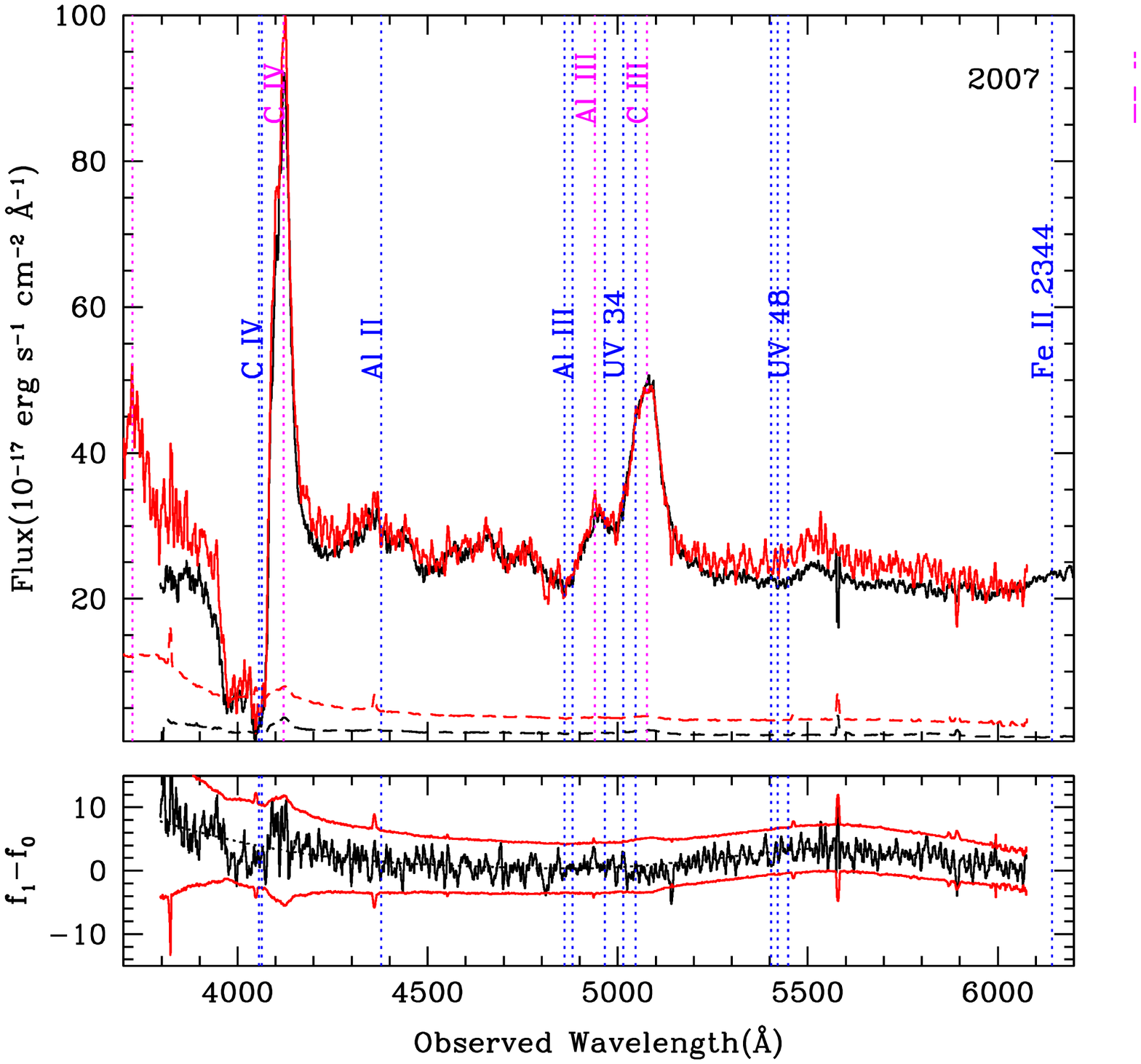,width=0.5\linewidth,height=0.4\linewidth,bbllx=18bp,bblly=144bp,bburx=592bp,bbury=718bp,clip=yes}\\
\psfig{figure=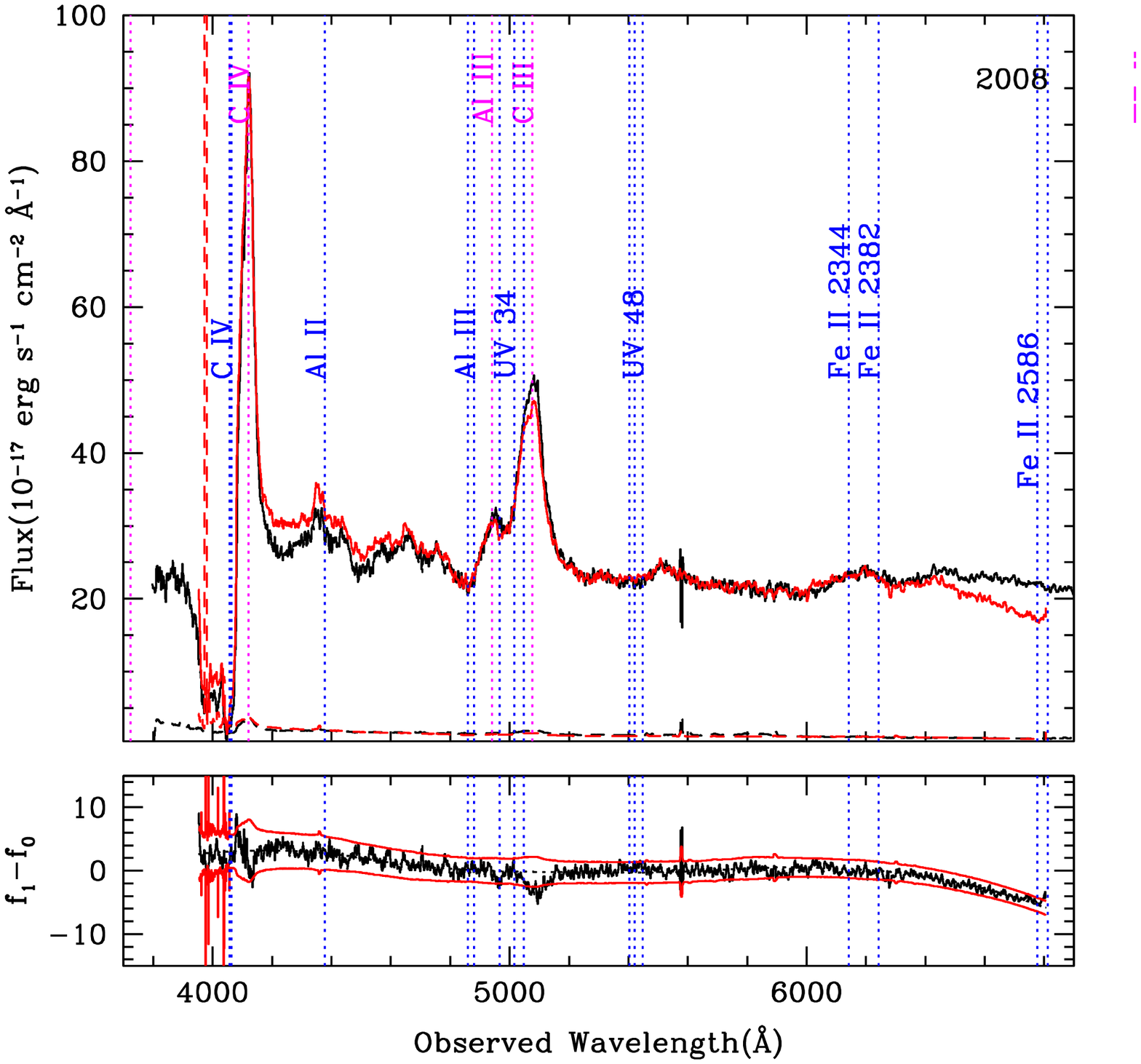,width=0.5\linewidth,height=0.4\linewidth,bbllx=18bp,bblly=144bp,bburx=592bp,bbury=718bp,clip=yes}&
\psfig{figure=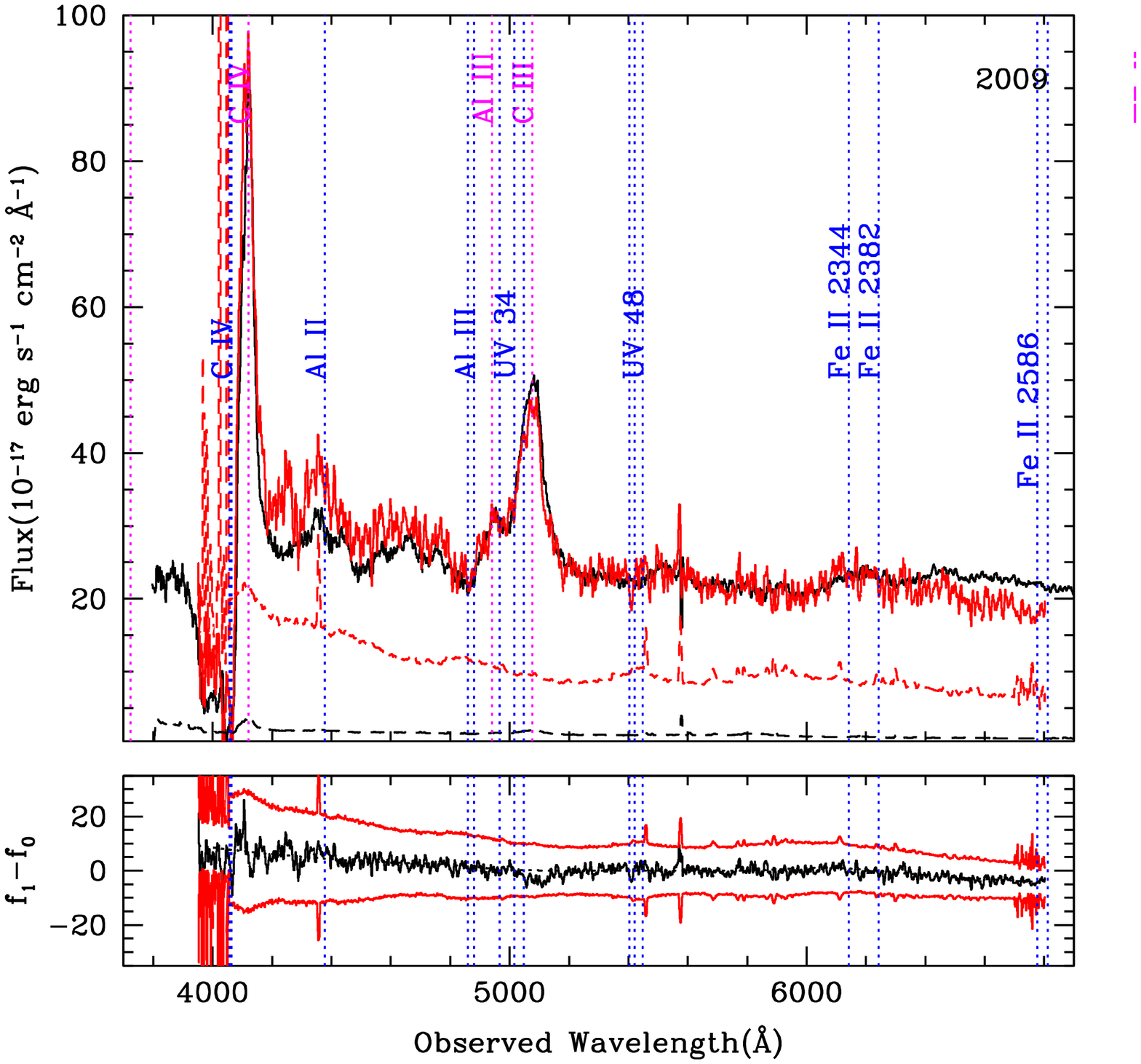,width=0.5\linewidth,height=0.4\linewidth,bbllx=18bp,bblly=144bp,bburx=592bp,bbury=718bp,clip=yes}\\

\end{tabular}
\caption{IGO Spectra of SDSS J0823+4334  observed on MJD  51959, 54447, 54806 and 554575  (in red/grey) are overplotted with the reference SDSS spectrum (black) observed on MJD 52207. The comparison of two IGO spectra are shown in the lower panel. The flux scale applies to the reference SDSS spectrum and all other spectra are scaled in flux to match the reference spectrum. In each plot, the error spectra are also shown. The difference spectrum for the corresponding MJDs is plotted in the lower panel of each plot. 1$\sigma$ error is plotted above and below the mean. }
\label{0823_diff}
\end{figure*}

\begin{figure*}
 \centering
\begin{tabular}{c c}
\psfig{figure=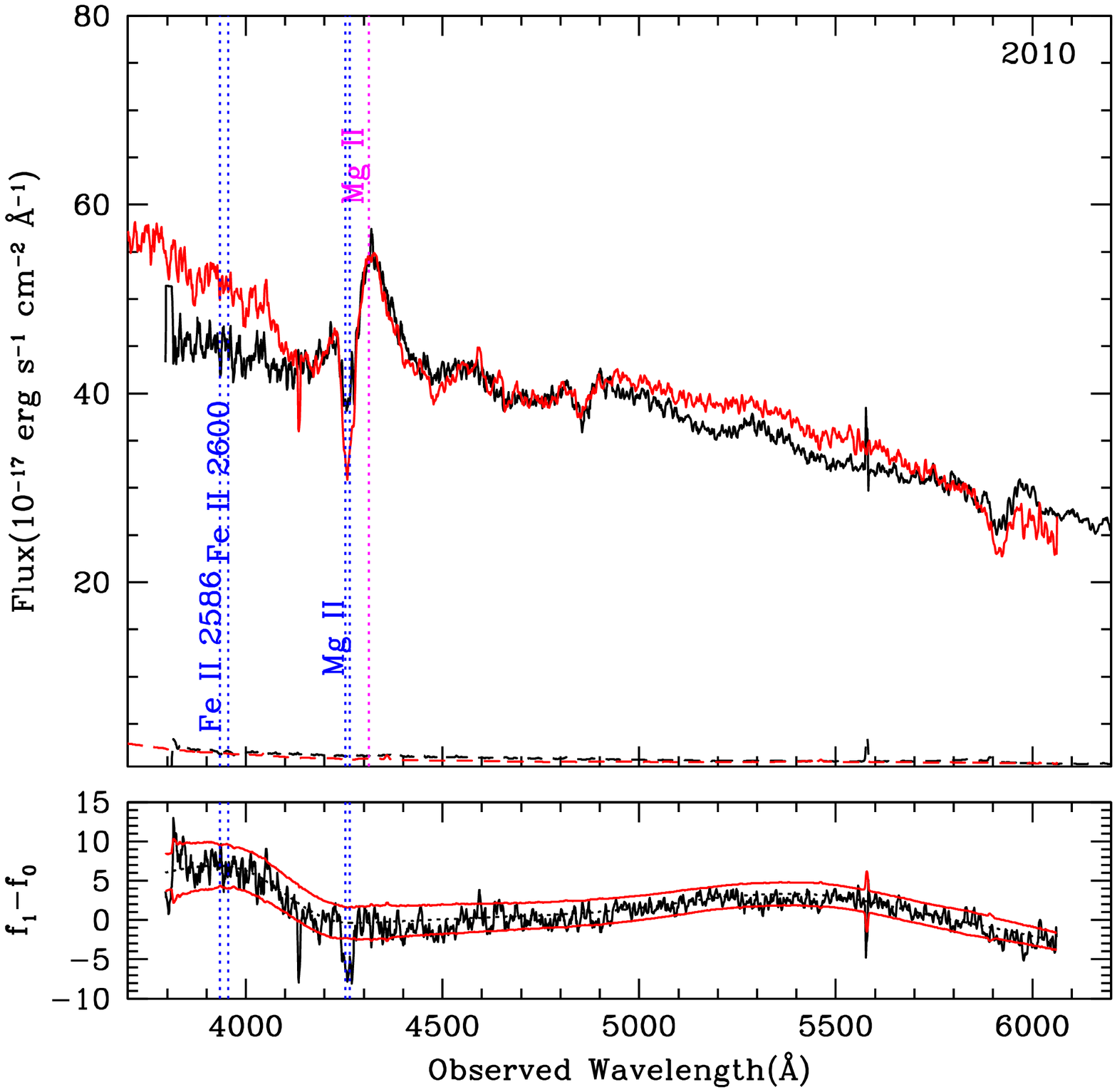,width=0.5\linewidth,height=0.4\linewidth,bbllx=18bp,bblly=144bp,bburx=592bp,bbury=718bp,clip=yes}&
\psfig{figure=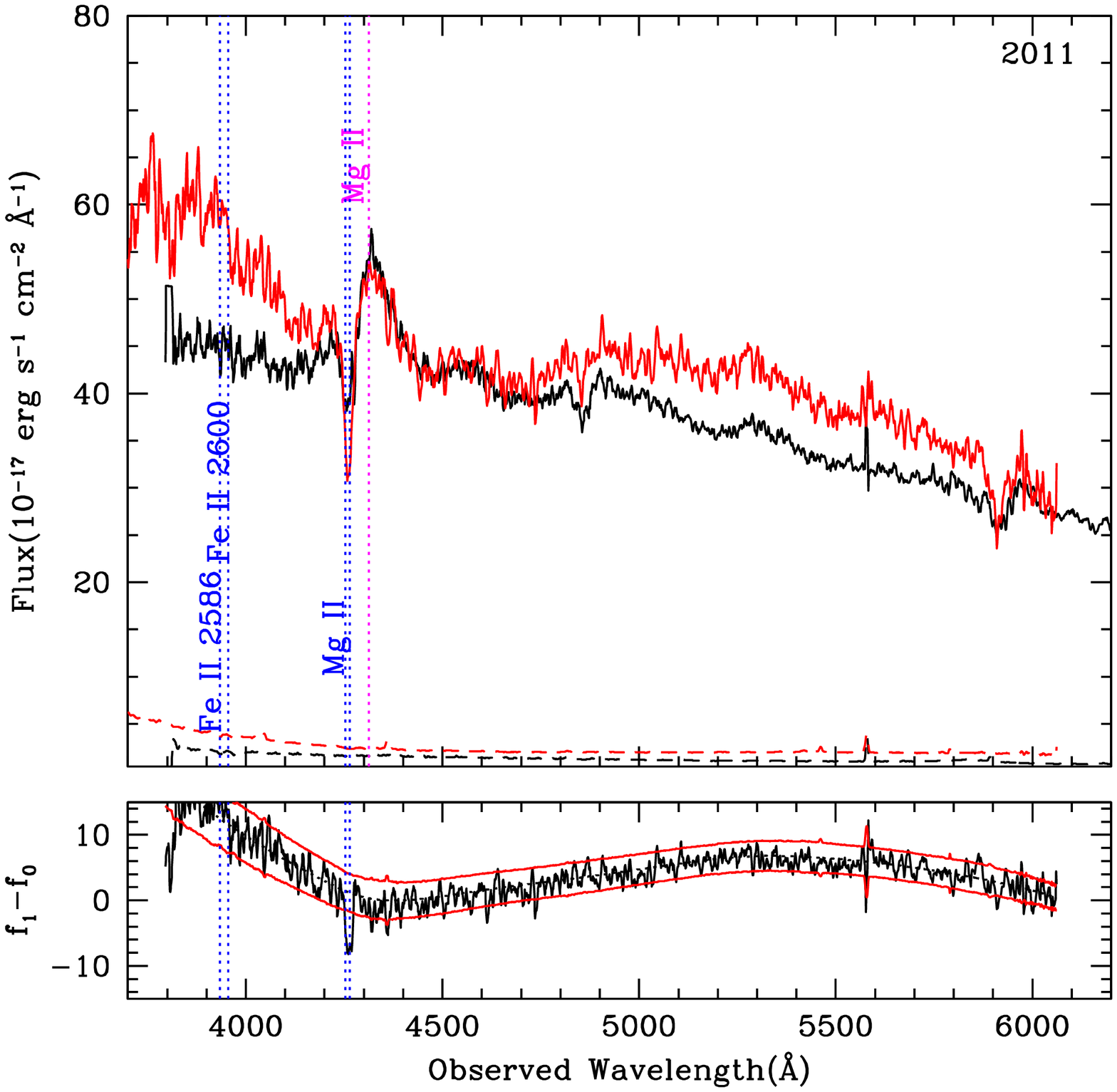,width=0.5\linewidth,height=0.4\linewidth,bbllx=18bp,bblly=144bp,bburx=592bp,bbury=718bp,clip=yes}\\
\psfig{figure=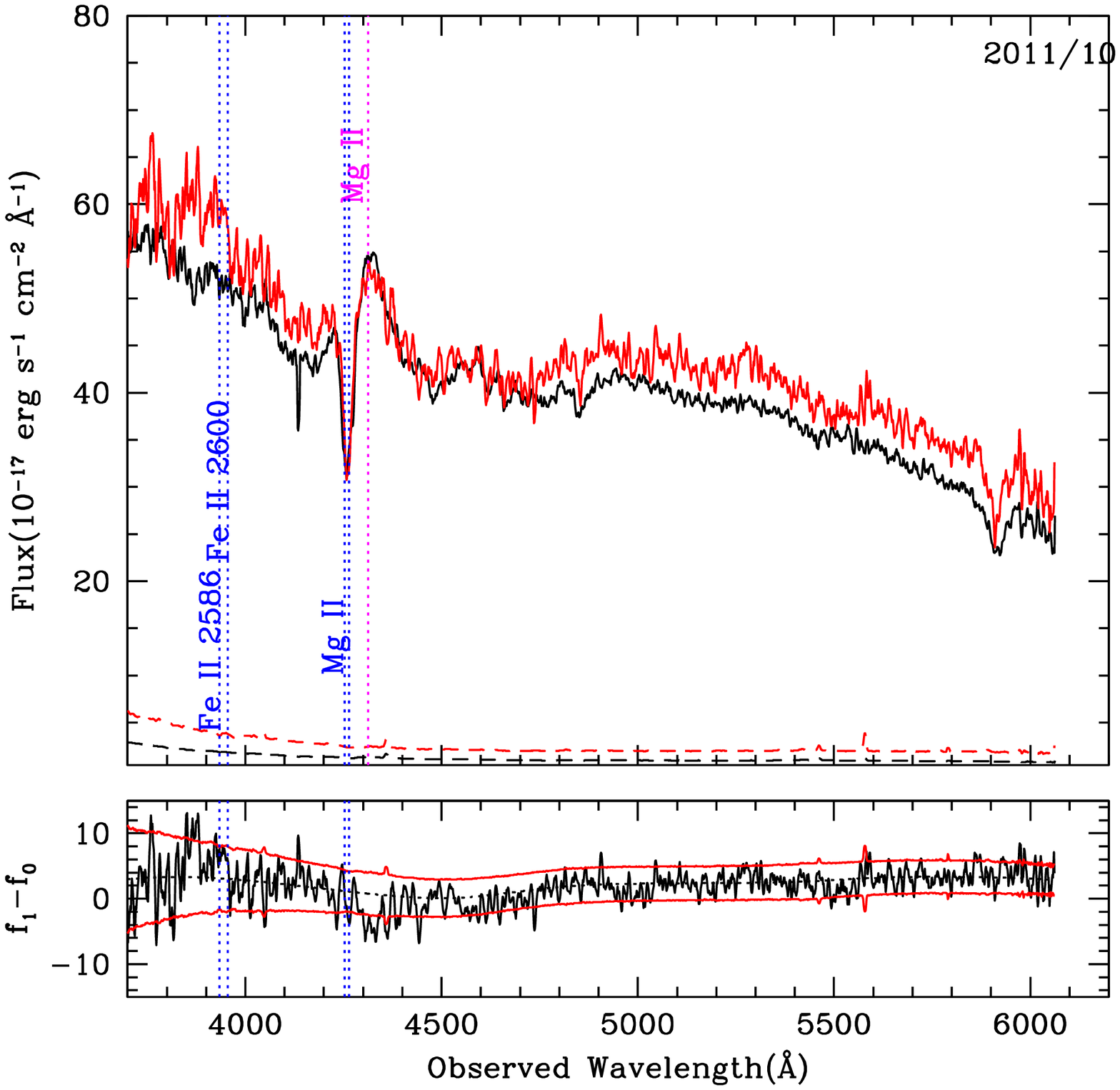,width=0.5\linewidth,height=0.4\linewidth,bbllx=18bp,bblly=144bp,bburx=592bp,bbury=718bp,clip=yes}\\
\end{tabular}
\caption{IGO Spectra of SDSS J0850$+$4451  observed on MJD  55215 and 55632  (in red/grey) are overplotted with the reference SDSS spectrum (black) observed on MJD 52605. The comparison of two IGO spectra are shown in the lower panel. The flux scale applies to the reference SDSS spectrum and all other spectra are scaled in flux to match the reference spectrum. In each plot, the error spectra are also shown. The difference spectrum for the corresponding MJDs is plotted in the lower panel of each plot. 1$\sigma$ error is plotted above and below the mean. }
\label{0850_diff}
\end{figure*}

\begin{figure*}
 \centering
\begin{tabular}{c c}
\psfig{figure=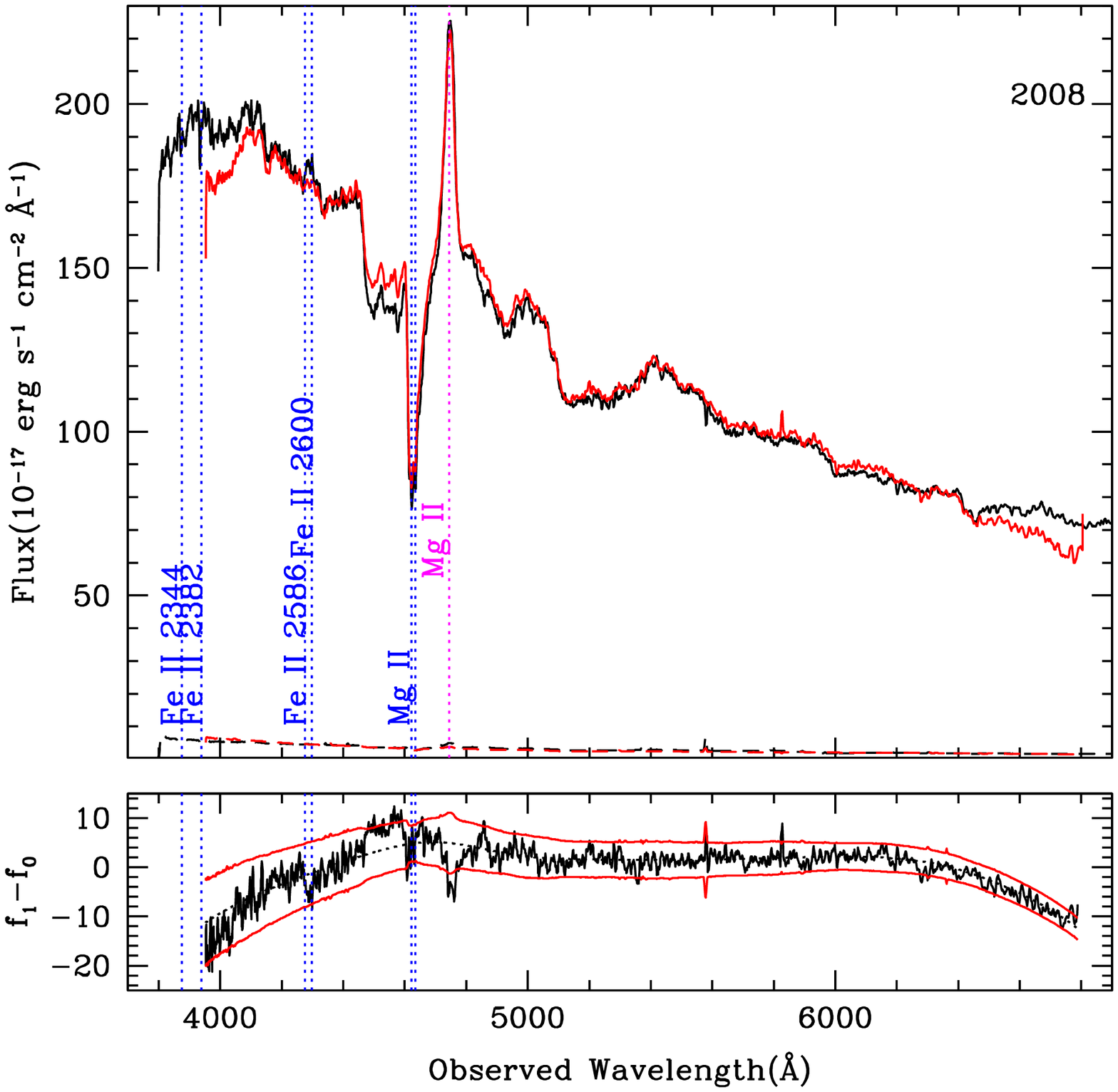,width=0.5\linewidth,height=0.4\linewidth,bbllx=18bp,bblly=144bp,bburx=592bp,bbury=718bp,clip=yes}&
\psfig{figure=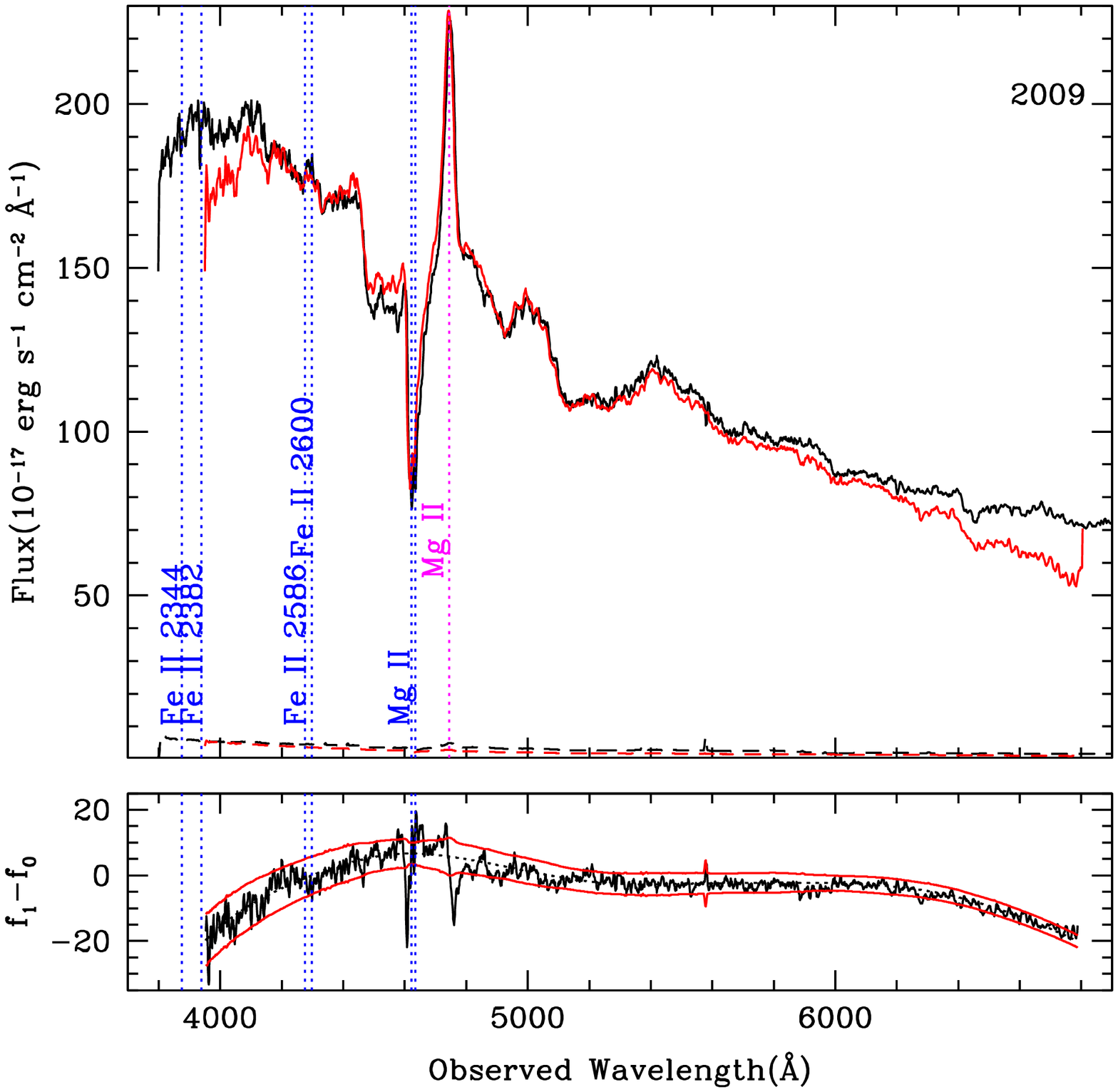,width=0.5\linewidth,height=0.4\linewidth,bbllx=18bp,bblly=144bp,bburx=592bp,bbury=718bp,clip=yes}\\
\psfig{figure=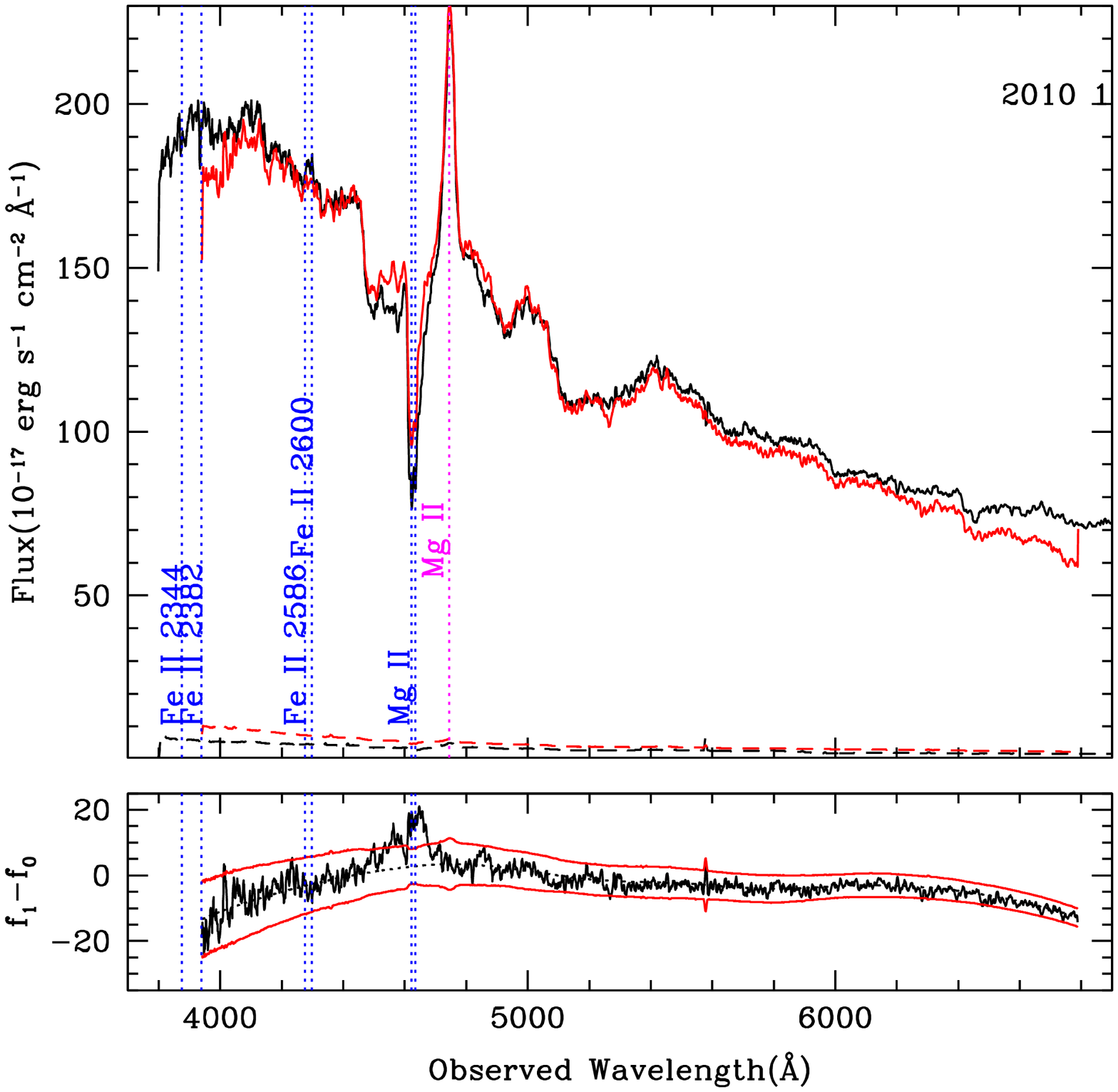,width=0.5\linewidth,height=0.4\linewidth,bbllx=18bp,bblly=144bp,bburx=592bp,bbury=718bp,clip=yes}&
\psfig{figure=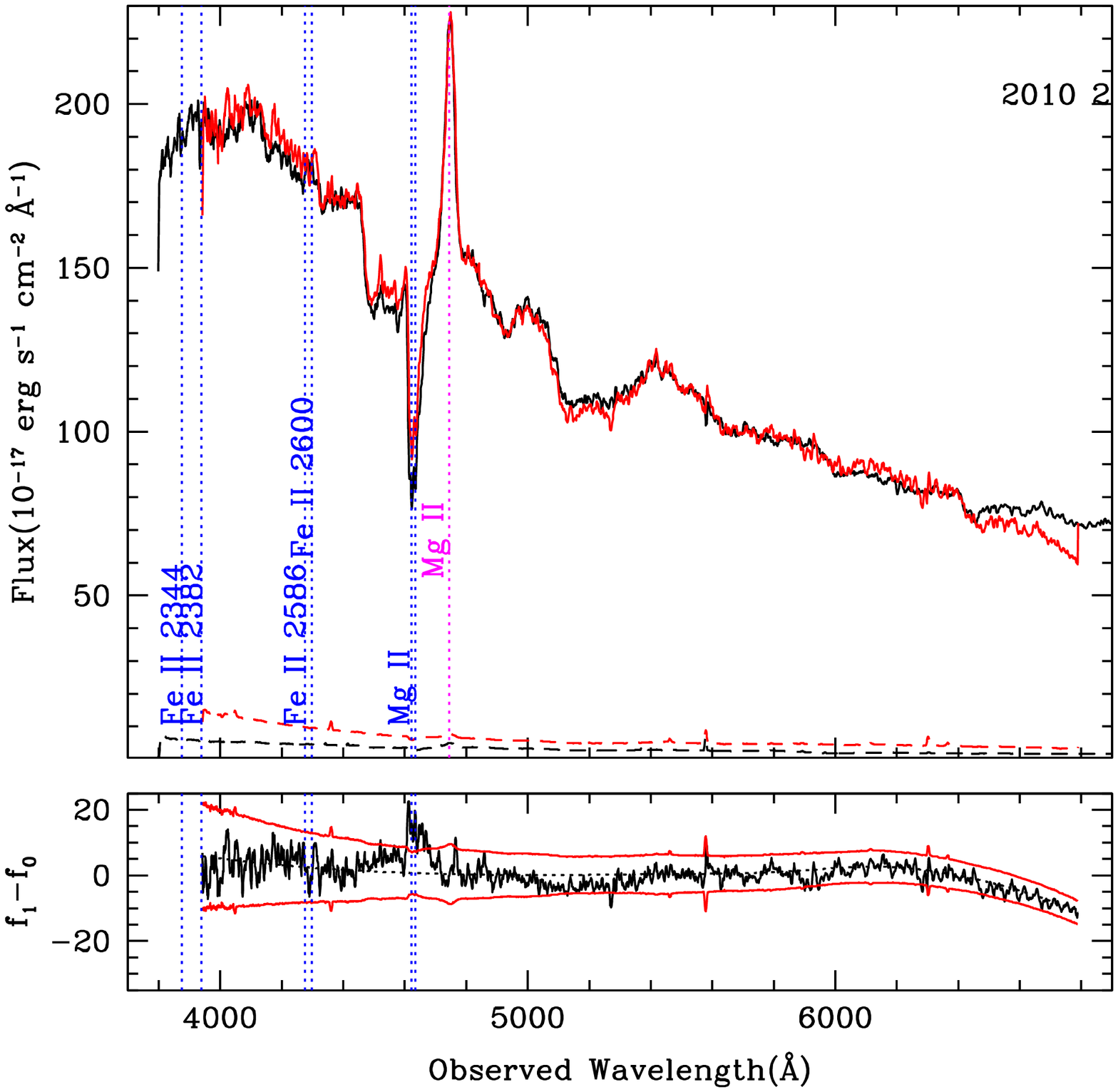,width=0.5\linewidth,height=0.4\linewidth,bbllx=18bp,bblly=144bp,bburx=592bp,bbury=718bp,clip=yes}\\
\psfig{figure=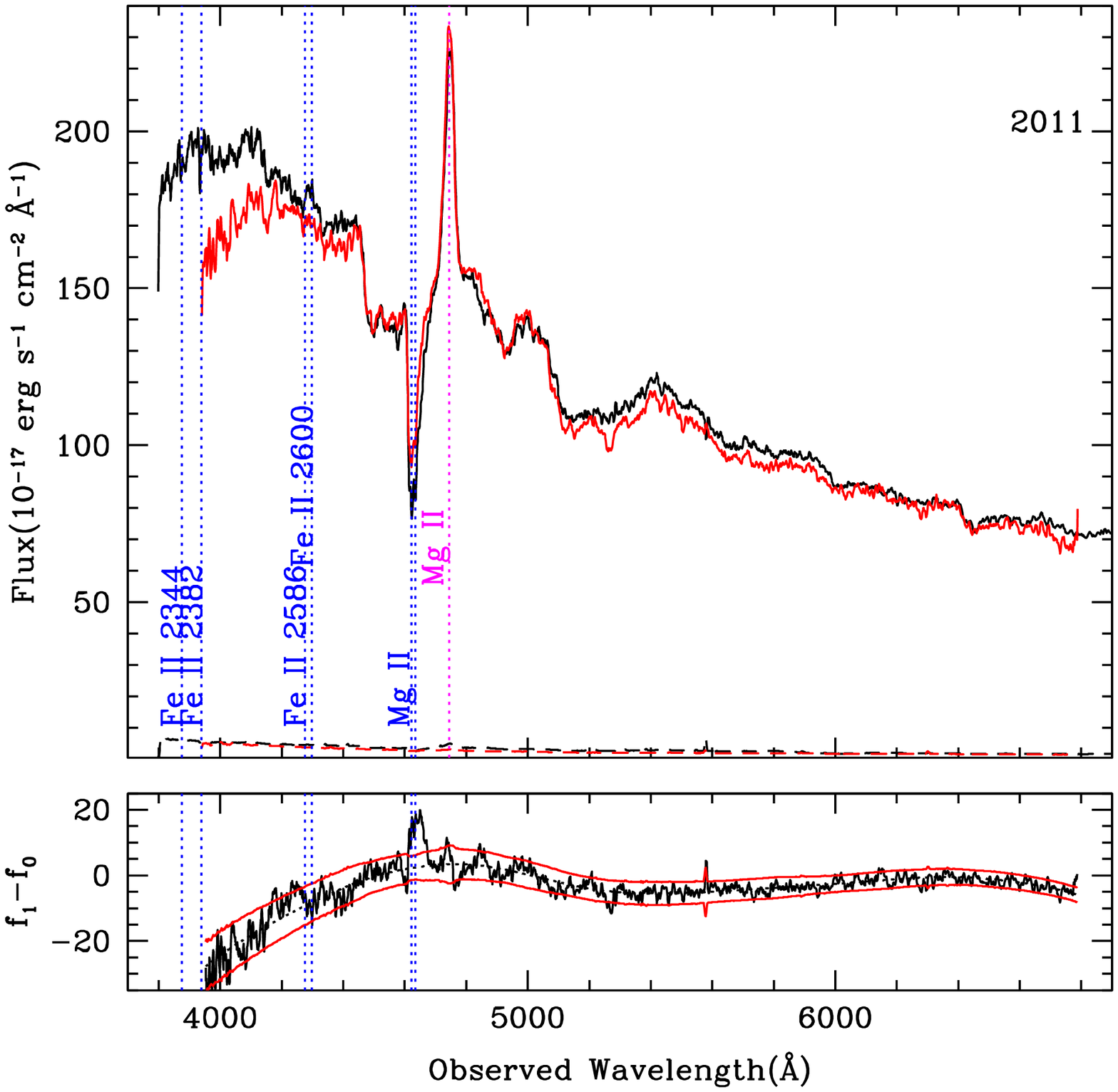,width=0.5\linewidth,height=0.4\linewidth,bbllx=18bp,bblly=144bp,bburx=592bp,bbury=718bp,clip=yes}&`\\
\end{tabular}
\caption{IGO Spectra of SDSS J0944$+$0625  observed on MJD 54559, 54861, 55220, 55297  and 55634   (in red/grey) are overplotted with the reference SDSS spectrum (black) observed on MJD 52710. The comparison of two IGO spectra are shown in the lower panel. The flux scale applies to the reference SDSS spectrum and all other spectra are scaled in flux to match the reference spectrum. In each plot, the error spectra are also shown. The difference spectrum for the corresponding MJDs is plotted in the lower panel of each plot. 1$\sigma$ error is plotted above and below the mean. }
\label{0944_diff}
\end{figure*}
\begin{figure*}
 \centering
\begin{tabular}{c c}
\psfig{figure=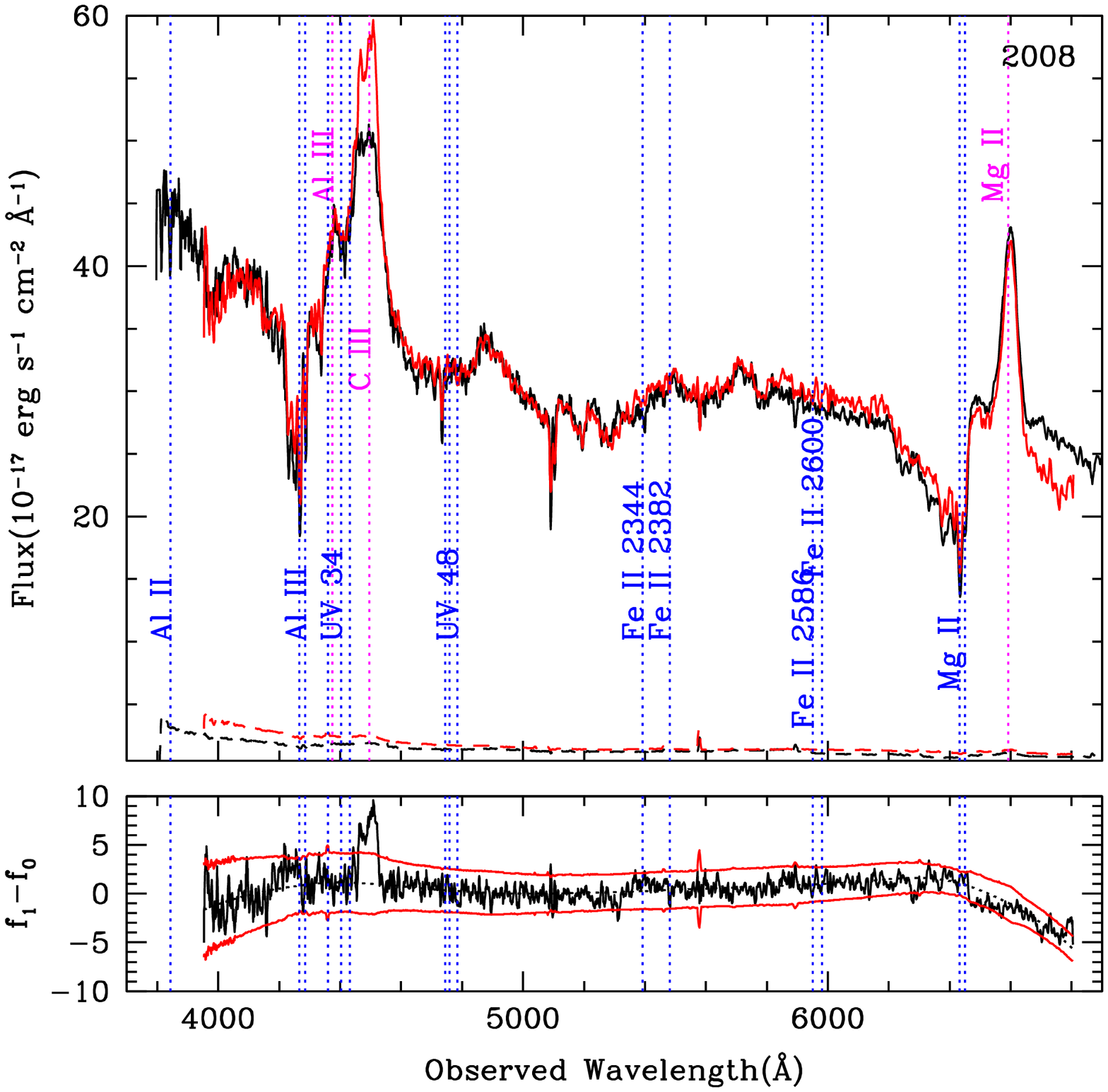,width=0.5\linewidth,height=0.4\linewidth,bbllx=18bp,bblly=144bp,bburx=592bp,bbury=718bp,clip=yes}&
\psfig{figure=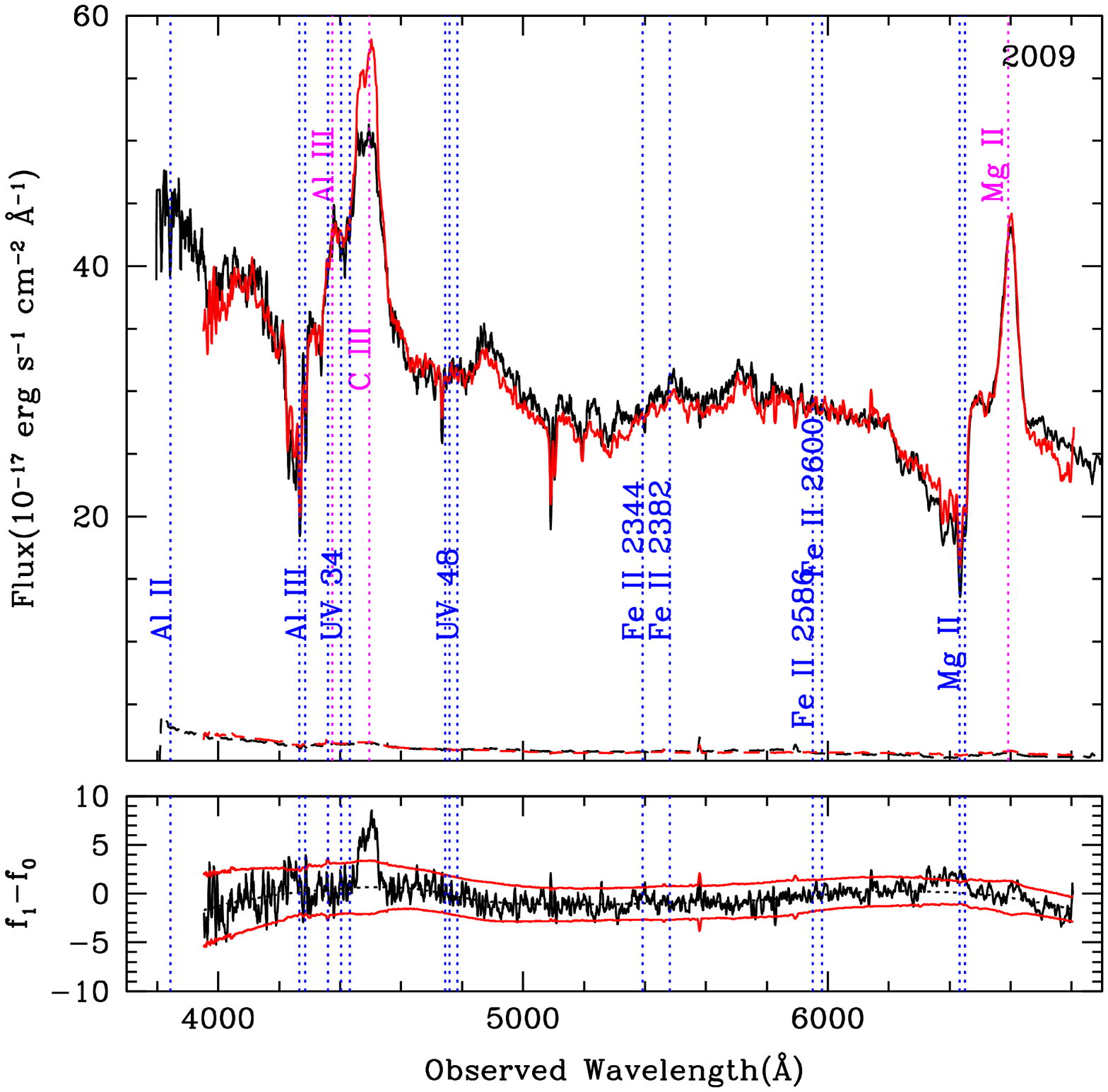,width=0.5\linewidth,height=0.4\linewidth,bbllx=18bp,bblly=144bp,bburx=592bp,bbury=718bp,clip=yes}\\
\psfig{figure=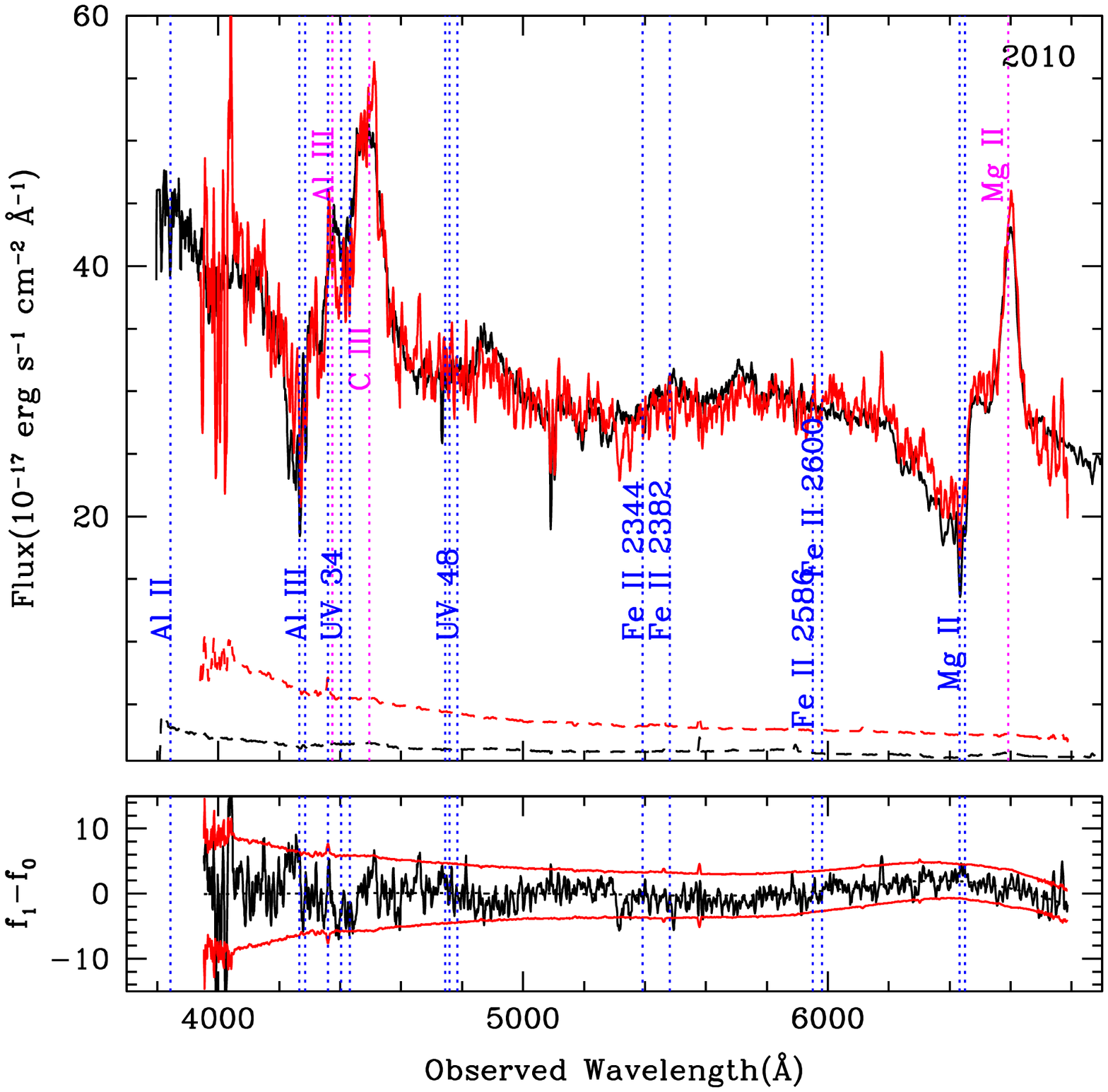,width=0.5\linewidth,height=0.4\linewidth,bbllx=18bp,bblly=144bp,bburx=592bp,bbury=718bp,clip=yes}&
\psfig{figure=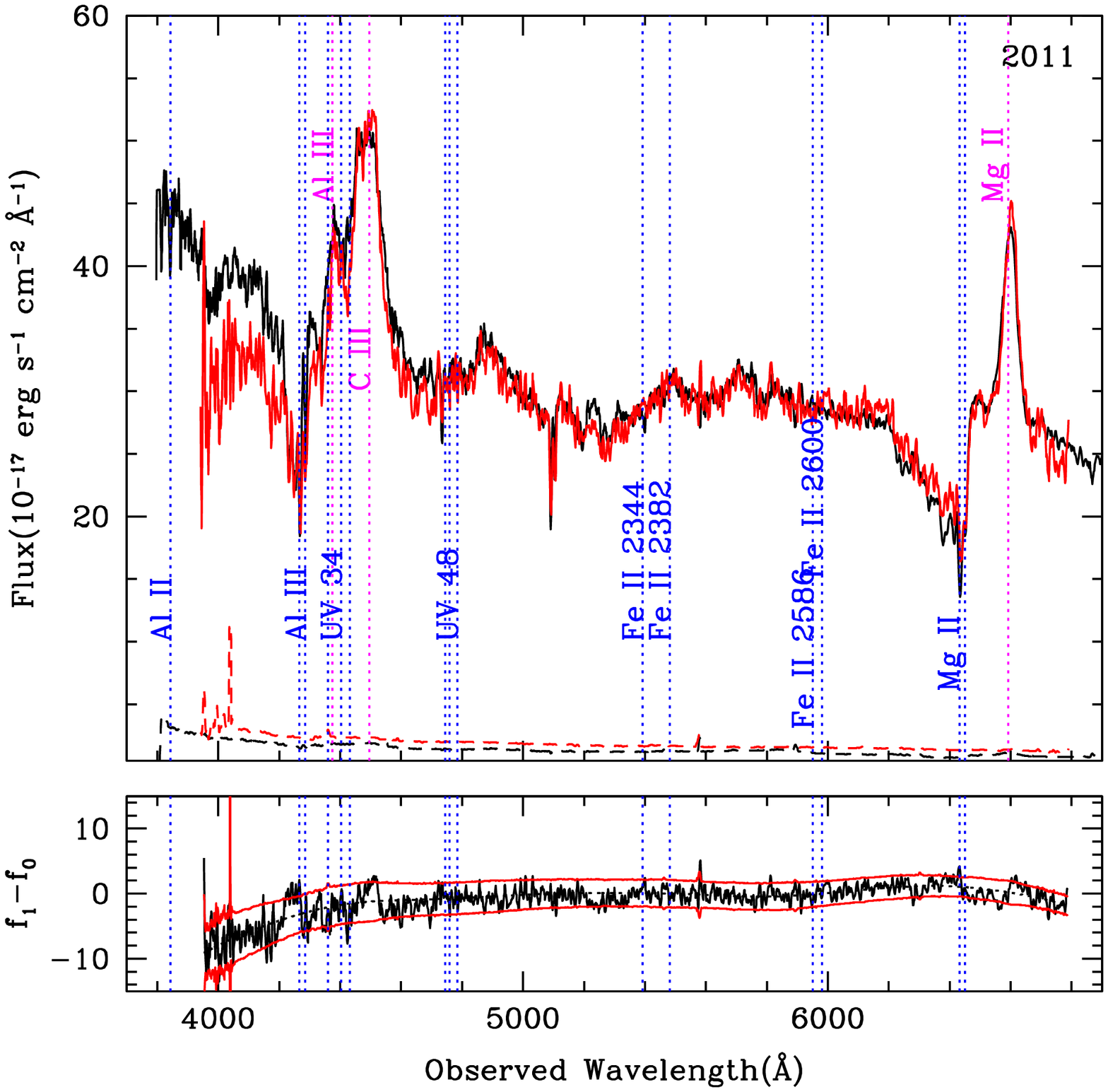,width=0.5\linewidth,height=0.4\linewidth,bbllx=18bp,bblly=144bp,bburx=592bp,bbury=718bp,clip=yes}\\
\end{tabular}
\caption{IGO Spectra of SDSS J0952$+$0257  observed on MJD 54559,  54887, 55221 and   and 55635  (in red/grey) are overplotted with the reference SDSS spectrum (black) observed on MJD 51908. The comparison of two IGO spectra are shown in the lower panel. The flux scale applies to the reference SDSS spectrum and all other spectra are scaled in flux to match the reference spectrum. In each plot, the error spectra are also shown. The difference spectrum for the corresponding MJDs is plotted in the lower panel of each plot. 1$\sigma$ error is plotted above and below the mean. }
\end{figure*}                                             
                                                           
\begin{figure*}
 \centering
\begin{tabular}{c c}
\psfig{figure=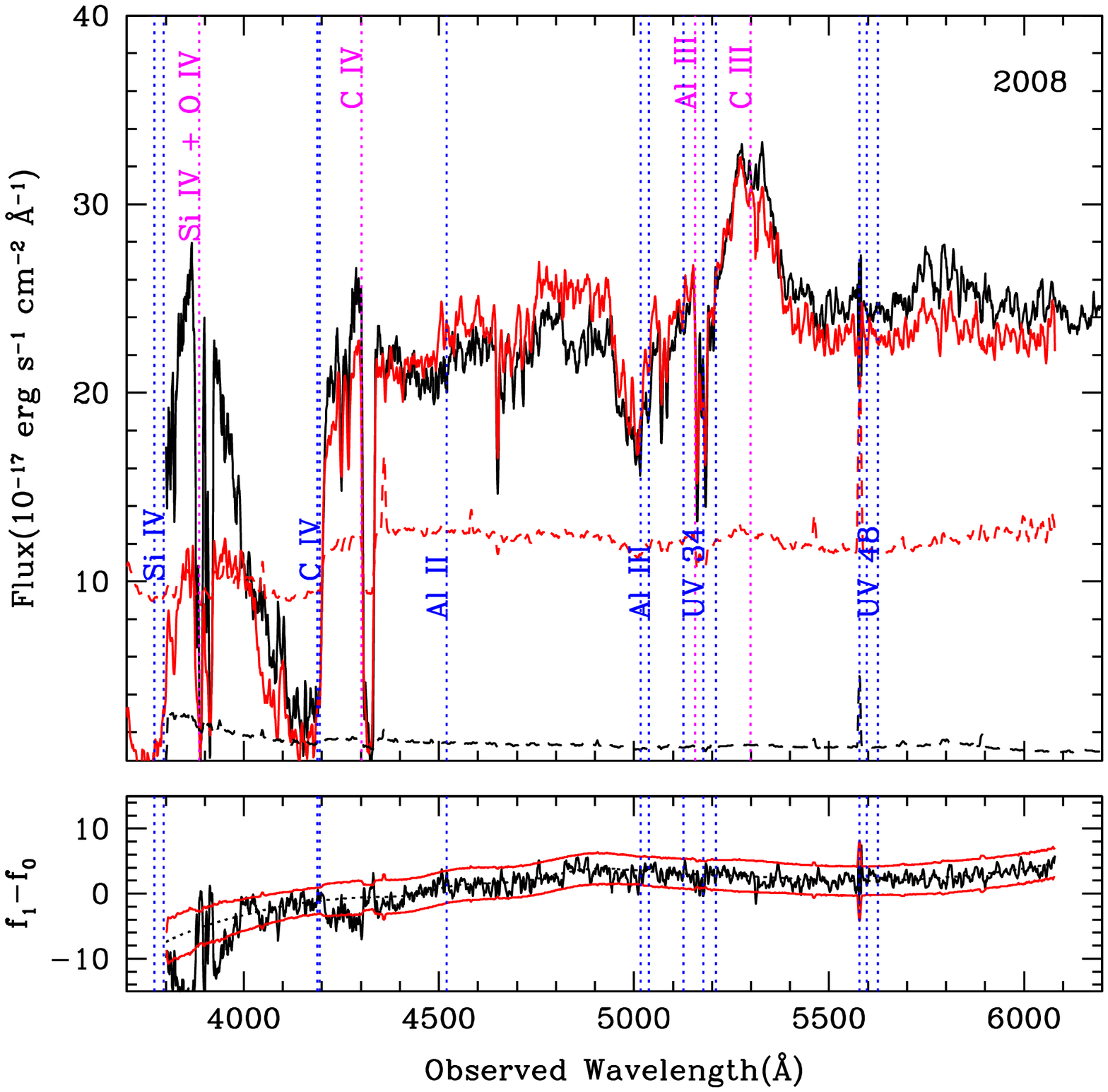,width=0.5\linewidth,height=0.4\linewidth,bbllx=18bp,bblly=144bp,bburx=592bp,bbury=718bp,clip=yes}&
\psfig{figure=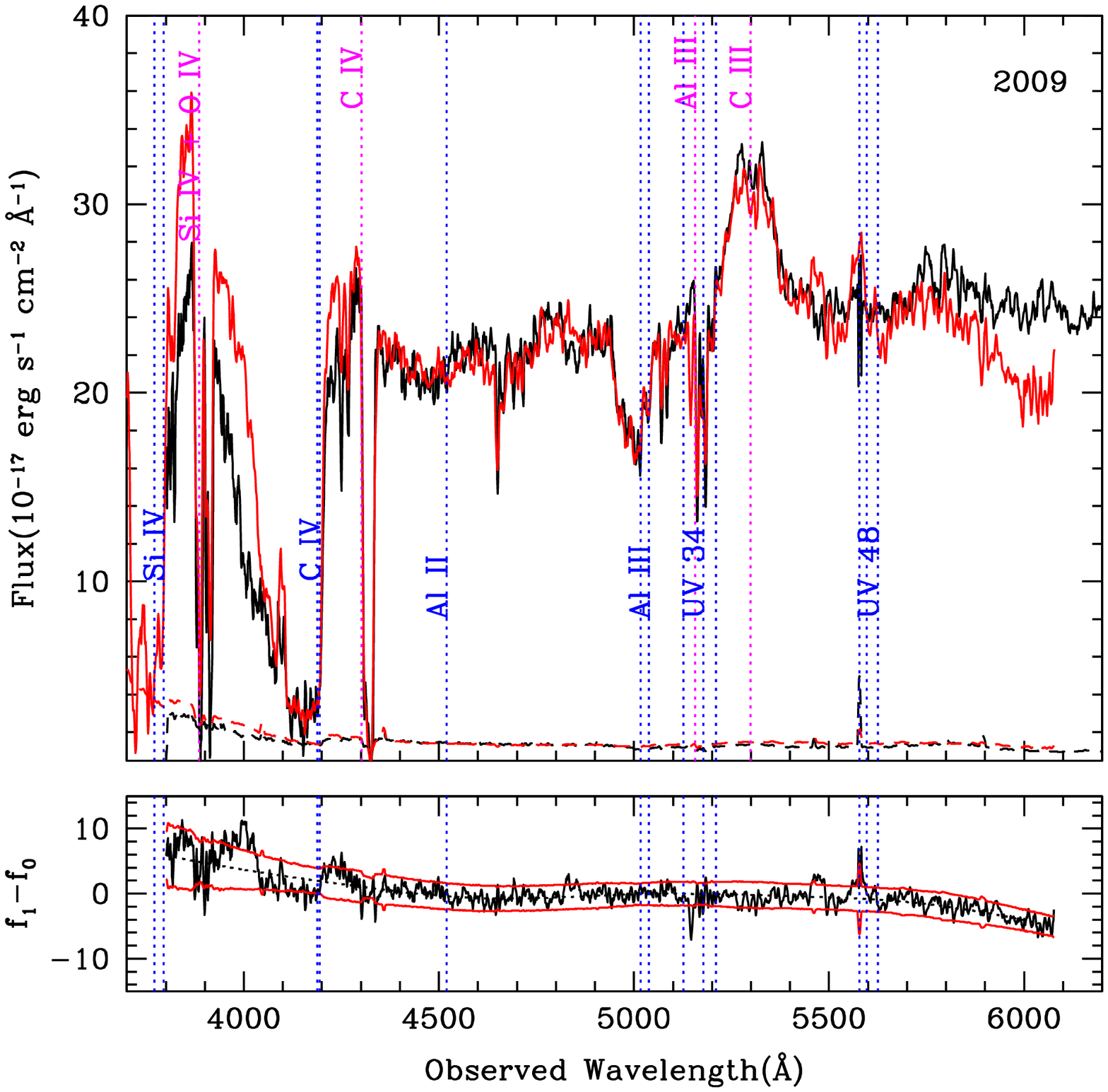,width=0.5\linewidth,height=0.4\linewidth,bbllx=18bp,bblly=144bp,bburx=592bp,bbury=718bp,clip=yes}\\
\psfig{figure=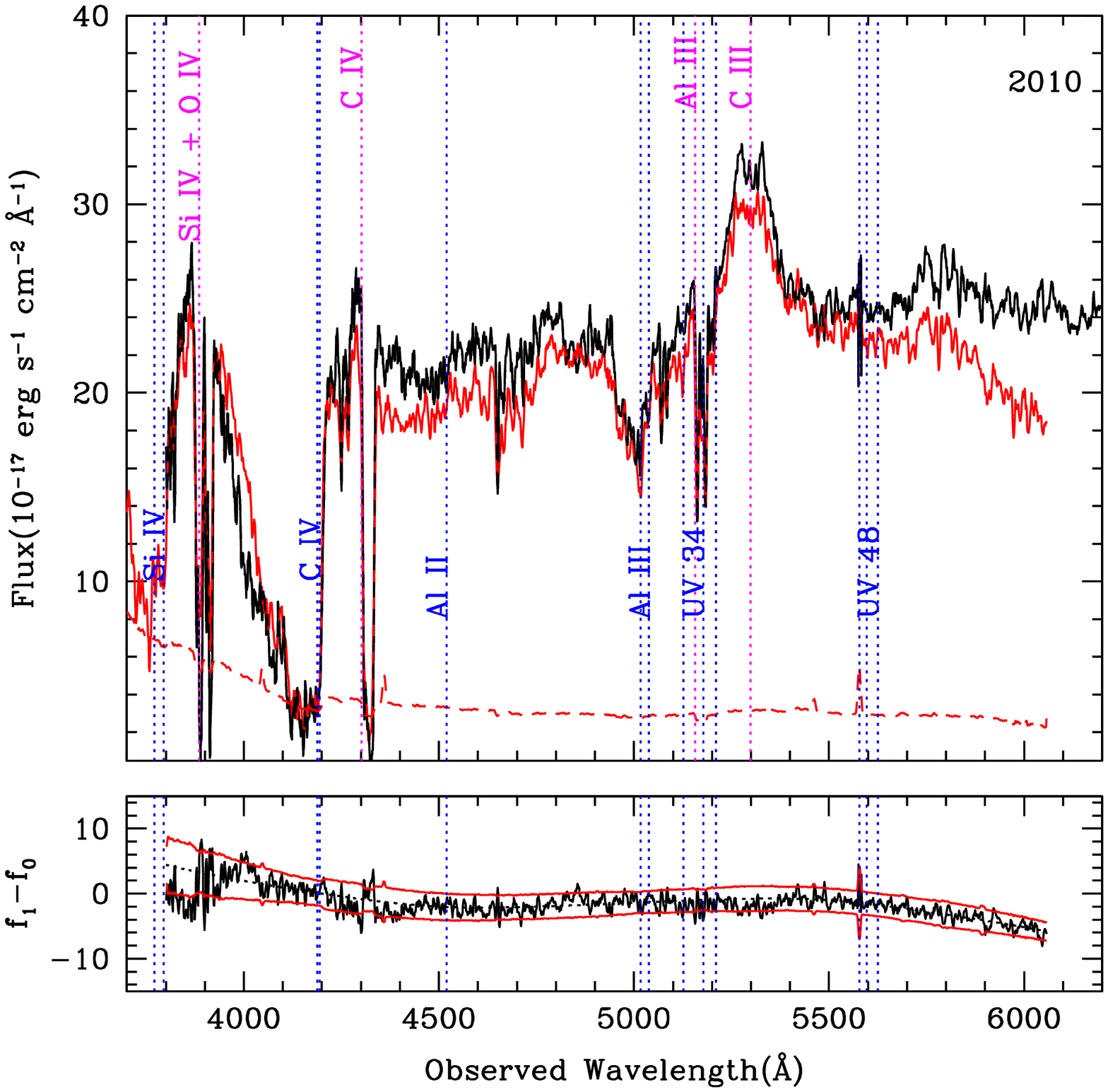,width=0.5\linewidth,height=0.4\linewidth,bbllx=18bp,bblly=144bp,bburx=592bp,bbury=718bp,clip=yes}\\
\end{tabular}	
\caption{IGO Spectra of SDSS J1010+4518  observed on MJD 54474, 54888  and 55220  (in red/grey) are overplotted with the reference SDSS spectrum (black) observed on MJD 52376 . The comparison of two IGO spectra are shown in the lower panel. The flux scale applies to the reference SDSS spectrum and all other spectra are scaled in flux to match the reference spectrum. In each plot, the error spectra are also shown. The difference spectrum for the corresponding MJDs is plotted in the lower panel of each plot. 1$\sigma$ error is plotted above and below the mean. }
\label{1010_diff}
\end{figure*}
\begin{figure*}
 \centering
\begin{tabular}{c c}
\psfig{figure=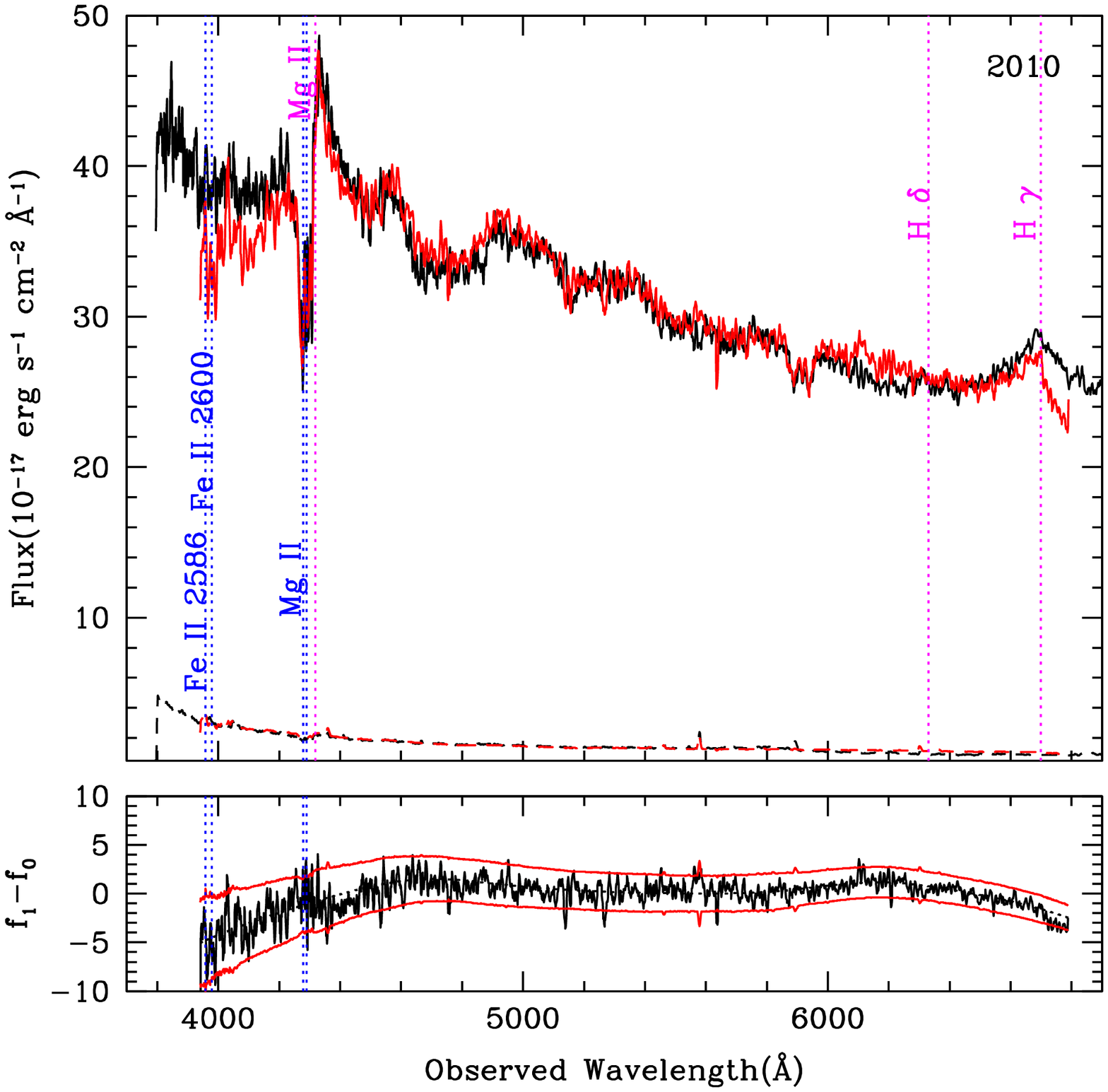,width=0.5\linewidth,height=0.4\linewidth,bbllx=18bp,bblly=144bp,bburx=592bp,bbury=718bp,clip=yes}&
\psfig{figure=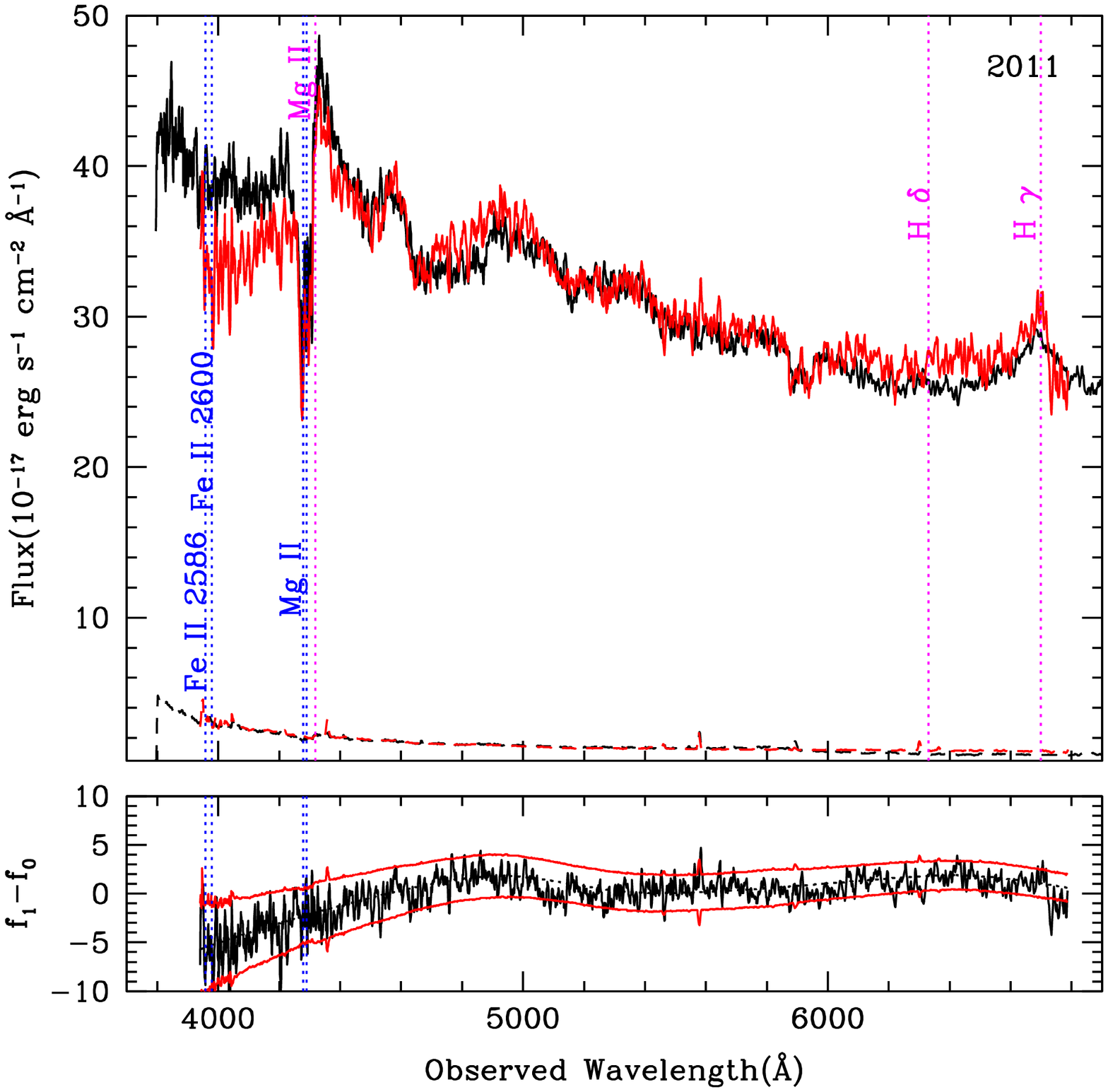,width=0.5\linewidth,height=0.4\linewidth,bbllx=18bp,bblly=144bp,bburx=592bp,bbury=718bp,clip=yes}\\
\end{tabular}	
\caption{IGO Spectra of SDSS J1128$+$4823  observed on MJD 55218  and 55634   (in red/grey) are overplotted with the reference SDSS spectrum (black) observed on MJD 52642 . The comparison of two IGO spectra are shown in the lower panel. The flux scale applies to the reference SDSS spectrum and all other spectra are scaled in flux to match the reference spectrum. In each plot, the error spectra are also shown. The difference spectrum for the corresponding MJDs is plotted in the lower panel of each plot. 1$\sigma$ error is plotted above and below the mean. }
\label{1128_diff}
\end{figure*}
\begin{figure*}
 \centering
\begin{tabular}{c c}
\psfig{figure=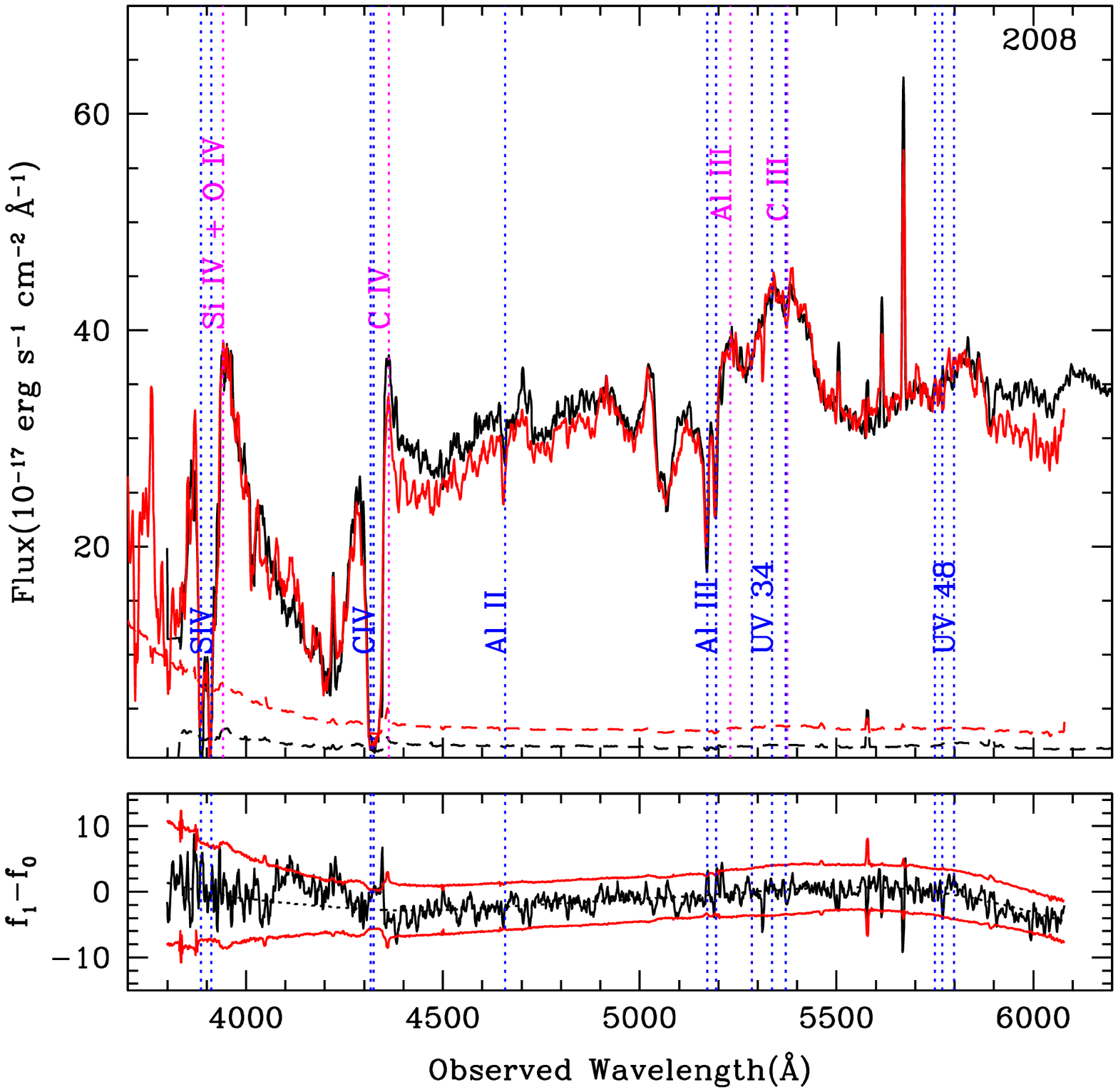,width=0.5\linewidth,height=0.4\linewidth,bbllx=18bp,bblly=144bp,bburx=592bp,bbury=718bp,clip=yes}&
\psfig{figure=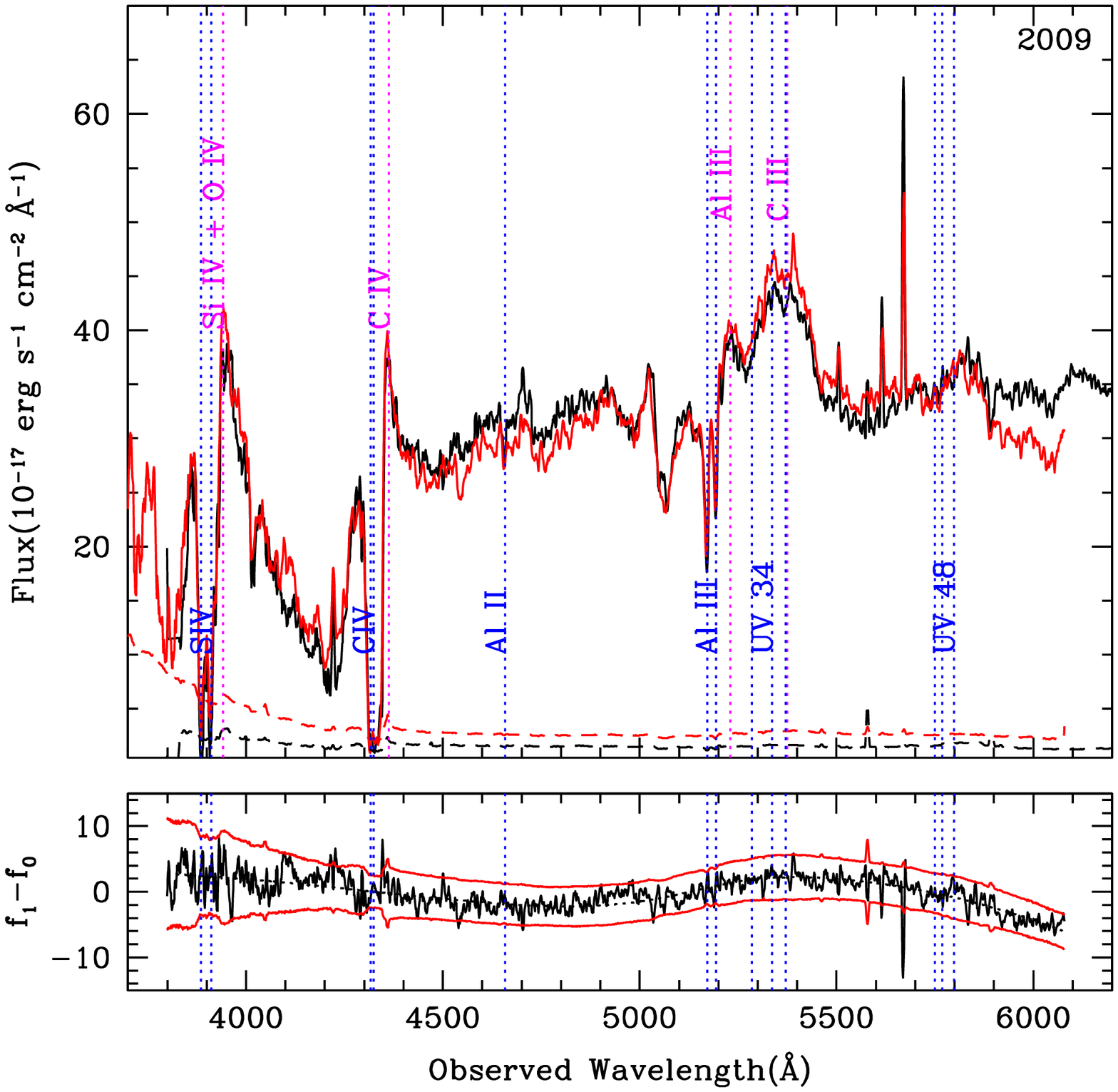,width=0.5\linewidth,height=0.4\linewidth,bbllx=18bp,bblly=144bp,bburx=592bp,bbury=718bp,clip=yes}\\
\end{tabular}	
\caption{IGO Spectra of SDSS J1143+5203  observed on MJD  54473 and 54859  (in red/grey) are overplotted with the reference SDSS spectrum (black) observed on MJD 52368. The comparison of two IGO spectra are shown in the lower panel. The flux scale applies to the reference SDSS spectrum and all other spectra are scaled in flux to match the reference spectrum. In each plot, the error spectra are also shown. The difference spectrum for the corresponding MJDs is plotted in the lower panel of each plot. 1$\sigma$ error is plotted above and below the mean. }
\label{1143_diff}
\end{figure*}

\begin{figure*}
 \centering
\begin{tabular}{c c}
\psfig{figure=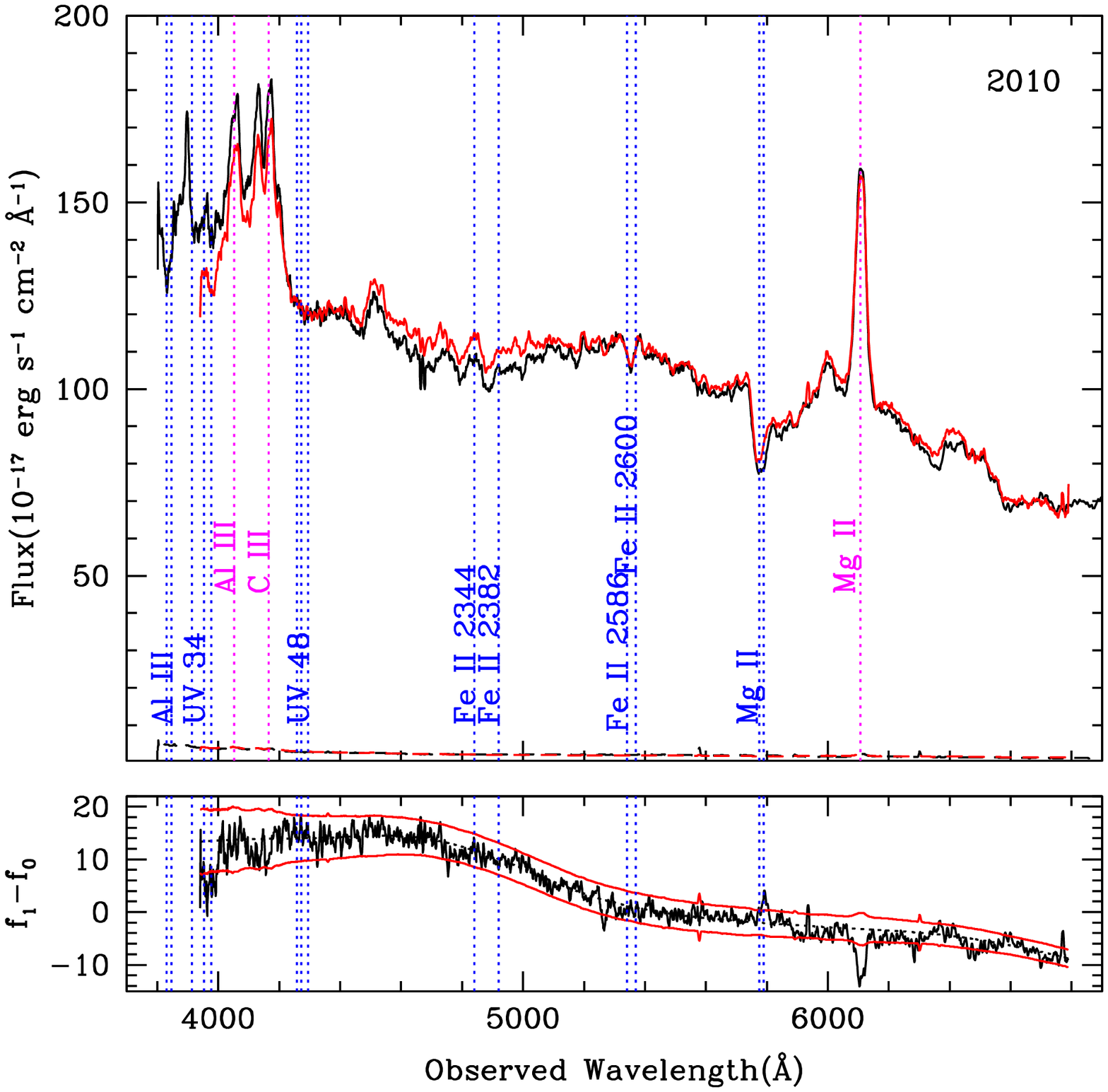,width=0.5\linewidth,height=0.4\linewidth,bbllx=18bp,bblly=144bp,bburx=592bp,bbury=718bp,clip=yes}&
\psfig{figure=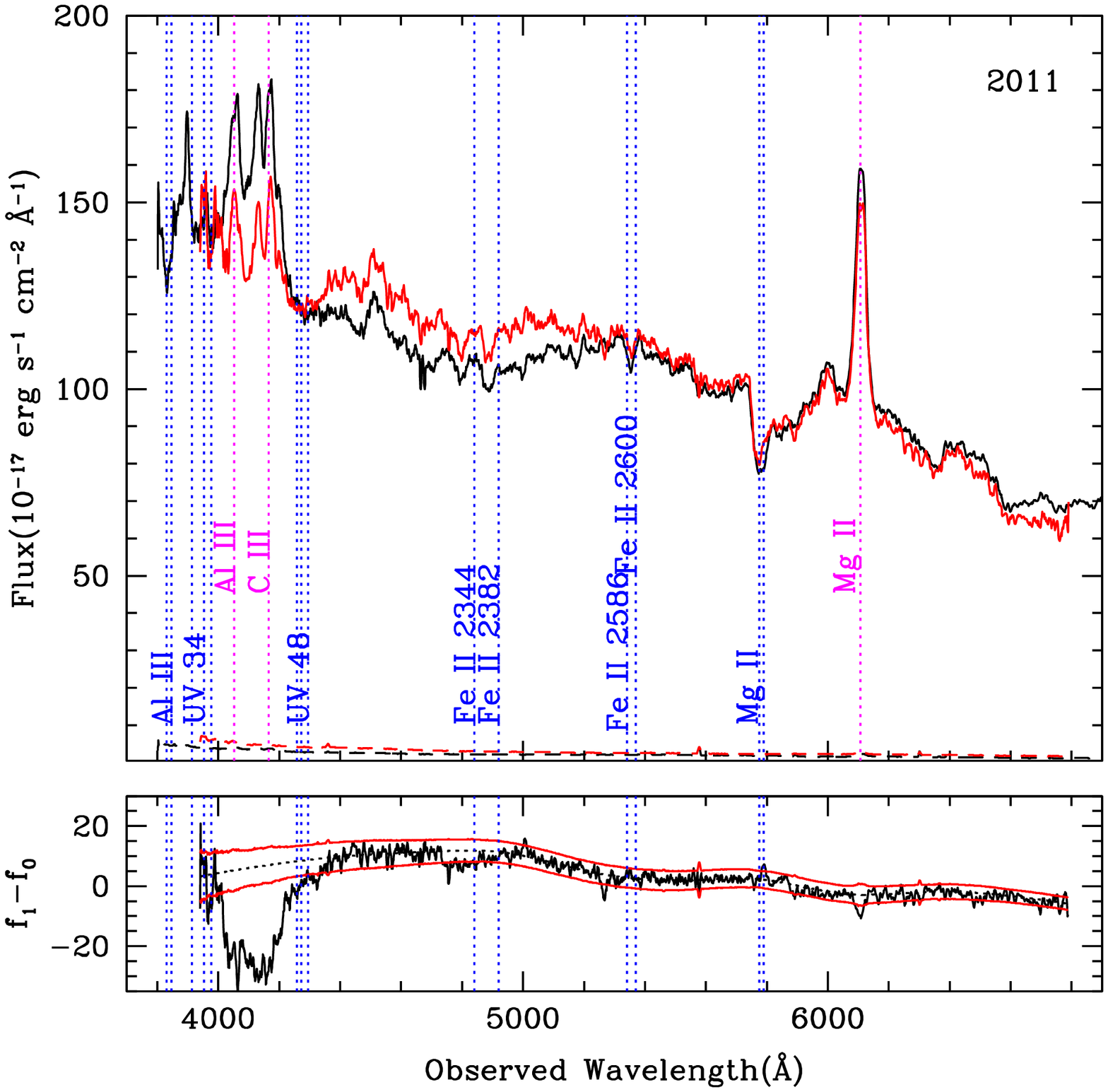,width=0.5\linewidth,height=0.4\linewidth,bbllx=18bp,bblly=144bp,bburx=592bp,bbury=718bp,clip=yes}\\
\end{tabular}
\caption{IGO Spectra of SDSS J1208$+$0230  observed on MJD  55219 and  55633 (in red/grey) are overplotted with the reference SDSS spectrum (black) observed on MJD 52024. The comparison of two IGO spectra are shown in the lower panel. The flux scale applies to the reference SDSS spectrum and all other spectra are scaled in flux to match the reference spectrum. In each plot, the error spectra are also shown. The difference spectrum for the corresponding MJDs is plotted in the lower panel of each plot. 1$\sigma$ error is plotted above and below the mean. }
\label{1208_diff}
\end{figure*}
\begin{figure*}
 \centering
\begin{tabular}{c c}
\psfig{figure=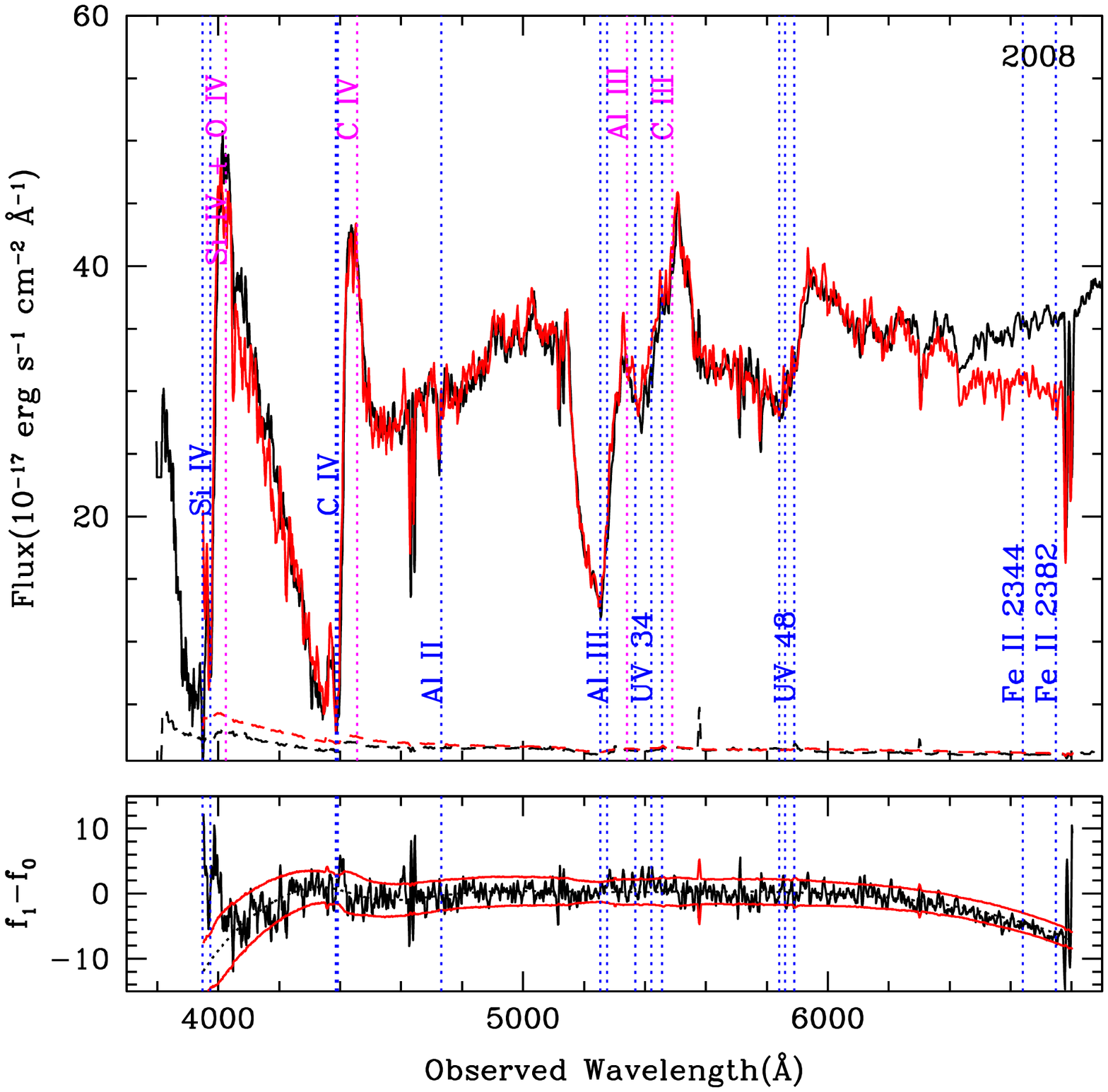,width=0.5\linewidth,height=0.4\linewidth,bbllx=18bp,bblly=144bp,bburx=592bp,bbury=718bp,clip=yes}&
\psfig{figure=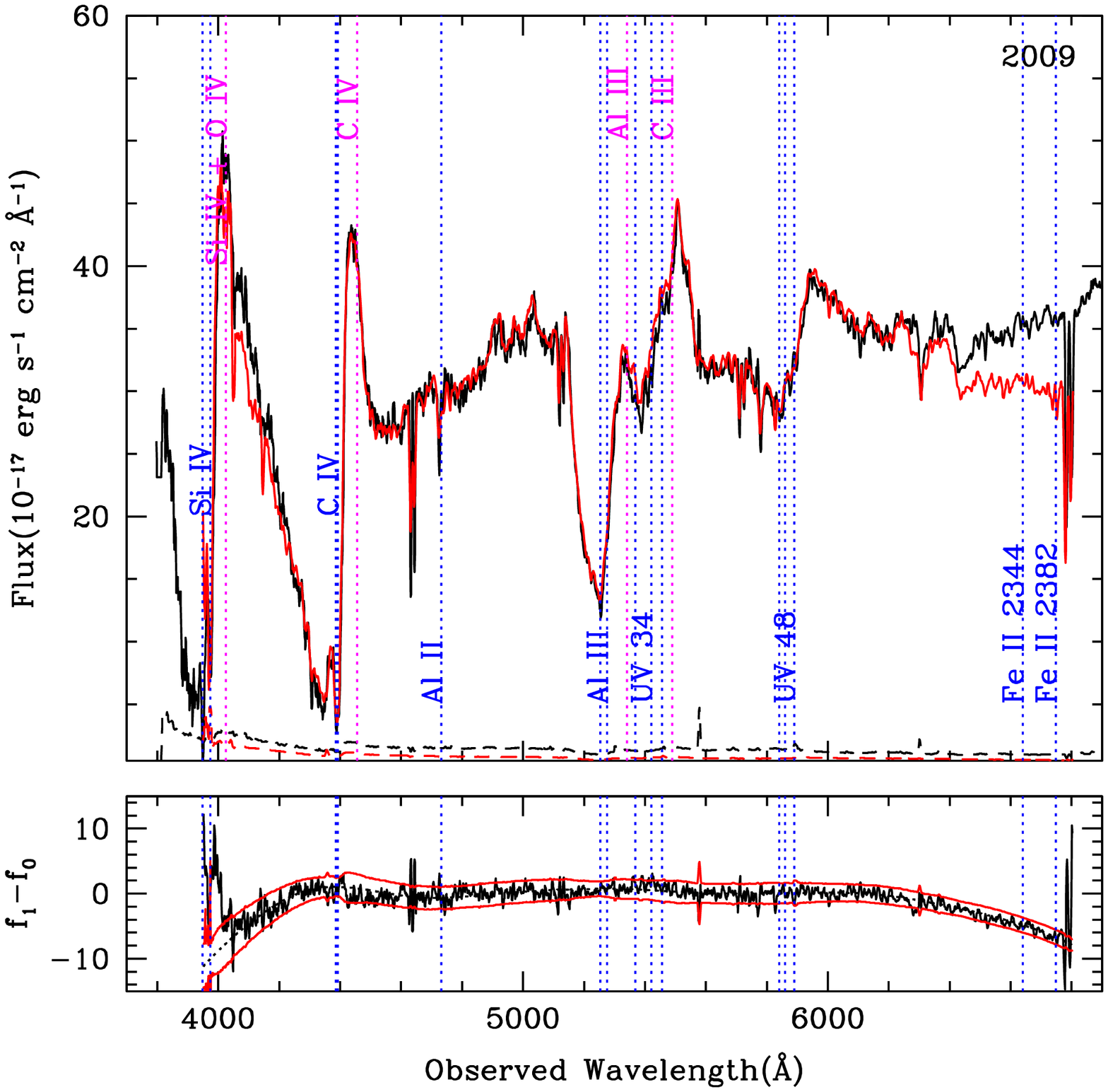,width=0.5\linewidth,height=0.4\linewidth,bbllx=18bp,bblly=144bp,bburx=592bp,bbury=718bp,clip=yes}\\
\psfig{figure=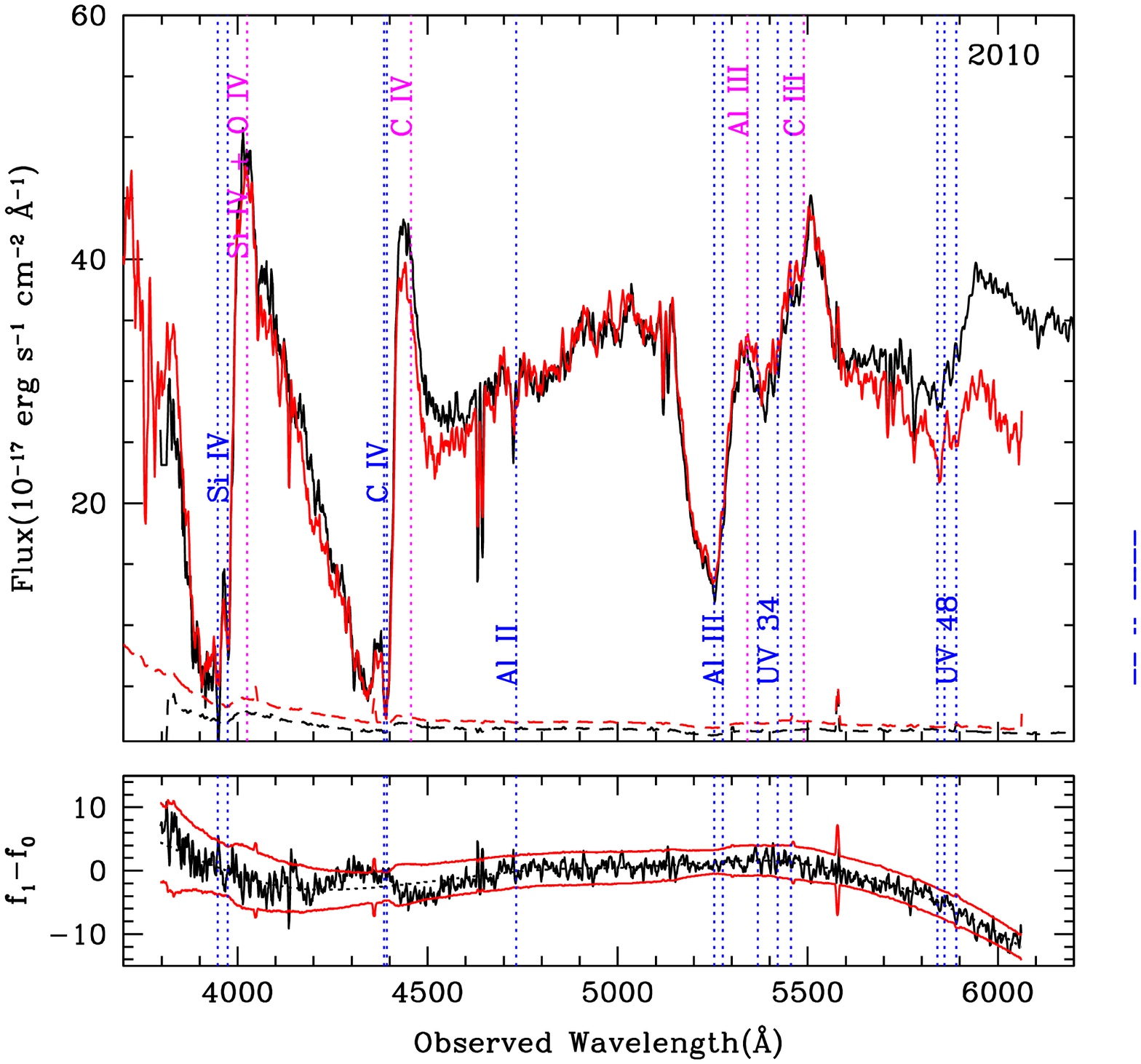,width=0.5\linewidth,height=0.4\linewidth,bbllx=18bp,bblly=144bp,bburx=592bp,bbury=718bp,clip=yes}&
\psfig{figure=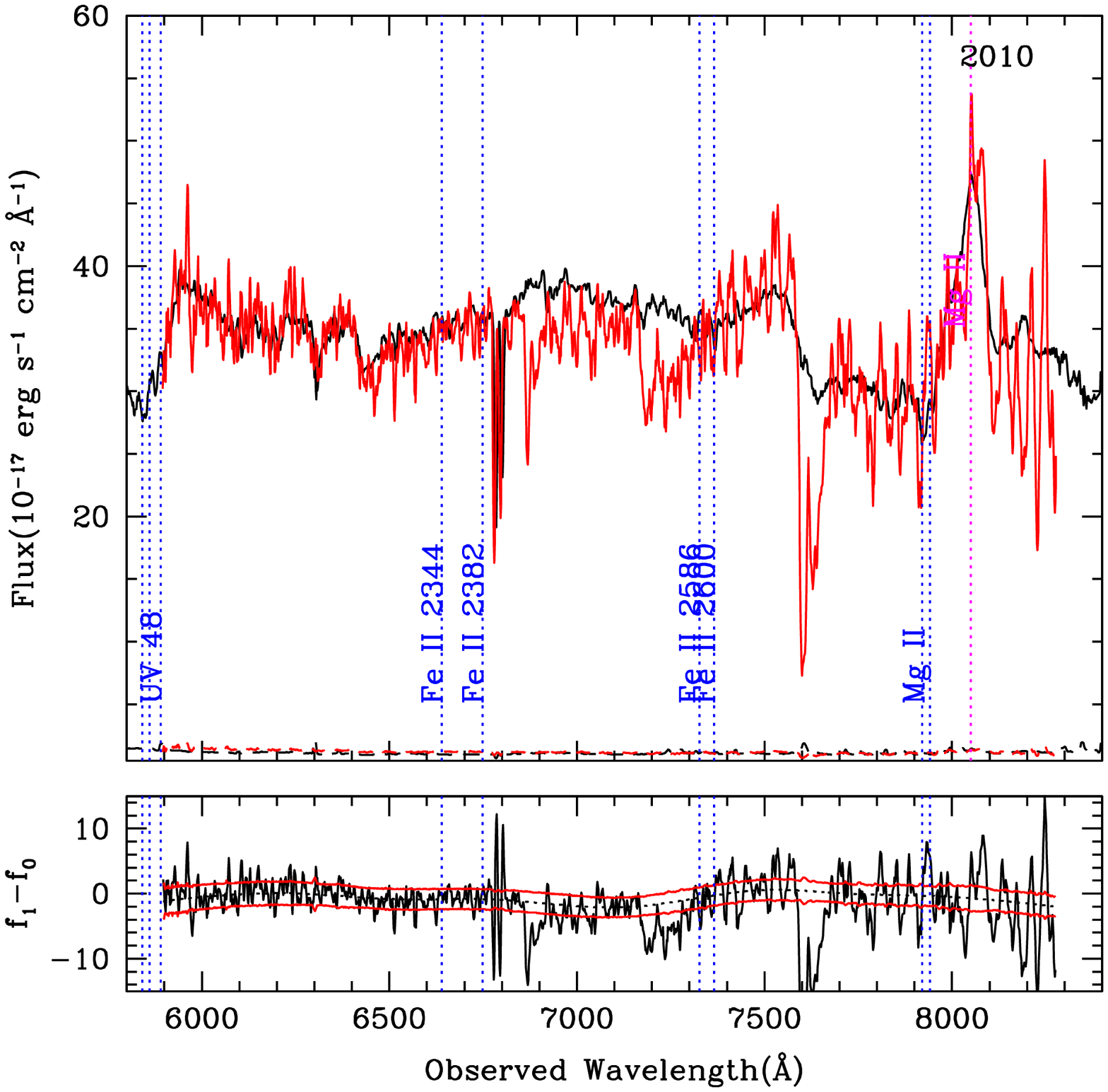,width=0.5\linewidth,height=0.4\linewidth,bbllx=18bp,bblly=144bp,bburx=592bp,bbury=718bp,clip=yes}\\
\end{tabular}
\caption{ Spectra of SDSS J1334-0123  observed on MJD 54560, 54823  and 55272  (in red/grey) are overplotted with the reference SDSS spectrum (black) observed on MJD 52426. The comparison of two IGO spectra are shown in the lower panel. The flux scale applies to the reference SDSS spectrum and all other spectra are scaled in flux to match the reference spectrum. In each plot, the error spectra are also shown. The difference spectrum for the corresponding MJDs is plotted in the lower panel of each plot. 1$\sigma$ error is plotted above and below the mean. }
\label{1334_diff}
\end{figure*}
\begin{figure*}
 \centering
\begin{tabular}{c c}
\psfig{figure=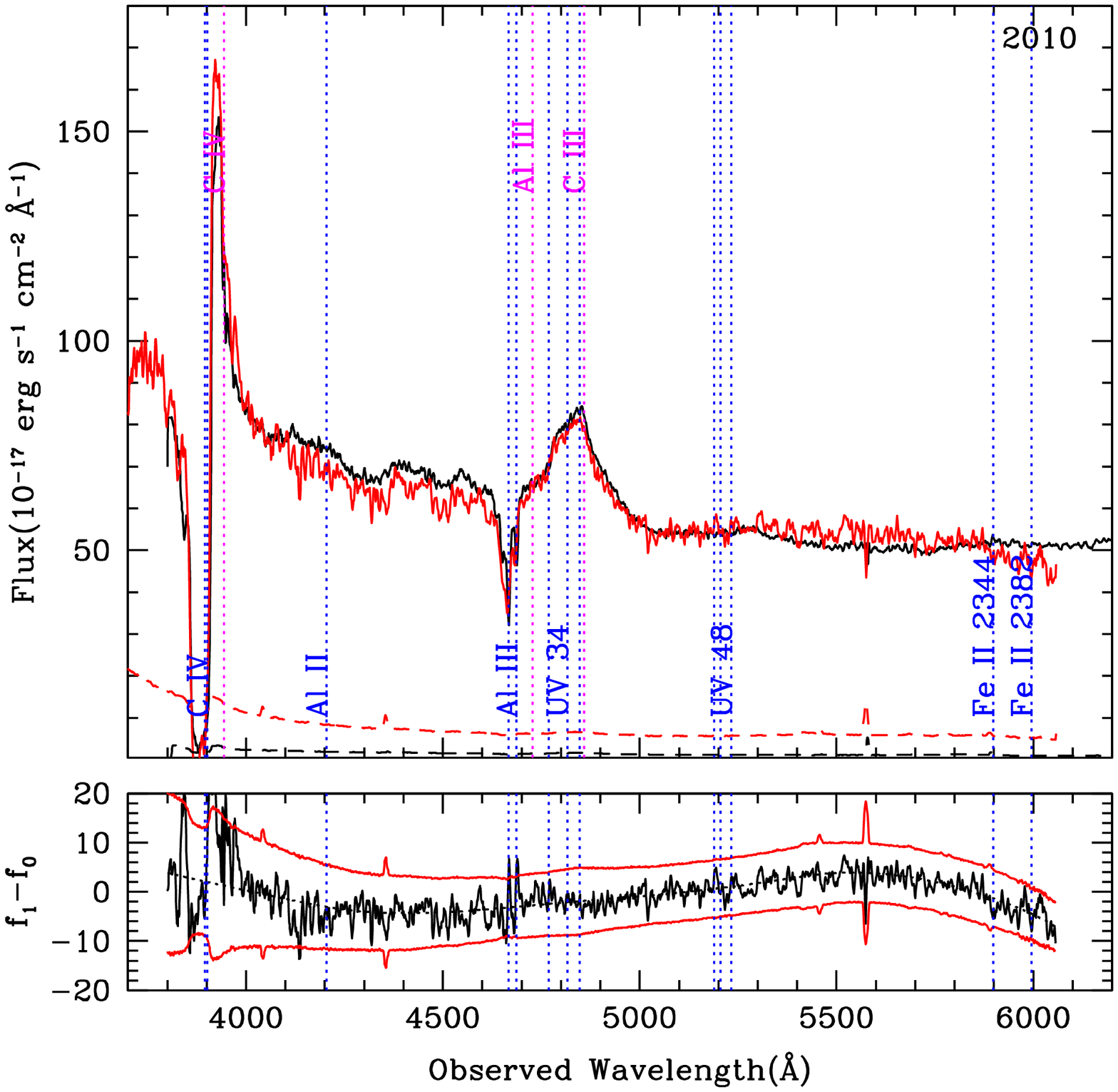,width=0.5\linewidth,height=0.4\linewidth,bbllx=18bp,bblly=144bp,bburx=592bp,bbury=718bp,clip=yes}&
\psfig{figure=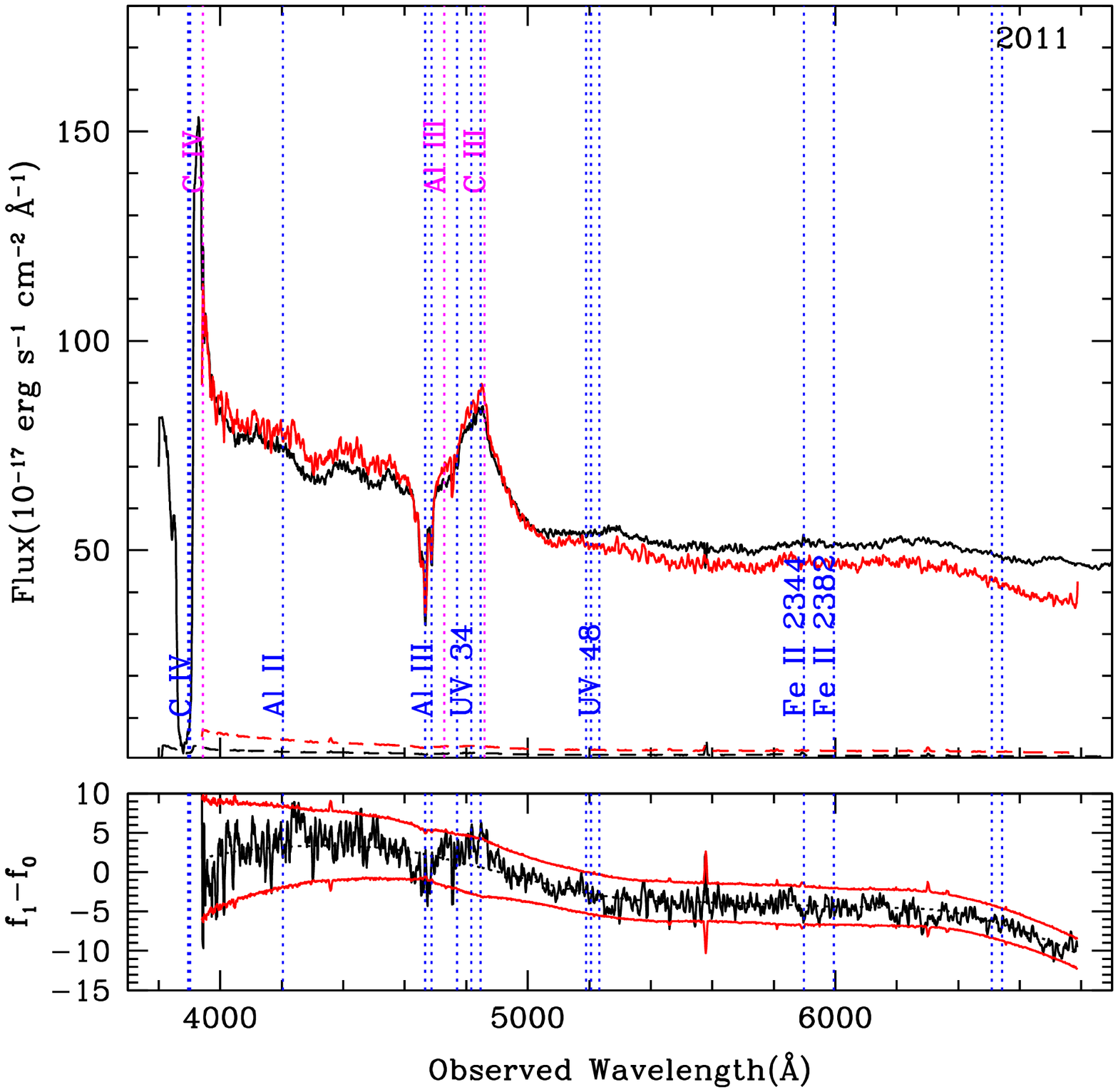,width=0.5\linewidth,height=0.4\linewidth,bbllx=18bp,bblly=144bp,bburx=592bp,bbury=718bp,clip=yes}\\
\psfig{figure=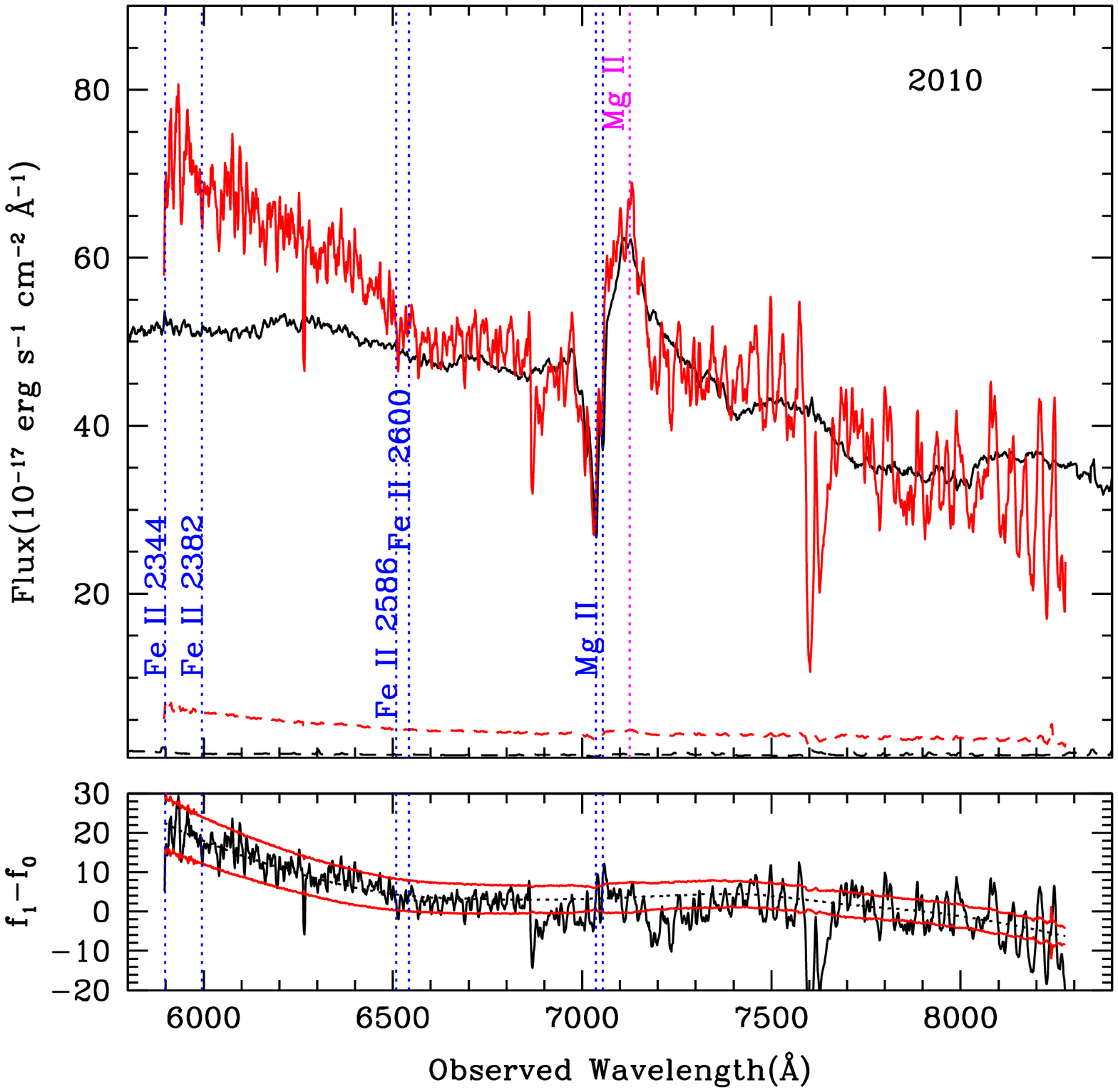,width=0.5\linewidth,height=0.4\linewidth,bbllx=18bp,bblly=144bp,bburx=592bp,bbury=718bp,clip=yes}&\\
\end{tabular}
\caption{IGO Spectra of SDSS J1448+0428  observed on MJD  55287, 55635 and 55277  (in red/grey) are overplotted with the reference SDSS spectrum (black) observed on MJD 52026 . The comparison of two IGO spectra are shown in the lower panel. The flux scale applies to the reference SDSS spectrum and all other spectra are scaled in flux to match the reference spectrum. In each plot, the error spectra are also shown. The difference spectrum for the corresponding MJDs is plotted in the lower panel of each plot. 1$\sigma$ error is plotted above and below the mean. }
\label{1448_diff}
\end{figure*}
\begin{figure*}
 \centering
\begin{tabular}{c c}
\psfig{figure=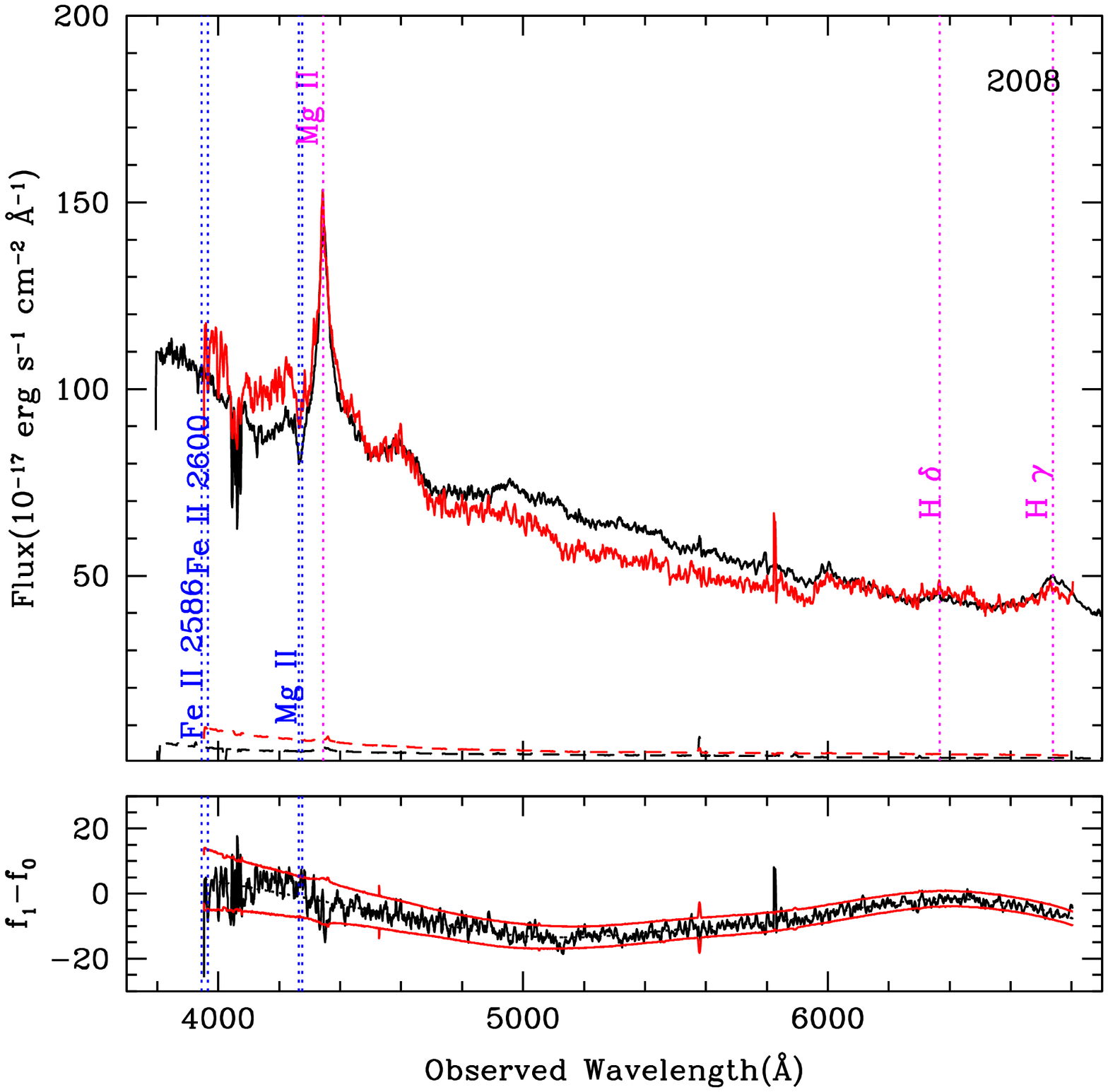,width=0.5\linewidth,height=0.4\linewidth,bbllx=18bp,bblly=144bp,bburx=592bp,bbury=718bp,clip=yes}&
\psfig{figure=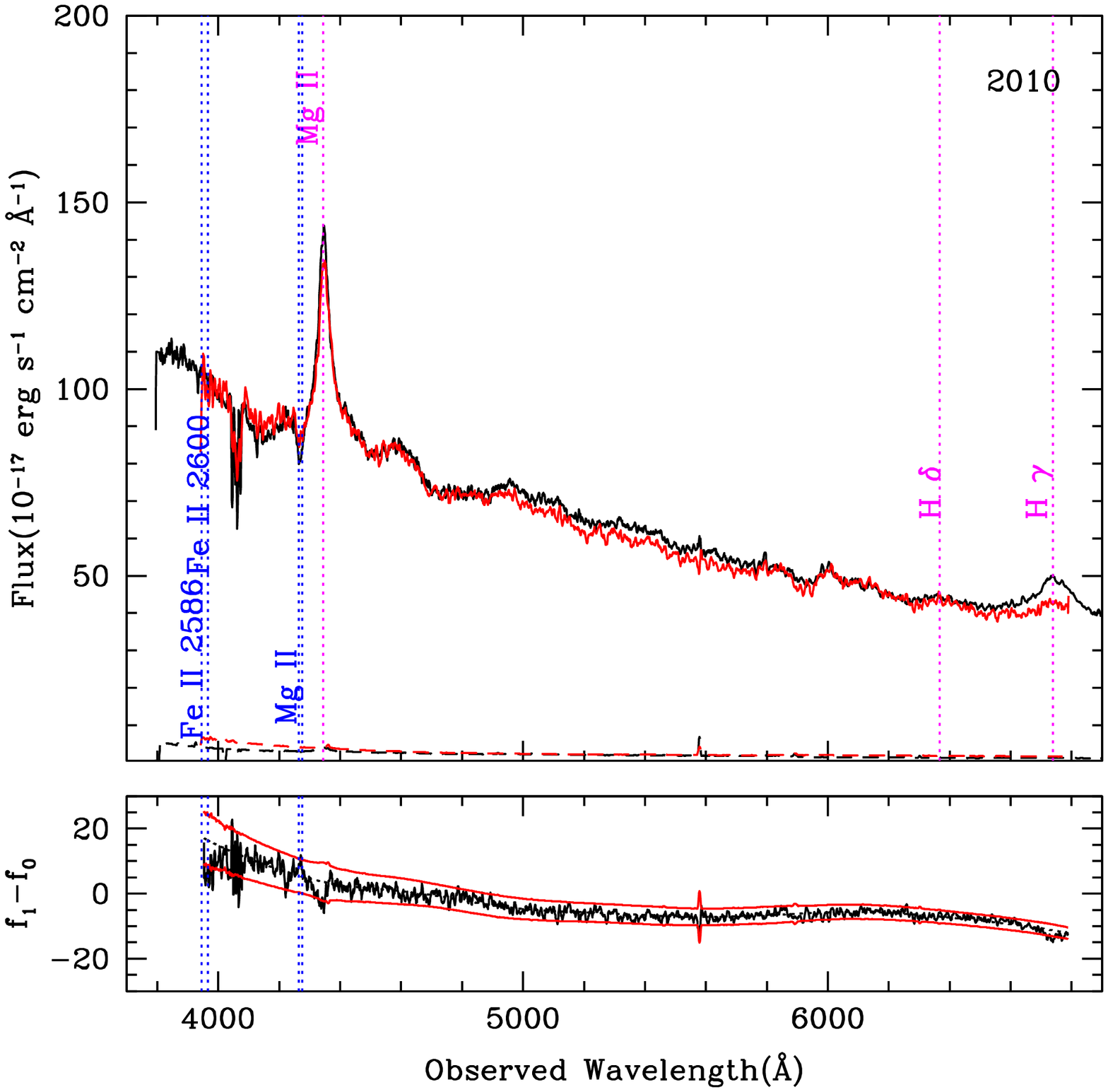,width=0.5\linewidth,height=0.4\linewidth,bbllx=18bp,bblly=144bp,bburx=592bp,bbury=718bp,clip=yes}\\
\psfig{figure=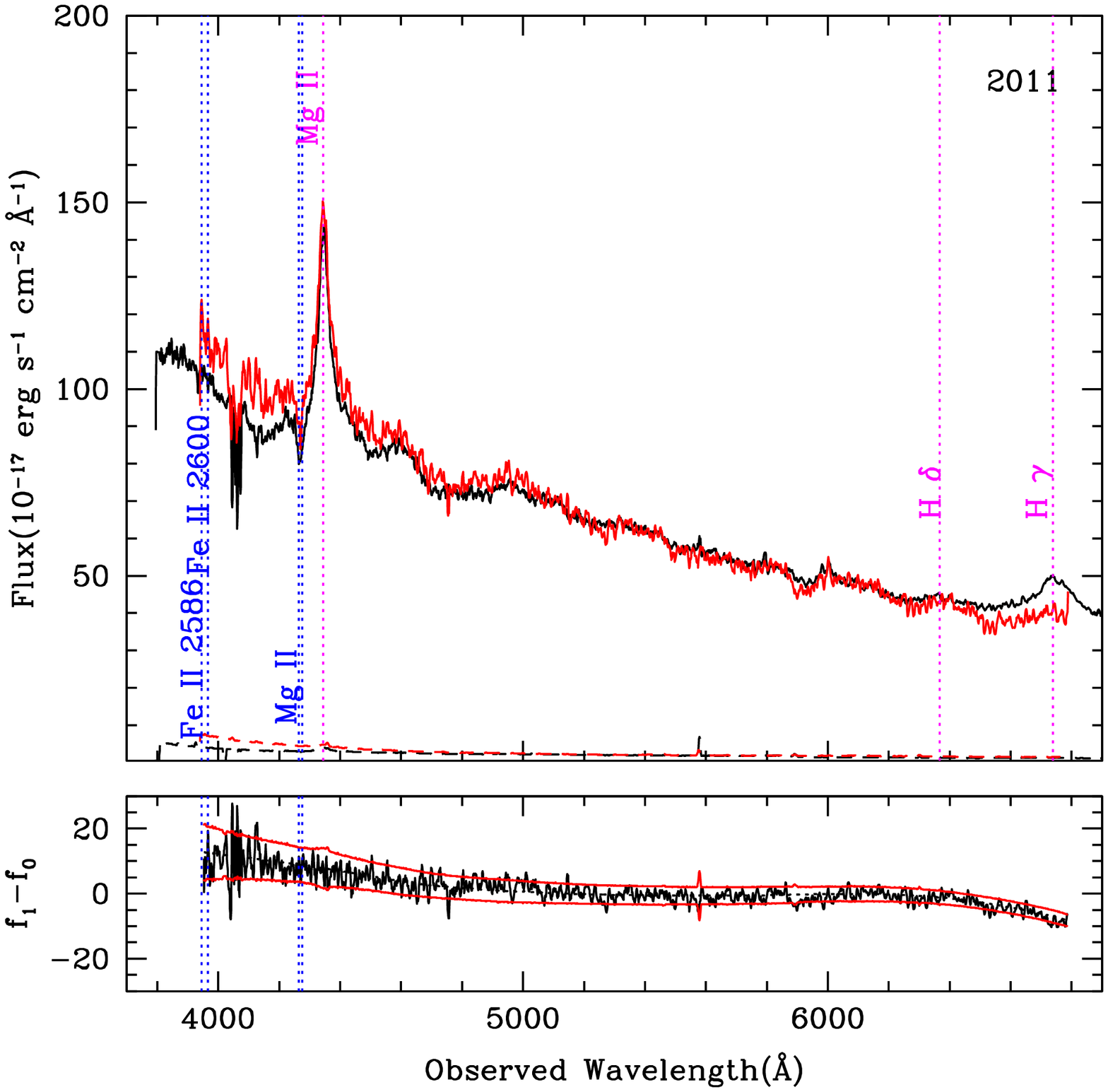,width=0.5\linewidth,height=0.4\linewidth,bbllx=18bp,bblly=144bp,bburx=592bp,bbury=718bp,clip=yes}&\\
\end{tabular}
\caption{IGO Spectra of SDSS J1614$+$3752  observed on MJD 54559, 55297  and 55658  (in red/grey) are overplotted with the reference SDSS spectrum (black) observed on MJD 52764. The comparison of two IGO spectra are shown in the lower panel. The flux scale applies to the reference SDSS spectrum and all other spectra are scaled in flux to match the reference spectrum. In each plot, the error spectra are also shown. The difference spectrum for the corresponding MJDs is plotted in the lower panel of each plot. 1$\sigma$ error is plotted above and below the mean. }
\label{1614_diff}
\end{figure*}
%
\begin{figure*}
 \centering
\begin{tabular}{c c}
\psfig{figure=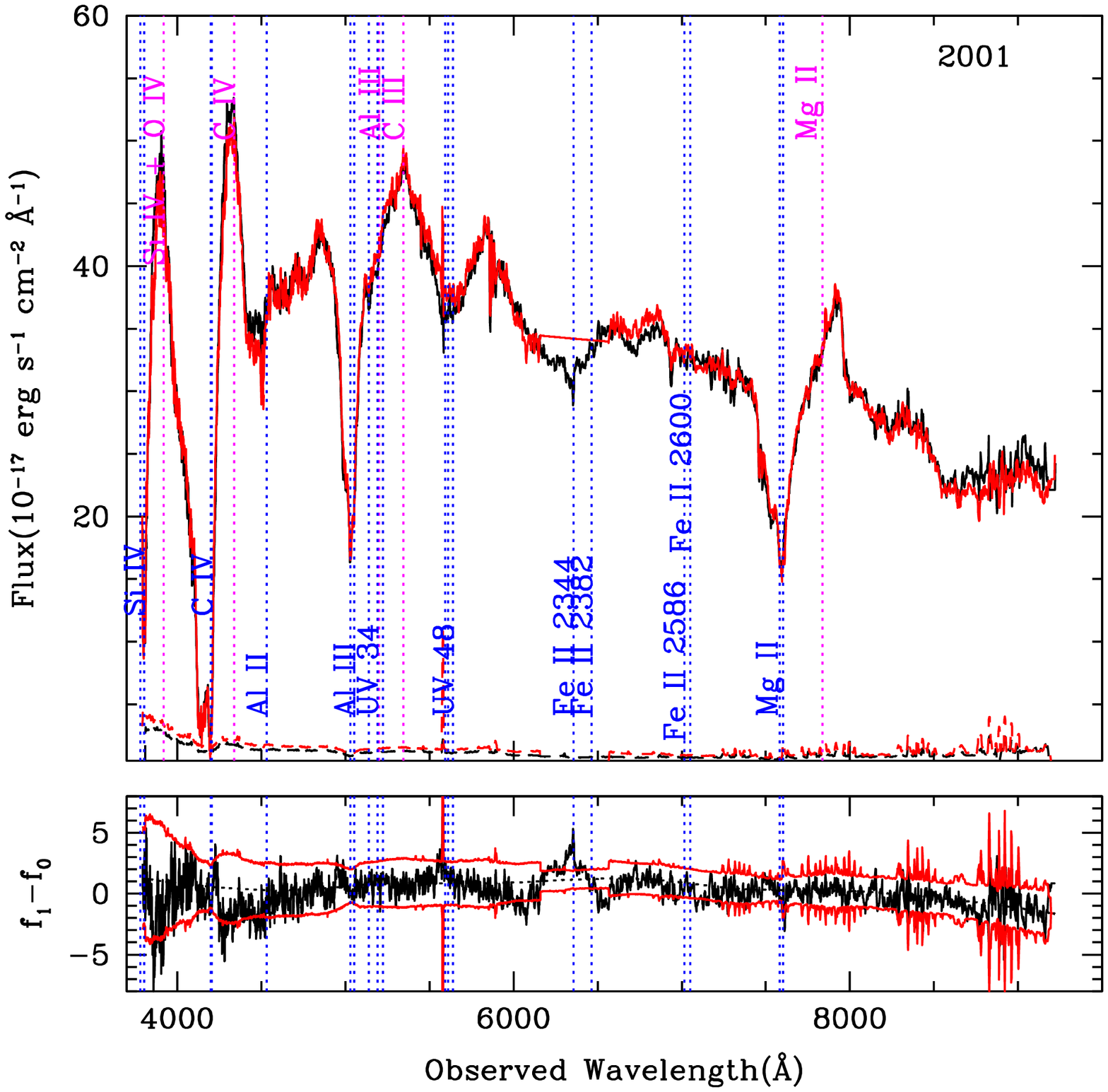,width=0.5\linewidth,height=0.4\linewidth,bbllx=18bp,bblly=144bp,bburx=592bp,bbury=718bp,clip=yes}&
\psfig{figure=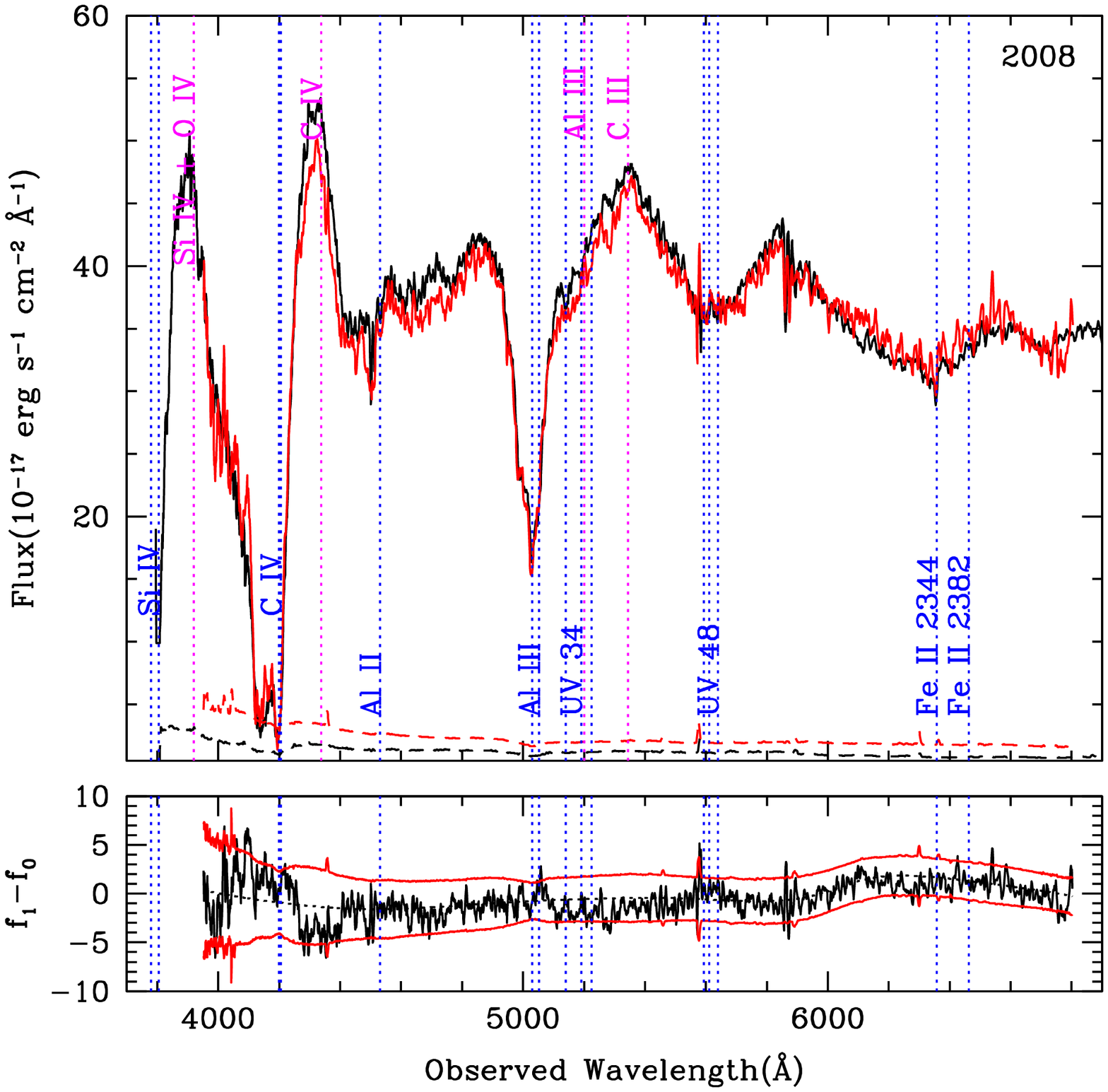,width=0.5\linewidth,height=0.4\linewidth,bbllx=18bp,bblly=144bp,bburx=592bp,bbury=718bp,clip=yes}\\
\psfig{figure=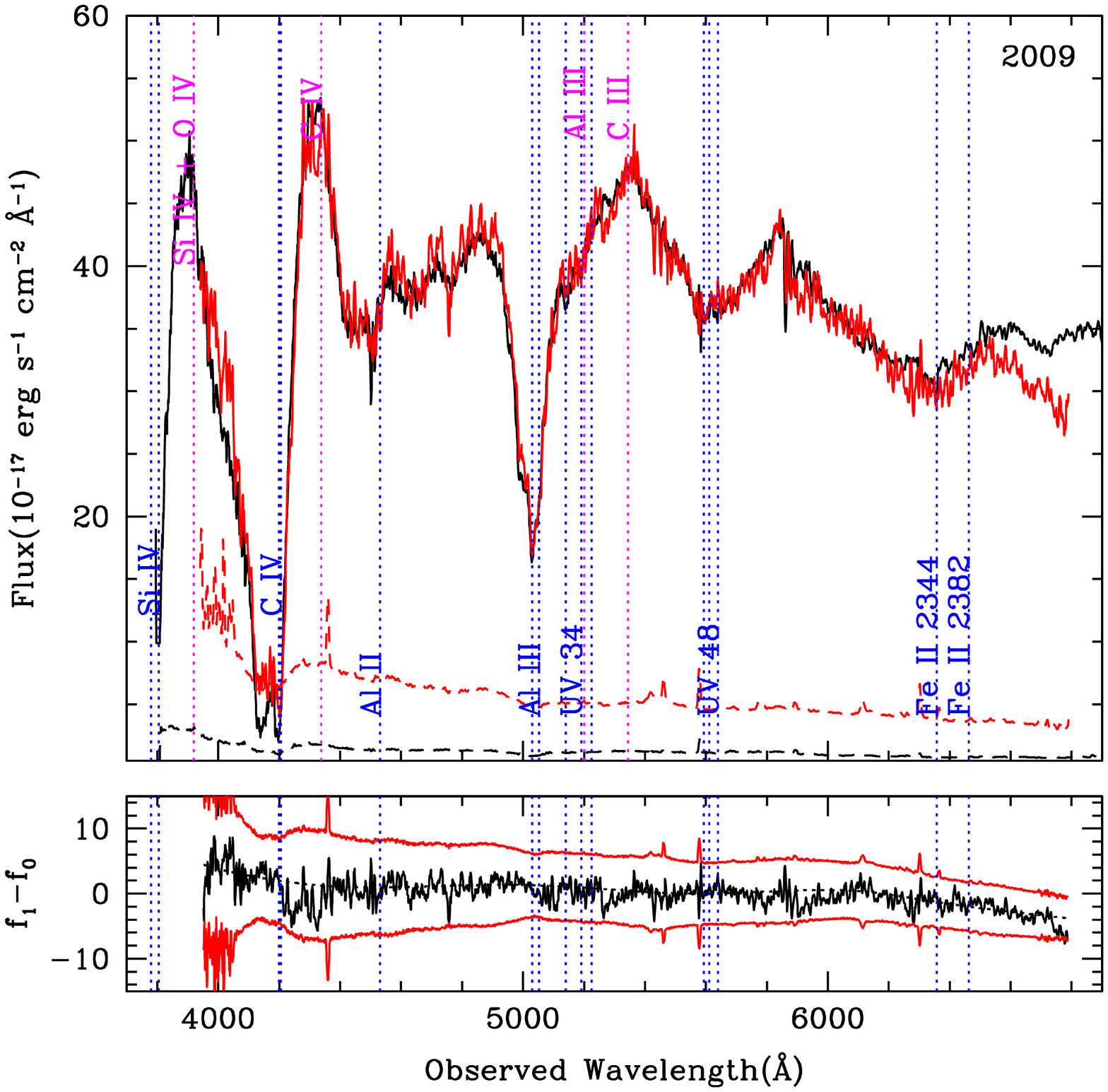,width=0.5\linewidth,height=0.4\linewidth,bbllx=18bp,bblly=144bp,bburx=592bp,bbury=718bp,clip=yes}&
\psfig{figure=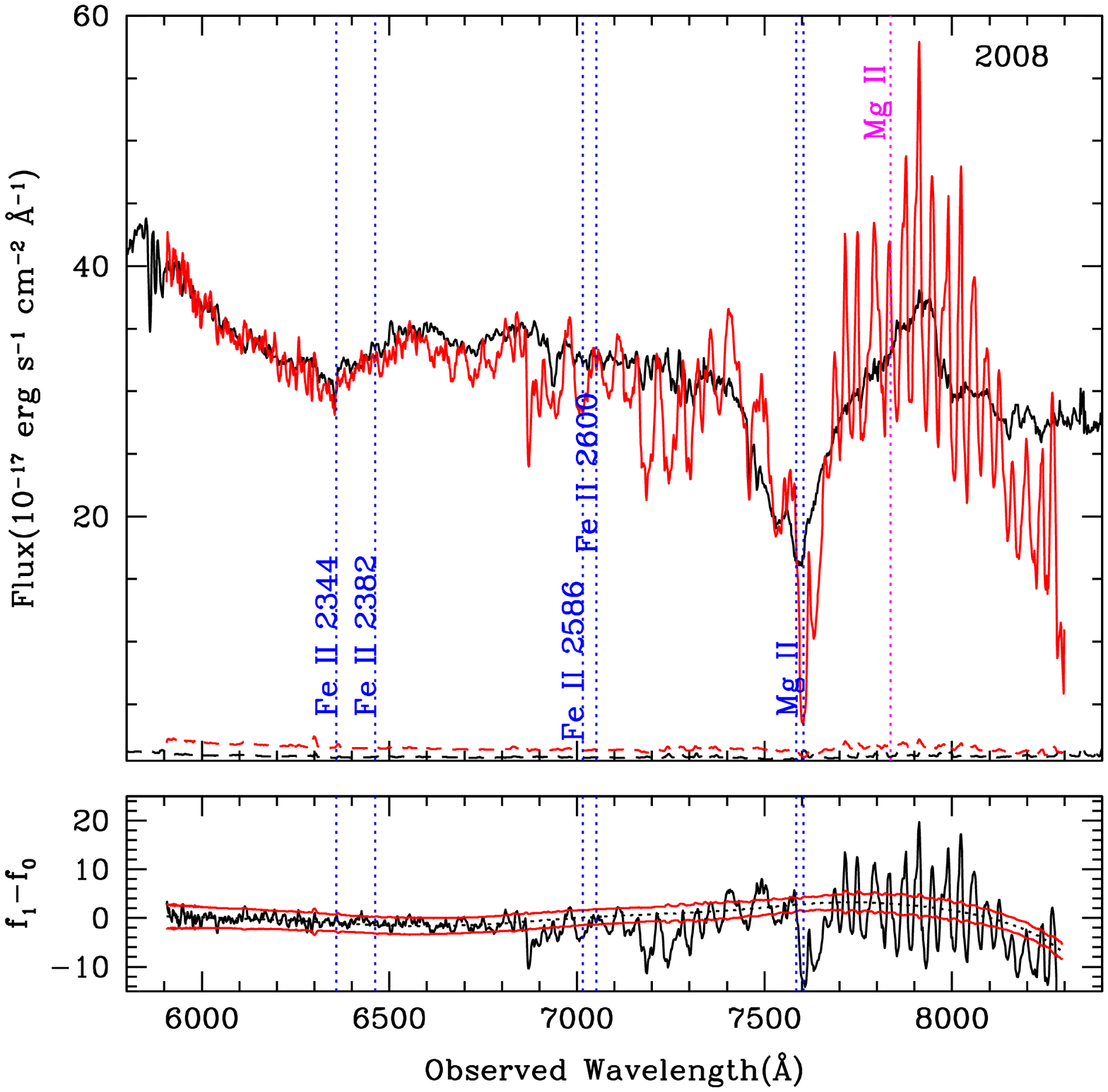,width=0.5\linewidth,height=0.4\linewidth,bbllx=18bp,bblly=144bp,bburx=592bp,bbury=718bp,clip=yes}\\
\end{tabular}
\caption{ Spectra of SDSS J2347-1037  observed on MJD 54331, 54770, 55180  and 54791  (in red/grey) are overplotted with the reference SDSS spectrum (black) observed on MJD 52559 . The comparison of two IGO spectra are shown in the lower panel. The flux scale applies to the reference SDSS spectrum and all other spectra are scaled in flux to match the reference spectrum. In each plot, the error spectra are also shown. The difference spectrum for the corresponding MJDs is plotted in the lower panel of each plot. 1$\sigma$ error is plotted above and below the mean. }
\label{2347_diff}
\end{figure*}

\end{document}